\documentclass[fleqn,usenatbib,useAMS]{mnras}

\usepackage{graphicx}	
\usepackage{amsmath}	
\usepackage{amssymb}	
\usepackage{multicol}        
\usepackage{bm}		
\usepackage{pdflscape}	
\usepackage{lmodern}
\usepackage{hyperref}

\title[Variable Stars in the Open Cluster NGC 6611]{A CCD Search for Variable Stars in the Open Cluster NGC 6611}

\author[G. Michalska et al.]{
G. Michalska,$^{1,2}$\thanks{E-mail: michalska@astro.uni.wroc.pl}
Z. Ko{\l}aczkowski$^{1,2}$,\thanks{Deceased}
R. Leiton,$^{3,4}$
O. Szewczyk$^{2}$,
K. Kinemuchi$^{2}$,
V.\,M. Kalari,$^{5}$
\\
$^1$Uniwersytet Wroc{\l}awski, Instytut Astronomiczny, Kopernika 11, 51-622 Wroc{\l}aw, Poland\\
$^2$Universidad de Concepci\'on, Departamento de Astronom\'ia, Chile\\
$^3$Las Campanas Observatory, Carnegie Institution for Science, Colina el Pino, Casilla 601 La Serena, Chile\\
$^4$CePIA, Departamento de Astronomía, Universidad de Concepción, Casilla 160-C, Concepci\'on, Chile\\
$^5$Gemini Observatory, Southern Operations Center, c/o AURA, Casilla 603, La Serena, Chile\\
}
\date{Accepted XXX. Received YYY; in original form ZZZ}

\pubyear{2015}
\begin{document}
\label{firstpage}
\pagerange{\pageref{firstpage}--\pageref{lastpage}}
\maketitle

\begin{abstract}
We present the results of the $UBVI_C$ variability survey in the young open cluster NGC 6611 based on observations obtained during 34 nights spanning one year. In total, we found 95 variable stars. Most of these stars are classified as periodic and irregular pre-main sequence (PMS) stars. 
The analysis of the $JHK_S$ 2MASS photometry and four-colour IRAC photometry revealed 165 Class II young stellar sources, 20 of which are irregular variables and one is an eclipsing binary.  
These classifications, complemented by $JHK$ UKIDSS photometry and $riH\alpha$ VPHAS photometry, were used to identify 24 candidates for classical T Tauri stars and 30 weak-lined T Tauri stars. 
In addition to the PMS variables, we discovered eight $\delta$ Scuti candidates. None of these were previously known. Furthermore, we detected 17 eclipsing binaries where two were previously known.
Based on the proper motions provided by the Gaia EDR3 catalogue, we calculated the cluster membership probabilities for 91 variable stars. For 61 variables, a probability higher than 80\% was determined, which makes them cluster members. Only 25 variables with a probability less than 20\% were regarded to be non-members.

\end{abstract}

\begin{keywords}
open clusters and associations: individual: NGC 6611 -- stars: pre-main sequence -- stars: variables: T Tauri, Herbig Ae/Be, $\delta$ Scuti stars, eclipsing binaries.
\end{keywords}



\section{Introduction}\label{sintro}

Many stars in very young open clusters are in the pre-main sequence (PMS) stage of evolution. Depending on their mass, PMS stars are divided into two groups: low mass ($<$\,2\,M$_{\sun}$) T Tauri stars (TTSs) with an outer convective zone and intermediate mass ($\sim$2$-$10\,M$_{\sun}$) Herbig Ae/Be stars (HAeBe). Both groups of PMS stars shows different types of photometric variability. 

As a class of variable stars, TTSs stars were first defined by \cite{1945ApJ...102..168J}, who selected a group of F5--G5 stars with fast irregular light variations.  These variations are indicated by emission lines resembling the solar chromosphere and/or are associated with a dark or bright nebulosity. 
Nowadays, such stars are characterized by a strong emission in the H and K Ca II and H$\alpha$ lines (with equivalent widths EW$_{H\alpha}$\,$>$\,10 \AA) and are known as classical TTSs (CTTSs). 
CTTSs are surrounded by a thick disk detected by infrared excess \citep{1987ApJ...323..714K}. 

The magnetic field generated by a solar-type dynamo plays an important role in the variability of CTTSs. A strong magnetic field interacts with the inner disk, truncates it at a few stellar radii, and channels the disk matter onto the stellar surface. The infalling gas produces hot spots and shocks. Dissipated X-ray and EUV radiation interact with the surrounding gas and cause emission at longer wavelengths, which gives the UV and optical continuum excesses. This process, known as magnetospheric accretion, has been modelled for many years \citep[see][and references therein]{1991ApJ...370L..39K,1994ApJ...429..781S,1998ApJ...492..743M,2007prpl.conf..479B,2012MNRAS.421...63R}. Unsteady accretion from the circumstellar disk causes irregular light variation of CTTSs with amplitudes up to 2.6 mag in $V$ \citep{herbst1994}. Non-periodic variability of PMS stars may be caused also by stellar flares, which are characterised by a fast rise and an exponential decay of the light lasting few minuts or hours (\citealt{2014ApJ...797..121H, 2018ApJ...862...44K}).  Additionally, the light curve of CTTSs can be modulated by rotating hot and cool spots. The periodic changes in the light curves also may be due to circumstellar matter surrounding CTTS.

Other TTSs characterized by a weak emission in the H$\alpha$ line (with EW$_{H\alpha}$\,$<$\,10 \AA) are called naked T Tauri stars \citep{1986ApJ...306..573W} or weak line TTSs \citep[WTTSs;][]{1989ARA&A..27..351B}.
They were first detected in X-ray \citep{1981ApJ...250..254W,1981ApJ...243L..89F}. For many years, these stars were thought to be more evolved than CTTSs \citep{1988AJ.....96..297W}. Some authors \citep[e.g.][]{2003Ap.....46..506P} postulated that the existence of an accretion disk depends on initial conditions at the star formation phase. \cite{2007A&A...473L..21B} analysed the mass and age distribution of TTSs in the Taurus-Auriga T association and inferred that younger CTTSs evolve into WTTSs when their disks are fully accreted by stars. \cite{2015A&A...580A..26G} came to the same conclusion after analysing TTSs in the Lupus association.
In general, WTTSs do not show the evidence of accretion. The small infrared excesses seen in some WTTSs come from the cold passive disk. The light variations of WTTSs are attributed to rotational modulation of cool spots with typical periods of 0.5-18 d and amplitudes less then 0.8 mag in $V$ \citep{herbst1994}.


The more massive counterparts of TTSs, HAeBe stars, were first defined by \citet{herbig1960}. These B, A or F-type stars with emission lines are often associated with nebulosity. They show an infrared excess due to circumstellar material (gas and dust).
Many HAeBe stars show photometric variability related to accretion processes and dust clumps in the surrounding material. \cite{1994A&AS..104..315T} 
found that HAeBe stars of spectral type earlier than A0 show small light variations ($\Delta V \leq 0.2$ mag), while those of later spectral types show larger variability ($0.65 \leq \Delta V \leq 3.0$ mag). This may be related to the different properties of Herbig Ae and Herbig Be stars. Many studies \citep[e.g.][]{2002MNRAS.337..356V,2011A&A...535A..99M,2017MNRAS.472..854A} have shown that Herbig Ae stars exhibit accretion features similar to those seen in CTTSs. These findings have been interpreted as evidence that magnetically controlled accretion also occurs in Herbig Ae stars.
In more massive Herbig Be stars, material from the disk is accreted directly onto the star through a so-called boundary layer \citep[see][and references therein]{2020Galax...8...39M}. 
Some Herbig Ae stars and early type TTSs show irregular, deep (1-3 mag in $V$) Algol-like drops in brightness. This group of stars is called UX~Orionis stars (UXors), after the prototype \citep{herbst1994}.

Other interesting types of PMS variable types are eruptive FU Orionis \citep[FUors;][]{herbig1977} and EX Lupi \citep[EXors;][]{herbig1989} stars. These objects show intermittent variability due to an unsteady rate of mass accretion. In the case of FUors, a change of brightness of $\sim$3\,--\,6~mag lasts several months, or even years, while EXors repeatedly increase their brightness by $\sim$1\,--\,4~mag, and return to their original brightness after several weeks or months. 

In this paper, we present the results of a variability survey in the young open cluster NGC 6611.
It is one of the youngest open clusters known in the Galaxy with estimated age of less than 3 Myr \citep[e.g.][]{2007A&A...462..245G}. 
The distance to the NGC 6611 was estimated in the relatively wide range between 1.6 kpc \citep{1974A&AS...16....1K} and 2.14 kpc \citep{2004MNRAS.353..991K}.
The difficulty in determining the distance accurately is related to the variable and anomalous reddening towards this cluster \citep{1993AJ....106.1906H,2007A&A...462..245G}.
The ratio of total extinction ($A_V$) to reddening ($E(B-V)$) was estimated between 3.27 \citep{2007A&A...462..245G} and 4.37 \citep{2021MNRAS.504.3203S}.
Recently, using the data of Gaia DR2, \cite{2019ApJ...870...32K} calculated the mean cluster parallax of 0.57$\pm$0.04 mas, corresponding to a distance of $1740^{+130}_{-120}$ pc. The newest distance determination of $1706\pm7$ pc, calculated from the Gaia EDR3 data, was published by \citet{2022arXiv220708452S}. The authors reported of two stellar populations in NGC 6611 -- the younger population has an age of 1.3 Myr and a $A_V=3.5$ mag, while the older has an age 7.5 Myr and $A_V=2.0$ mag.

A dozen O-type and 50 early B-type main-sequence stars have been identified among the brightest members. Recent studies have revealed several Be and Ae stars, some of which are Herbig Ae/Be stars \citep{2001PASP..113..195H,2007sf2a.conf..518M}.
A large fraction of the bright stars in NGC\,6611 are known visual and spectroscopic binaries \citep{2001A&A...379..147D,2005A&A...437..467E}.
The cluster contains a significant population of intermediate and low-mass PMS stars \citep{1993AJ....106.1906H,1995ApJ...454..151M,2005MNRAS.358L..21O}, which is typical for such a young and massive region. Moreover, it is associated with the famous H\,II and star-forming region, the Eagle Nebula.  
This region has been the subject of intensive spectroscopic \citep{1993AJ....106.1906H,2005A&A...437..467E}, multiwavelength optical \citep{2000A&A...358..886B} and infrared \citep{2005MNRAS.358L..21O,2007A&A...462..245G} surveys. No systematic study of the cluster photometric variability has ever been performed. 
The rich population of the O and B-type stars makes the cluster a very promising target for variability searches among young and massive stars. The immediate goal of the present work is to detect and characterize interesting objects showing photometric variations, especially on the short time-scale.

The paper is organized as follows. First, we described the photometric observations and transformation to the standard system (Sect.~\ref{ssubvi}). Next, the calculation of membership probabilities are presented (Sect.~\ref{smem}). In Sect.~\ref{svar} we provide a review of variable stars in NGC 6611, and in the following Sect.~\ref{syso}, we describe the supplementary photometry\footnote{The supplementary photometry for variable stars is available in electronic form.} used to the identification and classification of PMS stars. Finally, a brief summary and conclusions are given (Sect.~\ref{ssum}).

\begin{table*}
\centering
\caption{The coefficients of the photometric transformations with Eq.~\ref{eq_V}, \ref{eq_BV}, and \ref{eq_VI} determined from photometry of \protect\cite{2007A&A...462..245G}. The numbers in parentheses are the rms errors of
the transformation coefficients with leading zeros omitted.}
\label{ttranf1}
\footnotesize
\begin{tabular}{lrrrrrrrrrrrrrr}
\hline\noalign{\vskip1pt}
Telescope& \multicolumn{1}{c}{$a_1$} & \multicolumn{1}{c}{$b_1$}& \multicolumn{1}{c}{$\sigma_1$}& $N_1$& & \multicolumn{1}{c}{$a_3$} & \multicolumn{1}{c}{$b_3$}& \multicolumn{1}{c}{$\sigma_3$}& $N_3$ & &\multicolumn{1}{c}{$a_4$} & \multicolumn{1}{c}{$b_4$}& \multicolumn{1}{c}{$\sigma_4$}& $N_4$ \\
\noalign{\vskip1pt}\hline\noalign{\vskip1pt}
SWOPE & $-$0.058(2) &10.850(1) & 0.029 &920 &&1.220(10) &0.713(4) &0.080 &725& &0.918(3) & 1.199(1) &0.036 &920\\
CT-1m   &0.052(6) &10.929(3) &0.042 &718 &&0.880(7)\enspace &0.663(4) &0.053 &718& &\multicolumn{1}{c}{--} &\multicolumn{1}{c}{--} &\multicolumn{1}{c}{--} & \multicolumn{1}{c}{--}\\
CT-0.9m &0.056(5) &10.941(2) &0.043 &573 &&0.966(7)\enspace &0.685(4) & 0.039&389 && 1.054(4) &1.213(2) &0.039 &573\\
\hline\noalign{\vskip1pt}
\end{tabular}
\end{table*}

\begin{table*}
\centering
\caption{The coefficients of the photometric transformations with Eq.~\ref{eq_Vb}, \ref{eq_BV}, and \ref{eq_UB} determined from photometry of \protect\cite{1993AJ....106.1906H}.}
\label{ttranf2}
\footnotesize
\begin{tabular}{lrrrrrrrrrrrrrr}
\hline\noalign{\vskip1pt}
Telescope& \multicolumn{1}{c}{$a_2$} & \multicolumn{1}{c}{$b_2$}& \multicolumn{1}{c}{$\sigma_2$}& $N_2$& & \multicolumn{1}{c}{$a_3$} & \multicolumn{1}{c}{$b_3$}& \multicolumn{1}{c}{$\sigma_3$}& $N_3$ && \multicolumn{1}{c}{$a_5$} & \multicolumn{1}{c}{$b_5$}& \multicolumn{1}{c}{$\sigma_5$}& $N_5$ \\
\noalign{\vskip1pt}\hline\noalign{\vskip1pt}
CT-0.9m &0.044(11) &10.967(4)&0.047&88 &&0.938(10) &0.588(3)&0.033 &99&&1.143(10) &$-$0.479(8)&0.045&63\\
\hline\noalign{\vskip1pt}
\end{tabular}
\end{table*}

\section{\texorpdfstring{$\bmath{UBVI_{\rm C}}$ photometry }{}}\label{ssubvi}

\subsection{Observations and Reductions}\label{sobs}
\begin{figure}
\includegraphics[width=84mm]{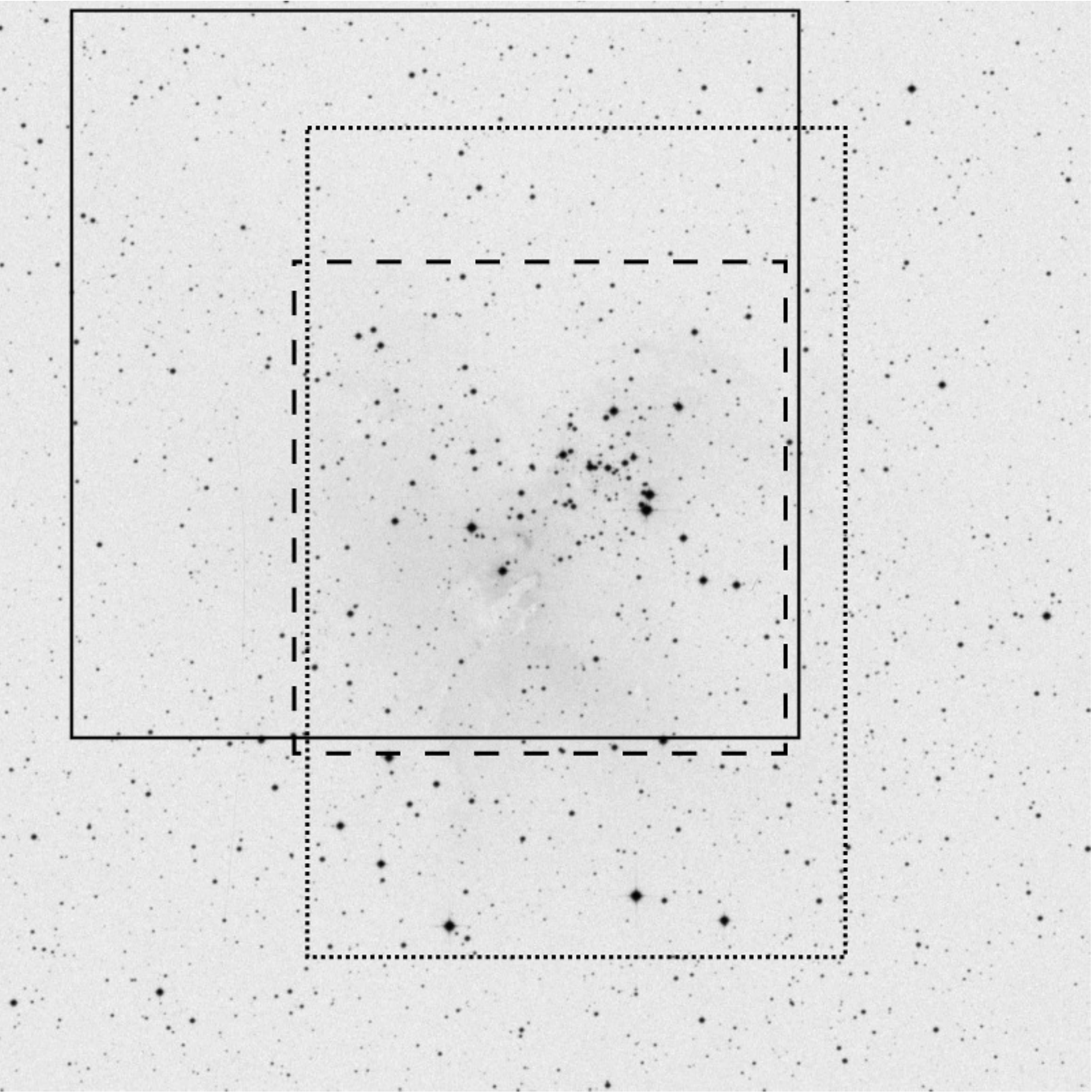}
\caption{A $30^{\prime}\times 30^{\prime}$ fragment of the DSS-1 plate containing the observed fields of NGC 6611. Solid line shows the field observed with the 1-m CTIO telescope, dashed line -- 0.9-m CTIO telescope, dotted line -- 1-m Swope telescope.} 
\label{dss}
\end{figure}

The observations of NGC 6611 were carried out with three telescopes located in Chile. Most of the observations were taken in no more than a few hours per night due to bad weather or as a background project. The observed fields are shown in Fig.~\ref{dss}.

We used the 1-m telescope at Cerro Tololo Inter-American Observatory (CTIO) operated by the SMARTS\footnote{Small and Moderate Aperture Research Telescope System} consortium from June 19 to July 18 2008. This telescope is equipped with a 4064$\times$4064 CCD camera that covers an area of the sky of $20^{\prime}\times 20^{\prime}$. We collected 380 frames in the $V$ filter and 170 frames in the $B$ filter. 

During three nights in July 2008, three nights in August 2008, and four nights in March 2009, NGC 6611 was observed with the SMARTS 0.9-m telescope at CTIO. The 2048$\times$2046 CCD camera covered $13.5^{\prime}\times 13.5^{\prime}$ area of the sky. We took 40 frames in the $U$ filter, 200 in the $B$ filter, 680 in the $V$ filter and 100 in the $I$ filter.

Finally, during several nights between May and July 2008 and between March and May 2009, we carried out observations with 1-m Swope telescope at Las Campanas Observatory, equipped with a 2048$\times$3150 CCD camera covering a $14.8^{\prime}\times 22.8^{\prime}$ area of the sky. With the Swope telescope, we collected 810 frames in the $V$ filter, 120 frames in the $B$ filter and 100 frames in the $I$ filter.

Observations from all telescopes were calibrated in a standard way. Profile and aperture photometry were derived by means of the DAOPHOT II package \citep{stet1987}, and the differential magnitudes were calculated. 

In total, we have detected 6038 stars but some have photometry of poor quality. Moreover, many stars were detected only on $I$ filter frames. 
\begin{figure*}
\centering
\includegraphics[width=140mm]{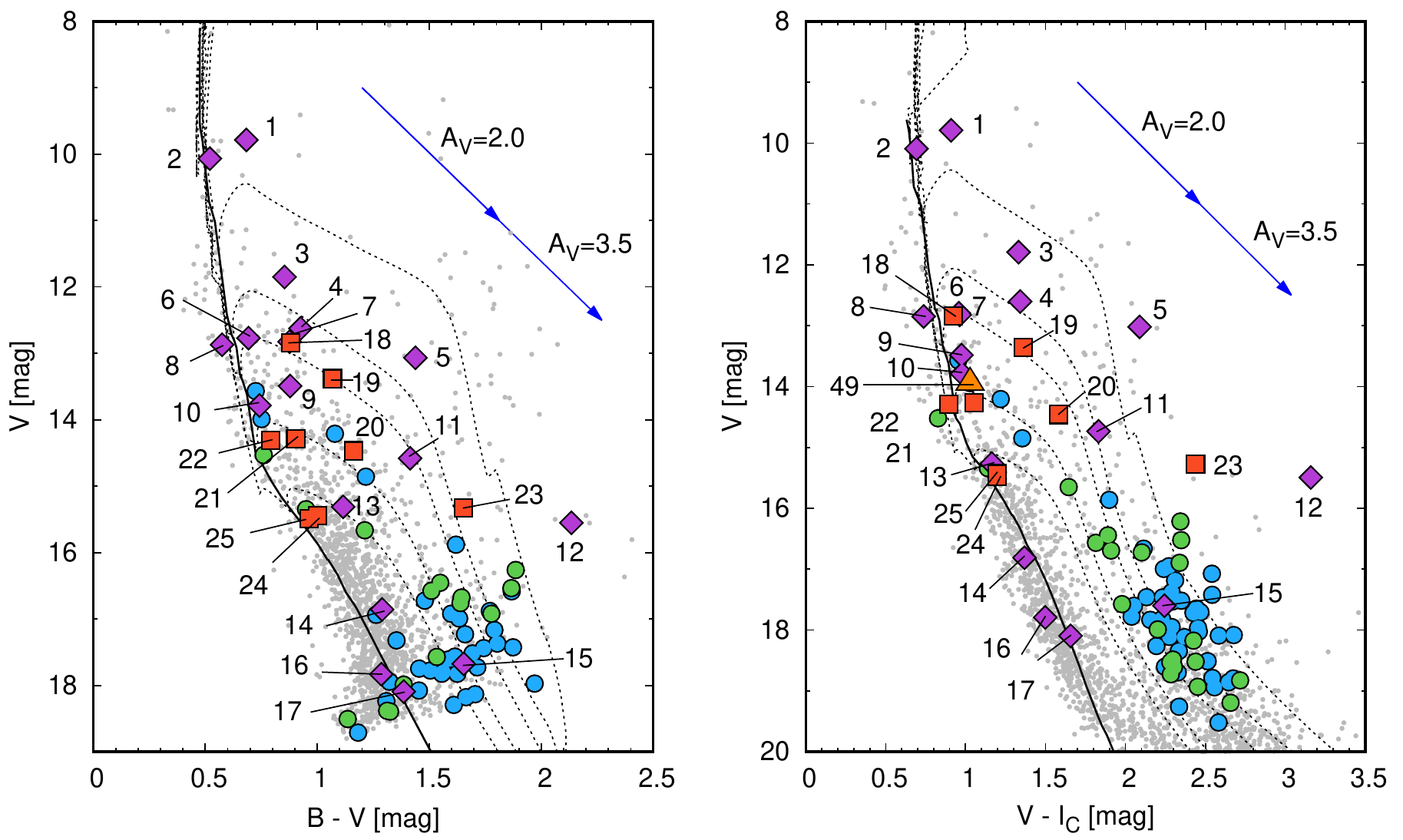}
\caption{Variable stars in $V$ vs. $(B-V)$ ({\it left}) and $V$ vs. $(V-I_{\rm C})$ ({\it right}) colour-magnitude diagrams for the observed field. Symbols denote the variable stars found in this paper: pulsating stars (red squares), eclipsing stars (pink diamonds), other periodic variables (green), the remaining variables (blue) and HAeBe star (orange triangle). Some variables discussed in text are labeled with a number. The zero-age main sequence (ZAMS) relation (thick line) was taken from \citet{pec2013}. The isochrones (dotted lines) for 0.1, 0.5, 1, 3, and 6 Myr were taken from \protect\cite{bress2012}. Following  \protect\cite{2007A&A...462..245G} we assumed a distance $d=1750$ pc, $A_V=2.6$ mag and $R_V=3.27$ and, which corresponds to $E(B-V)=0.795$ mag and $E(V-I)=0.994$ mag. Additionally, two extinction vectors for young ($A_V=3.5$ mag) and for old ($A_V=2.0$ mag) population \citep{2022arXiv220708452S} adopting $R_V=3.27$. and colour excess relation $E(B-V)/E(V-I)=0.8$ \citep{1996A&A...314..108M}} \label{cmd}
\end{figure*}

\begin{figure}
\centering
\includegraphics[width=76mm]{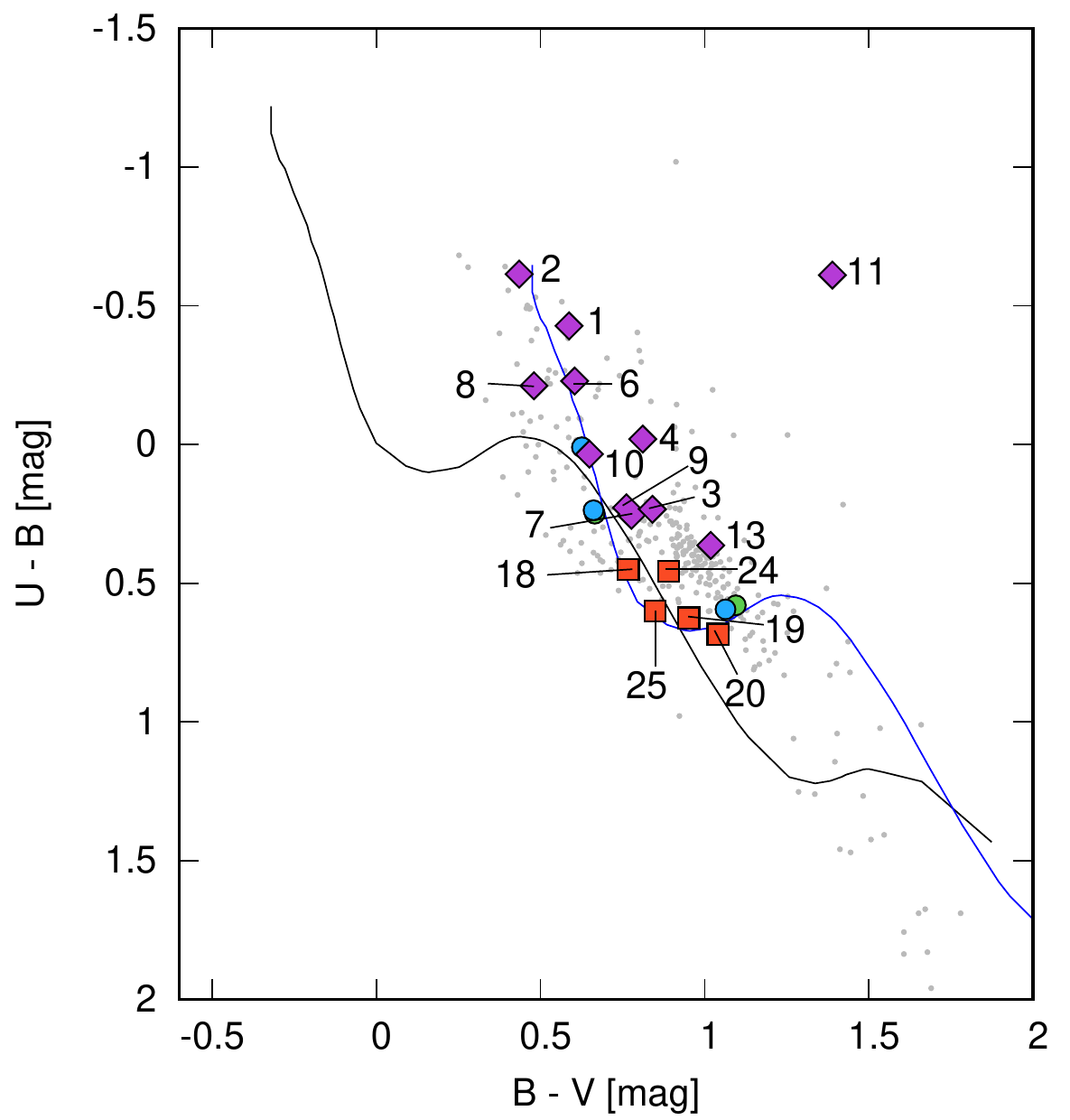}
\caption{Variable stars in the $(U-B)$ vs.~$(B-V)$ colour-colour diagram for the observed field. The symbols are the same as in Fig.~\ref{cmd}. The black line represent the intrinsic relation for unreddened main-sequence stars taken from \citet{pec2013}. 
The same relation for reddened stars with $E(B-V) = 0.795$ mag is plotted with the blue line. For the reddening line, $E(U-B)/E(B-V )=0.72$ was adopted.} 
\label{ccd}
\end{figure}
\subsection{\label{sstransf}Transformation to the standard system}

$UBV$ photometry of NGC 6611 was provided by \cite{1993AJ....106.1906H}, while $BVI_{\rm C}$ photometry was published by \cite{2007A&A...462..245G}. Unfortunately, there is no good agreement between $B$ and $V$ photometry for common stars. This problem was described by \cite{2007A&A...462..245G}, who found the difference between their own magnitudes and the ones of \cite{1993AJ....106.1906H} equal to $-$0.065 mag in the $V$ and 0.018 mag in the $B$ band. Since our observations were made in $UBVI_{\rm C}$ bands, we used both photometric data to transform mean instrumental magnitudes and colours to the standard system.
We used the following transformation equations:
\begin{equation}
\label{eq_V}
V-v=a_1\times(v-i)+b_1,
\end{equation}
\begin{equation}
\label{eq_Vb}
V-v=a_2\times(b-v)+b_2,
\end{equation}
\begin{equation}
\label{eq_BV}
B-V=a_3\times(b-v)+b_3,
\end{equation}
\begin{equation}
\label{eq_VI}
V-I_{\rm C}=a_4\times(v-i)+b_4,
\end{equation}
\begin{equation}
\label{eq_UB}
U-B=a_5\times(u-b)+b_5,
\end{equation}
where $v$, $b$, $i$, and $u$ denote the mean instrumental magnitudes from our photometry, and  $V$, $B$, $I_{\rm C}$, and $U$ are the standard magnitudes. The coefficients obtained by means of the least-squares method are shown in Tables \ref{ttranf1} and \ref{ttranf2}. In addition, these tables include the standard deviation of the residuals ($\sigma$) and the number of stars ($N$) used for the transformation. Transformations based on the photometry of \cite{2007A&A...462..245G} were performed for each CCD camera independently (Table \ref{ttranf1}) after which the magnitude and colours of each star were averaged. 
In this way, we calculated $V$ magnitudes and $(V-I_{\rm C})$ colour indices for 1900 stars and $(B-V)$ colour indices for 1598 stars. The colour-magnitude diagrams are shown in Fig.~\ref{cmd}.

Photometry in the $U$ band was taken only with the 0.9-m CTIO telescope. The coefficient transformations defined in Eq.~(\ref{eq_Vb}), (\ref{eq_BV}) and (\ref{eq_UB}), were obtained from photometry of \cite{1993AJ....106.1906H} and are shown in Table \ref{ttranf2}. The colour-colour diagram for 319 stars is shown in Fig.~\ref{ccd}.
The symbols in Fig.~\ref{cmd} and Fig.~\ref{ccd} represent the variable stars we found, which are described in Sect.~\ref{svar}.


\begin{figure}
\includegraphics[width=84mm]{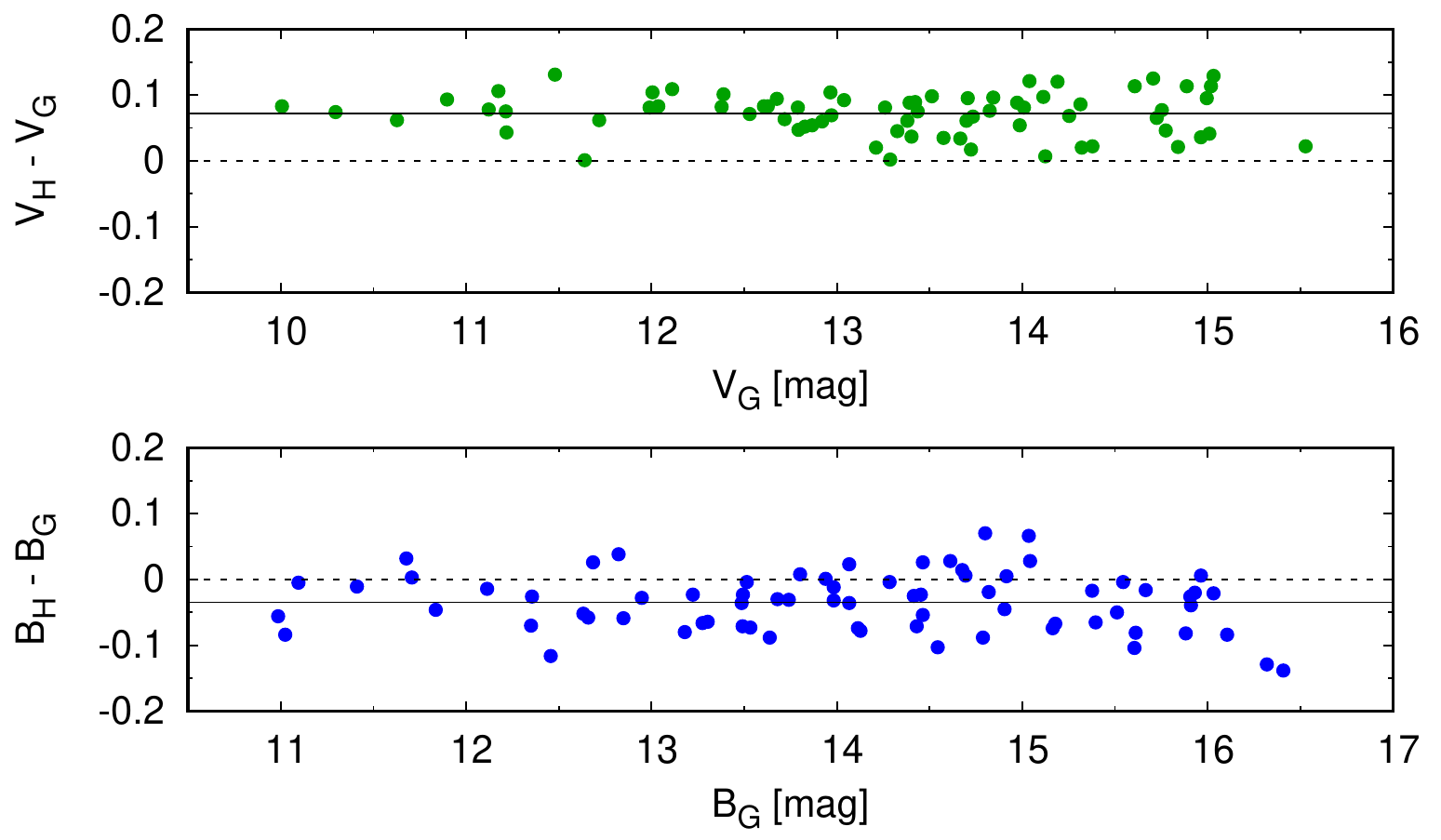}
\caption{Difference between the magnitudes of \citet{1993AJ....106.1906H} (intex 'H') and \citet{2007A&A...462..245G} (index 'G') for the $V$ (top) and $B$ (bottom) magnitudes for common stars in the field of NGC 6611. The continuous lines represent the mean magnitude offset between the two samples.} 
\label{hg}
\end{figure}

The resulting magnitudes and colours of the variable stars are given in Table \ref{tubvi} in the Appendix. It is easy to see there are some differences between the $(B-V)$ colours derived from the photometry of \cite{2007A&A...462..245G} (column 6) and \cite{1993AJ....106.1906H} (column 7). This is mainly due to some inconsistency in the $V$ photometry published by both groups of authors. The differences in $V$ and $B$ magnitudes of the common stars, shown in Fig.~\ref{hg}, are 0.071 and $-$0.035 mag, respectively. Although, as a consequence of these differences, the magnitudes and colours we obtained are subject to systematic errors, and the colour-magnitude and colour-colour diagrams proved useful in identifying the variability type of some stars.

\begin{figure*}
\includegraphics[width=14.6cm]{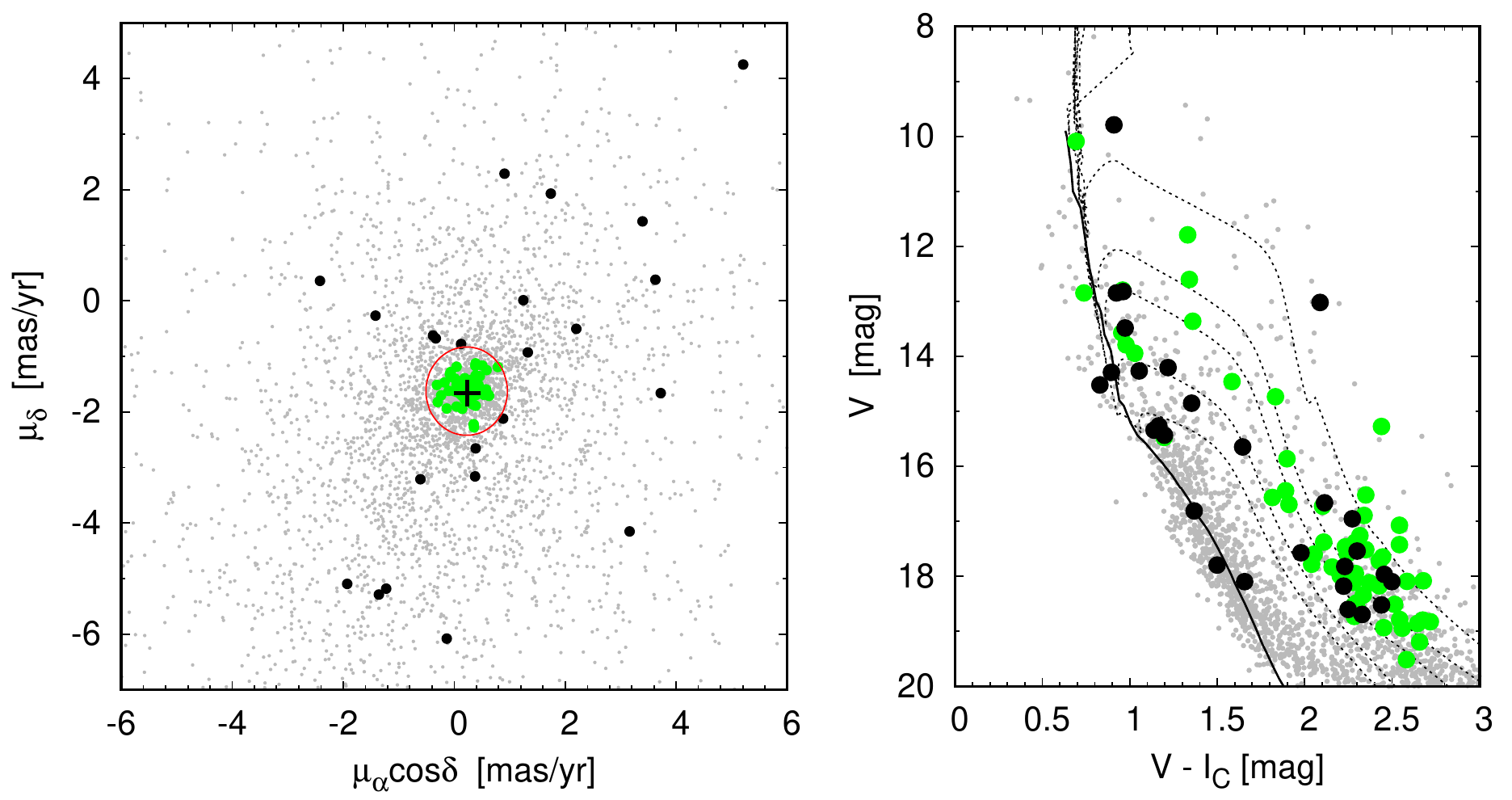}
\caption{{\it Left}: Distribution of proper motions for about 4000 stars in the observed field. {\it Right}: Colour-magnitude diagram for about 4000 stars in the observed field. Green circles represent variable stars falling within the oval area around the mean proper motion of NGC 6611 (black cross), black circles -- the remaining variable stars. The isochrones (dotted lines) for 0.1, 0.5, 1, 3, and 6 Myr were taken from \protect\cite{bress2012}.} 
\label{pm}
\end{figure*}

\section{Membership}
\label{smem}

The accurate determination of the probability of cluster membership is an important step in the study of cluster properties. 
The selection of cluster members based on proper motions for cluster and field stars was proposed by \citet{vas1958}. This method has been developed over the years (\citealt{sanders1971}, \citealt{zhao1990}, \citealt{1994Ap&SS.214..171U}).
The probability that the $i$-th star belongs to the cluster can be calculated from the following formula:
{\setlength\arraycolsep{2pt}  
\begin{eqnarray}
P_{\mu}(i)& =& \frac{n_c~.~\phi_c^{\nu}(i)}{n_c~.~\phi_c^{\nu}(i)+n_f~.~\phi_f^{\nu}(i)},\\\nonumber
\end{eqnarray}}

\noindent where $n_{c}$ and $n_{f}$ are, respectively, the normalized numbers of cluster and field stars 
($n_c + n_f = 1$), and $\phi_c^{\nu}$ and $\phi_f^{\nu}$ are the frequency distribution functions for the cluster and field stars, defined as: 
{\setlength\arraycolsep{2pt}  
\begin{eqnarray}
    \phi_c^{\nu} &=&\frac{1}{2\pi\sqrt{{(\sigma_{xc}^2 + \epsilon_{xi}^2 )} {(\sigma_{yc}^2 + \epsilon_{yi}^2 )}}}\nonumber\\
   &&\times \exp\left\{{-\frac{1}{2}\left[\frac{(\mu_{xi} - \mu_{xc})^2}{\sigma_{xc}^2 + \epsilon_{xi}^2 } + \frac{(\mu_{yi} - \mu_{yc})^2}{\sigma_{yc}^2 + \epsilon_{yi}^2}\right] }\right\}
\end{eqnarray}}
and
{\setlength\arraycolsep{2pt}  
\begin{eqnarray}
\Phi_f^{\nu}&=&\frac{1}{2\pi \sqrt{(1-\gamma^2)}\sqrt{ (\sigma^2_{xf}+\epsilon^2_{xi})
(\sigma^2_{yf}+\epsilon^2_{yi})}}\nonumber\\
  && \times\exp \left\{-\frac{1}{2(1-\gamma^2)} \left[ \frac{(\mu_{xi}-\mu_{xf})^2}{\sigma^2_{xf}+\epsilon^2_{xi}}- \right. \right.\nonumber \\
 &&\left. \left. \frac{2\gamma(\mu_{xi}-\mu_{xf})(\mu_{yi}-\mu_{yf})} {\sqrt{(\sigma^2_{xf}+\epsilon^2_{xi})(\sigma^2_{yf}+\epsilon^2_{yi})}} + \frac{(\mu_{yi}-\mu_{yf})^2}{\sigma^2_{yf}+\epsilon^2_{yi}} \right]\right\}, 
\end{eqnarray}}

\noindent where $\mu_{xi}$ and $\mu_{yi}$ are the proper motions of the
$i$-th star ($\mu_{xi}=\mu_{\alpha}\cos\delta$ and $\mu_{yi}=\mu_{\delta}$ of the $i$-th star) with errors $\epsilon_{xi}$ and $\epsilon_{yi}$, respectively. The parameters $\mu_{xc}$ and $\mu_{yc}$ denote the cluster proper motion centre with dispersions $\sigma_{xc}$ and $\sigma_{yc}$, while $\mu_{xf}$ and $\mu_{yf}$ stand for the proper motion centre for field stars with dispersions $\sigma_{xf}$ and $\sigma_{yf}$, respectively. The correlation coefficient $\gamma$ was calculated by\\
{\setlength\arraycolsep{2pt}  
\begin{eqnarray}
\gamma &=& \frac{(\mu_{xi} - \mu_{xf})(\mu_{yi} - \mu_{yf})}{\sigma_{xf}\sigma_{yf}}
\end{eqnarray}}
\citep{sariya2012}
 
The proper motions of observed stars were taken from the Gaia Early Data Release 3 (Gaia EDR3; \citealt{gaia2016}, \citealt{gaia2020}). They are shown in the left panel of Fig.\,\ref{pm}.
The mean proper motion of the cluster was calculated from the 30 stars with membership probability above 80\% \citep{1999A&AS..134..525B}. We obtained $\mu _{\alpha}\cos\delta=0.241$ mas yr$^{-1}$ and $\mu _{\delta}=-1.658$ mas yr$^{-1}$ with standard deviation $\sigma_{\mu _{\alpha}}=0.179 $ and $\sigma_{\mu _{\delta}}=0.231$, respectively. We assume that the cluster stars are within an ellipse with semi-axis equal to $3\sigma_{\mu _{\alpha}}$ and $3\sigma_{\mu _{\delta}}$ around the calculated centre (black cross in Fig.~\ref{pm}). Then, using the 995 stars located in this area, we calculated the mean proper motion of the cluster $\mu_{xc}=0.224$ and $\mu_{yc}=-1.627$ mas\,yr$^{-1}$ with dispersions $\sigma_{xc}=0.244$ and $\sigma_{yc}=0.265$ mas\,yr$^{-1}$. Using the remaining 3122 stars, we calculated the mean proper motion of field stars $\mu_{xf}=-0.164$ and $\mu_{yf}=-2.536$ mas\,yr$^{-1}$ with dispersions $\sigma_{xf}=2.339$ and $\sigma_{yf}=3.023$ mas\,yr$^{-1}$. From these values, we calculated the frequency distribution functions $\phi_c^{\nu}$ and $\phi_f^{\nu}$ and the membership probability $P_{\mu}(i)$ for each star. 
The probabilities for the variable stars are shown in Table \ref{tubvi}. As can be seen in the left panel of Fig.~\ref{pm}, most of our variable stars are located near the cluster centre and many of them are younger than 3 Myr (right panel of Fig.~\ref{pm}).

The Gaia EDR3 catalogue also contains parallaxes of stars. Using stars with membership probabilities greater than 80\%, we derived the weighted average parallax of the cluster to be 0.56 mas. This corresponds to the distance of 1780 pc, which is consistent with the value obtained by \cite{2019ApJ...870...32K}.

\begin{figure}
\centering
\includegraphics[width=86mm]{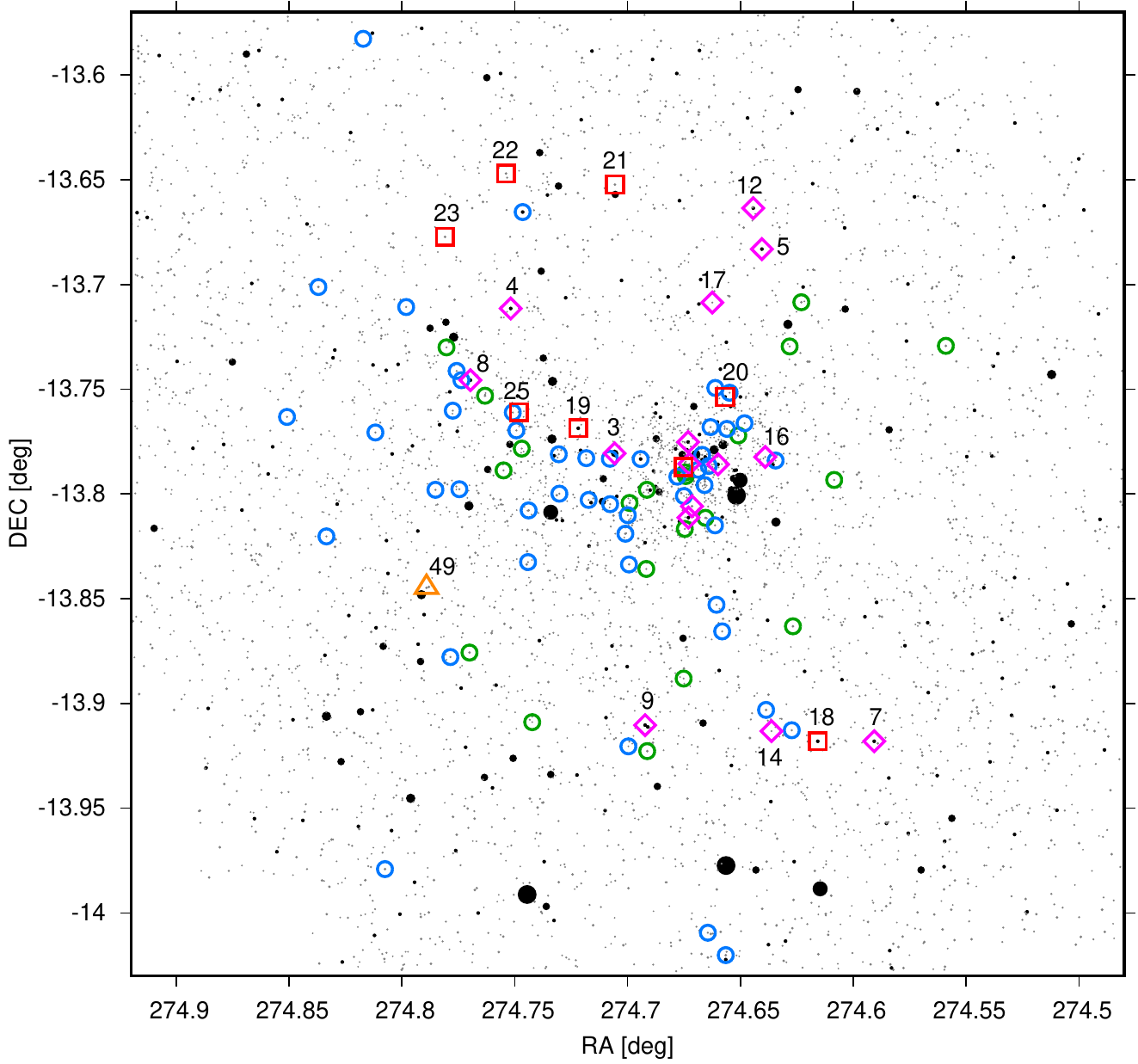}
\caption{Schematic map of the observed field. Variable stars are colour-coded: pulsating stars (red squares), eclipsing stars (pink diamonds), other periodic variables (green circles), the remaining variables (blue circles) and HAeBe star (orange triangle). Some variables discussed in the text are labelled.} 
\label{map}
\end{figure}

\section{Variable Stars}
\label{svar}
The search for variable stars was performed on merged data. Prior to merging, data sets from each observatory were prepared: the outliers were removed using $3\sigma$-clipping and the mean magnitude was subtracted. The data-sets were combined for each filter separately. In the next step, Fourier spectra up to 80 d$^{-1}$ were calculated from the $V$-filter data, as they have the best time coverage. In addition, the light curves, periodgrams, and phased light curves for all stars were visually inspected. In total, we found 95 variable stars. Their location in schematic map of the observed field is
shown in Fig.~\ref{map}. Some variable stars discussed in the paper are labelled with our identification number. The cross-identification with the numbering provided by the WEBDA\footnote{The WEBDA database is available on a web page \url{http://webda.physics.muni.cz}.} database is shown in Table \ref{tubvi} of the Appendix. If we refer to the WEBDA number in the paper, we preceded it with letter 'W'.

\subsection{\label{spuls}Pulsating stars}

Among 95 variables found in the observed field, eight are candidates for $\delta$ Scuti-type stars. 
The parameters of the sinusoidal terms (frequency, $f_i$, semi-amplitude, $A_i$, and phase, $\phi_i$), listed in Table \ref{tpar}, were derived by fitting the formula:
\begin{equation}
\label{eqmag}
\langle m \rangle + \sum_{i=1}^{n} A_i \sin (2 \pi f_i (t-T_0) + \phi _i)
\end{equation}
to the analyzed differential magnitudes. In Eq.~(\ref{eqmag}) $\langle m \rangle$ is the mean differential magnitude, $n$ is the number of terms, $t$ is the time elapsed from the initial epoch $T_0 =HJD2454000$.  Additionally, for each frequency, we calculated signal-to-noise ($S/N$) where $S$ is the amplitude of the frequency and $N$ is the average noise level in an amplitude spectrum of the residuals.

\begin{figure}
\centering
\includegraphics[width=76mm]{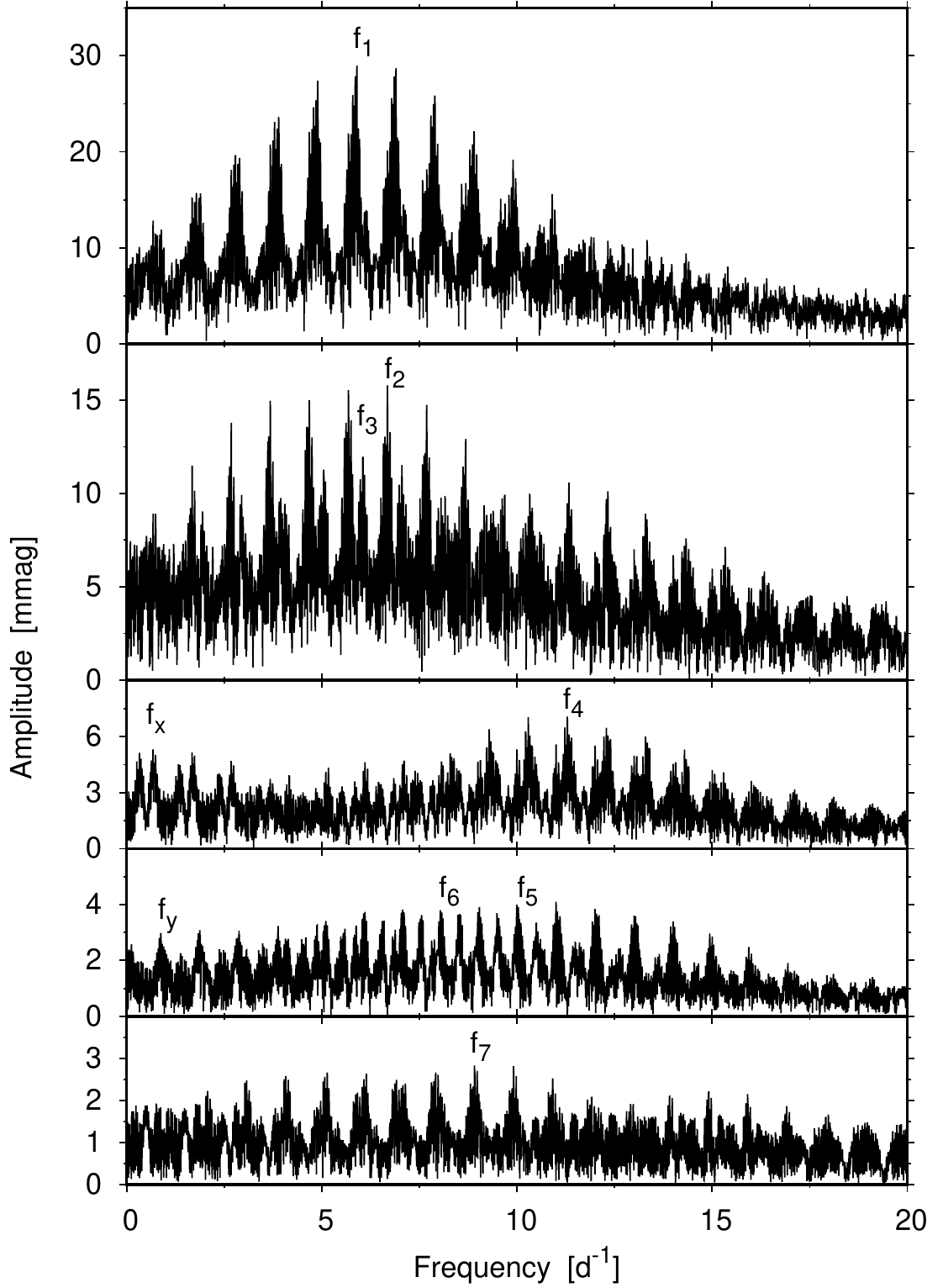}
\caption{Fourier frequency spectrum of the $V$-filter data of the $\delta$ Scuti candidate, star \#18, at four steps of prewhitening.} 
\label{f153}
\end{figure}

The most luminous object among $\delta$ Scuti candidates, classified as F0 III-V star \citep{1943ApJ....97..300N}, is star \#18 (W135). It is not a member of the cluster (\citealt{1986A&AS...66..311T}, \citealt{1999A&AS..134..525B}, Table \ref{tubvi}). The Fourier spectrum (Fig.~\ref{f153}) reveals a frequency of $5.89911$ d$^{-1}$. This result is consistent with a periodicity of 0.169567 d reported by \citet{2018AJ....156..241H}. We found six other terms in successive steps of prewhitening. The two with the lowest frequencies are presumably not due to pulsations but instrumental effects. They are denoted as $f_x$ and $f_y$. Four other frequencies $f_4-f_7$ are almost equally high as their corresponding alias frequencies. It means that the true frequency may differ by 1 d$^{-1}$. The multiperiodicity and spectral type clearly indicate a $\delta$ Scuti-type star. 

\begin{figure}
\centering
\includegraphics[width=76mm]{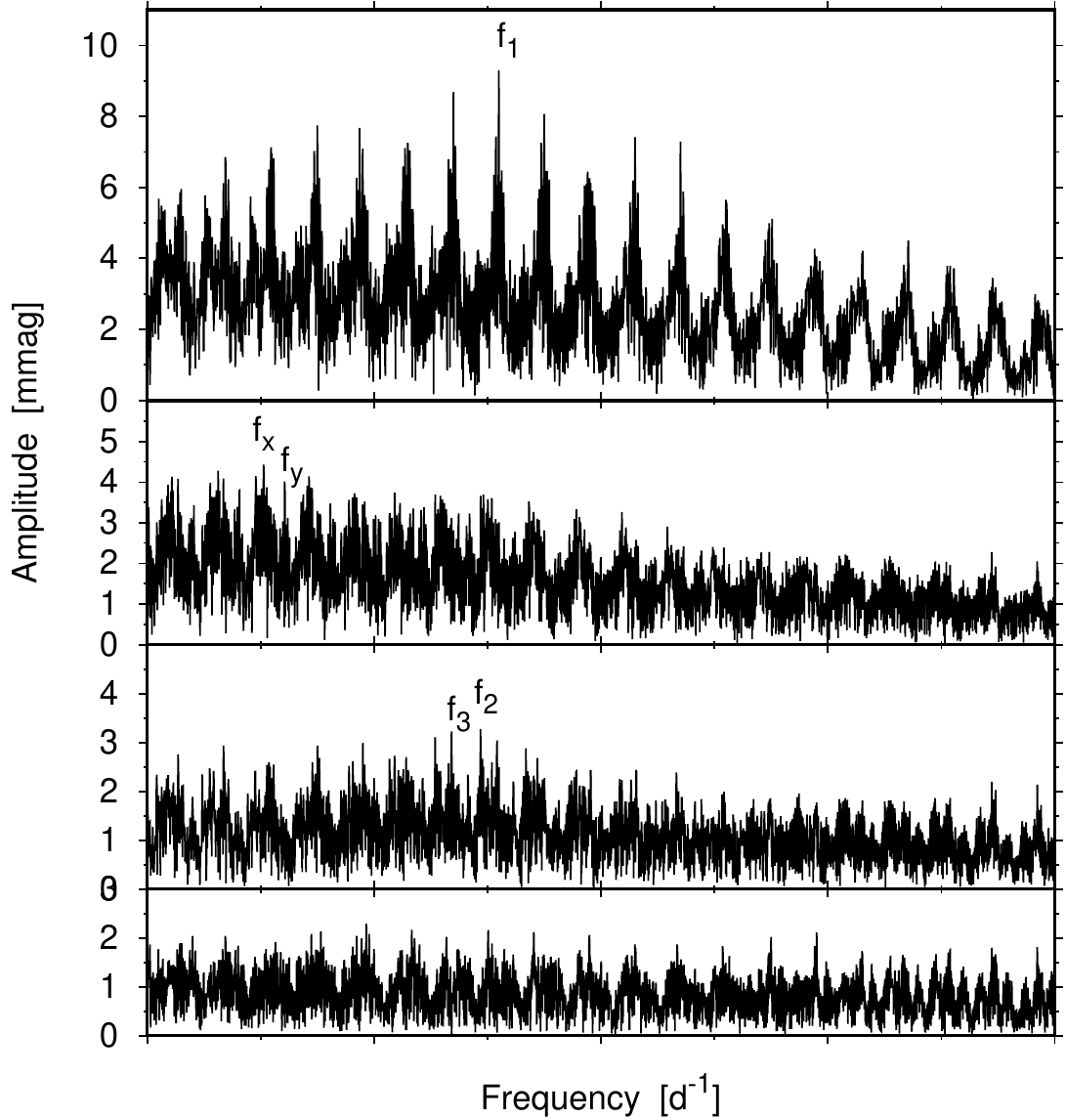}
\caption{Fourier frequency spectrum of the $V$-filter data of the $\delta$ Scuti candidate, star \#19: original data (top) and three steps of prewhitening.} 
\label{f184}
\end{figure}
\begin{figure}
\centering
\includegraphics[width=76mm]{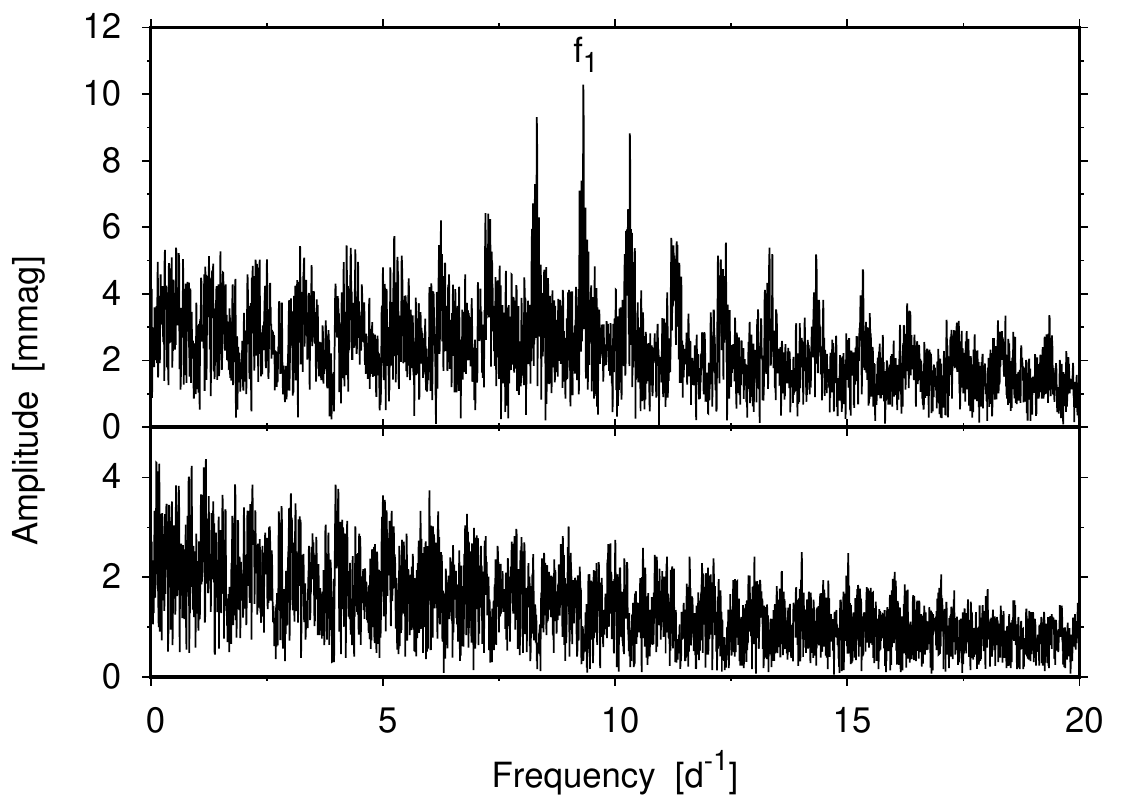}
\caption{Fourier frequency spectrum of the original $V$-filter data of the $\delta$ Scuti candidate, star \#20 (upper panel), and after prewhitening with $f_1$ and its three harmonics (lower panel).} 
\label{f355}
\end{figure}
\begin{figure}
\centering
\includegraphics[width=76mm]{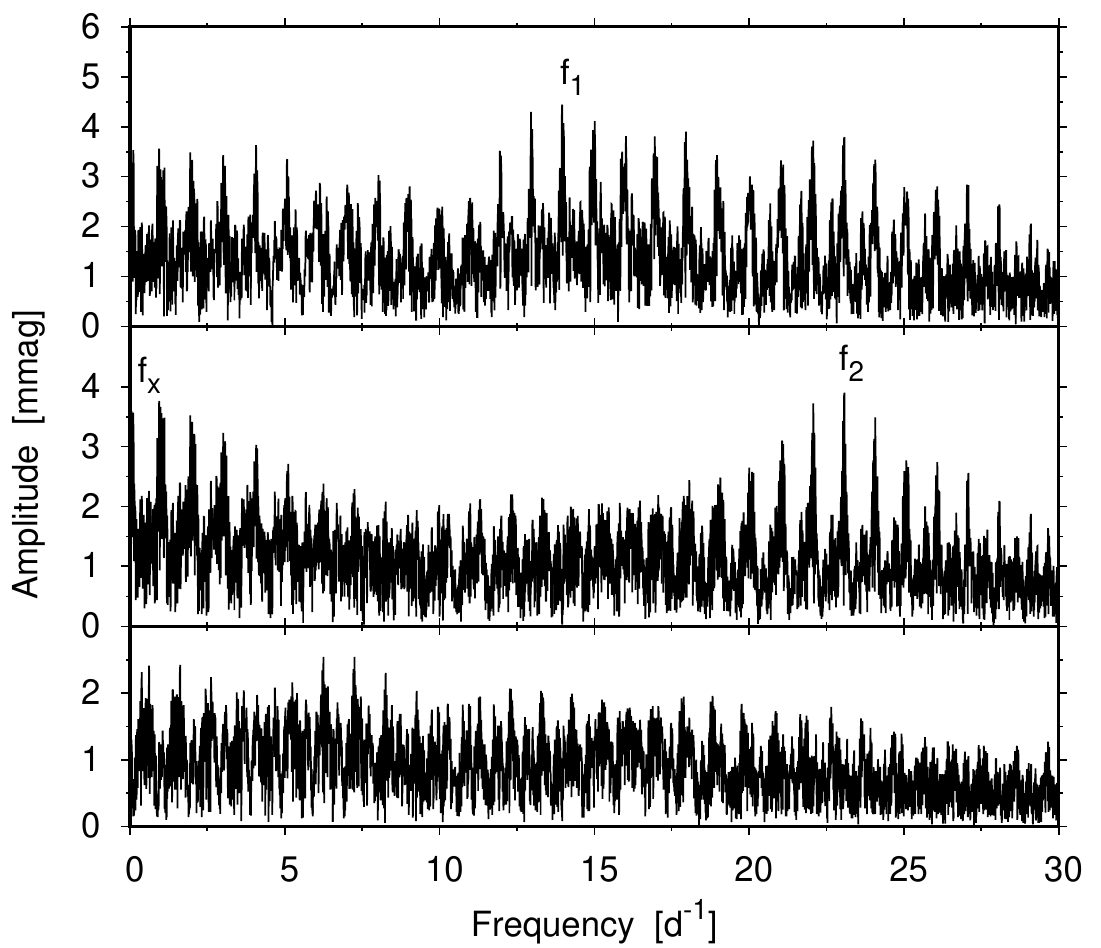}
\caption{Fourier frequency spectrum of the $V$-filter data of the $\delta$ Scuti candidate, star \#21: original data (top), after removing $f_1$ (middle), after removing  $f_1$, $f_2$ and $f_x$ (bottom).} 
\label{f372}
\end{figure}

Another candidate for a $\delta$ Scuti type star is \#19 (W374). The variability was reported by \cite{1997A&AS..121..223D}, who classified the star as F2e.
After prewhitening with a dominant frequency ($f_1=7.747645$), two other frequencies of 2.56 and 3 d$^{-1}$ (marked as $f_x$ and $f_y$ in Fig.~\ref{f184}) are clearly seen in the Fourier spectrum. The corresponding phase diagrams, however, indicate that these frequencies are not caused by pulsations. The removal of these frequencies led us to find another two peaks with frequencies of $f_2=7.34435$ and $f_3=6.70819$ d$^{-1}$. 
Based on the Gaia proper motions, we calculated a membership probability of 94.8\%, which is inconsistent with previous determinations of 0\% derived by \cite{1986A&AS...66..311T} and \cite{1999A&AS..134..525B}.

An interesting star appears to be \#20 (W221). The Fourier spectrum (Fig.~\ref{f355}a) reveals a single periodicity with a frequency of 9.318127 d$^{-1}$, typical of $\delta$ Scuti-type stars. The spectral type of this star estimated by \cite{1961ApJ...133..438W} is G. The star was later found to be B8Ve type with broad lines \citep{1993AJ....106.1906H}. This spectral type, however, seems to be too early for a $\delta$ Scuti star. The membership probability we found is equal to 97.0\%. The position in the colour-colour and colour-magnitude diagrams indicates that if the star is a member of NGC 6611, it cannot be B8 V type.

The next $\delta$ Scuti candidate is star \#21 (W347) classified as A7\,III \citep{2008A&A...489..459M}. Two frequencies are clearly seen in the Fourier spectrum (Fig.~\ref{f372}). The low-frequency peak labelled as $f_x$ probably comes from instrumental effects. This star is not a member of the cluster (Table \ref{tubvi}).

\begin{figure}
\centering
\includegraphics[width=76mm]{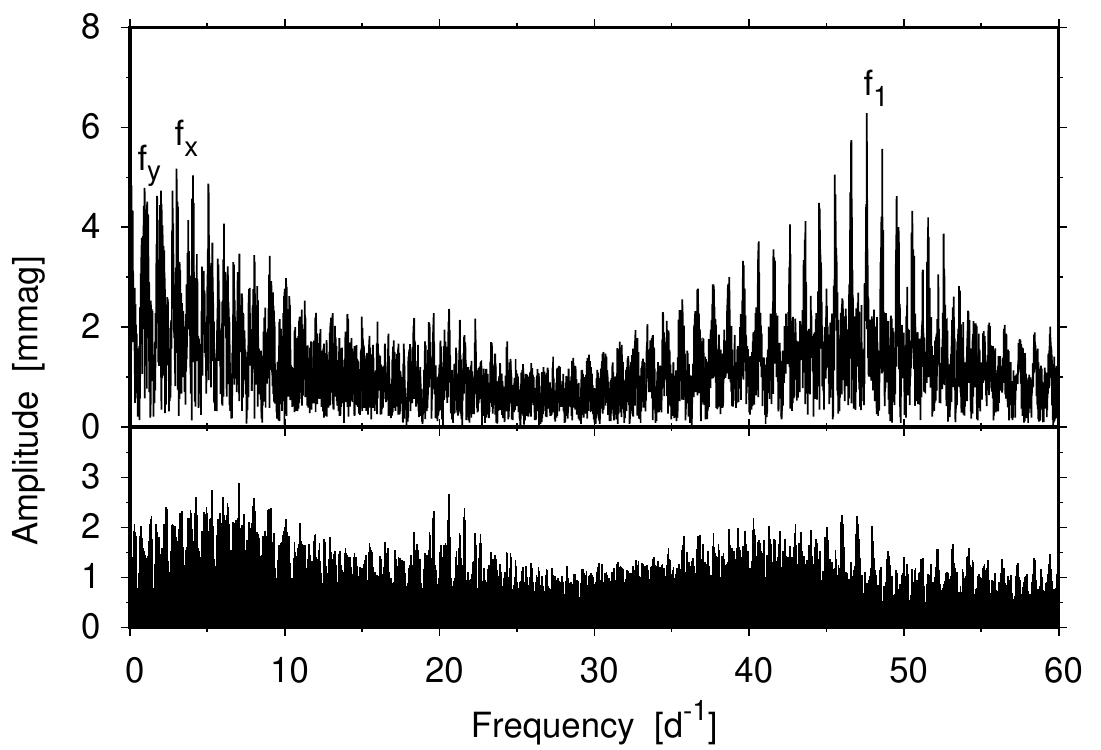}
\caption{Fourier frequency spectrum of the $V$-filter data of the $\delta$ Scuti candidate, star \#22: original data (top), after removing $f_1$,$f_x$ and $f_y$ (bottom).} 
\label{f392}
\end{figure}

\begin{figure}
\centering
\includegraphics[width=76mm]{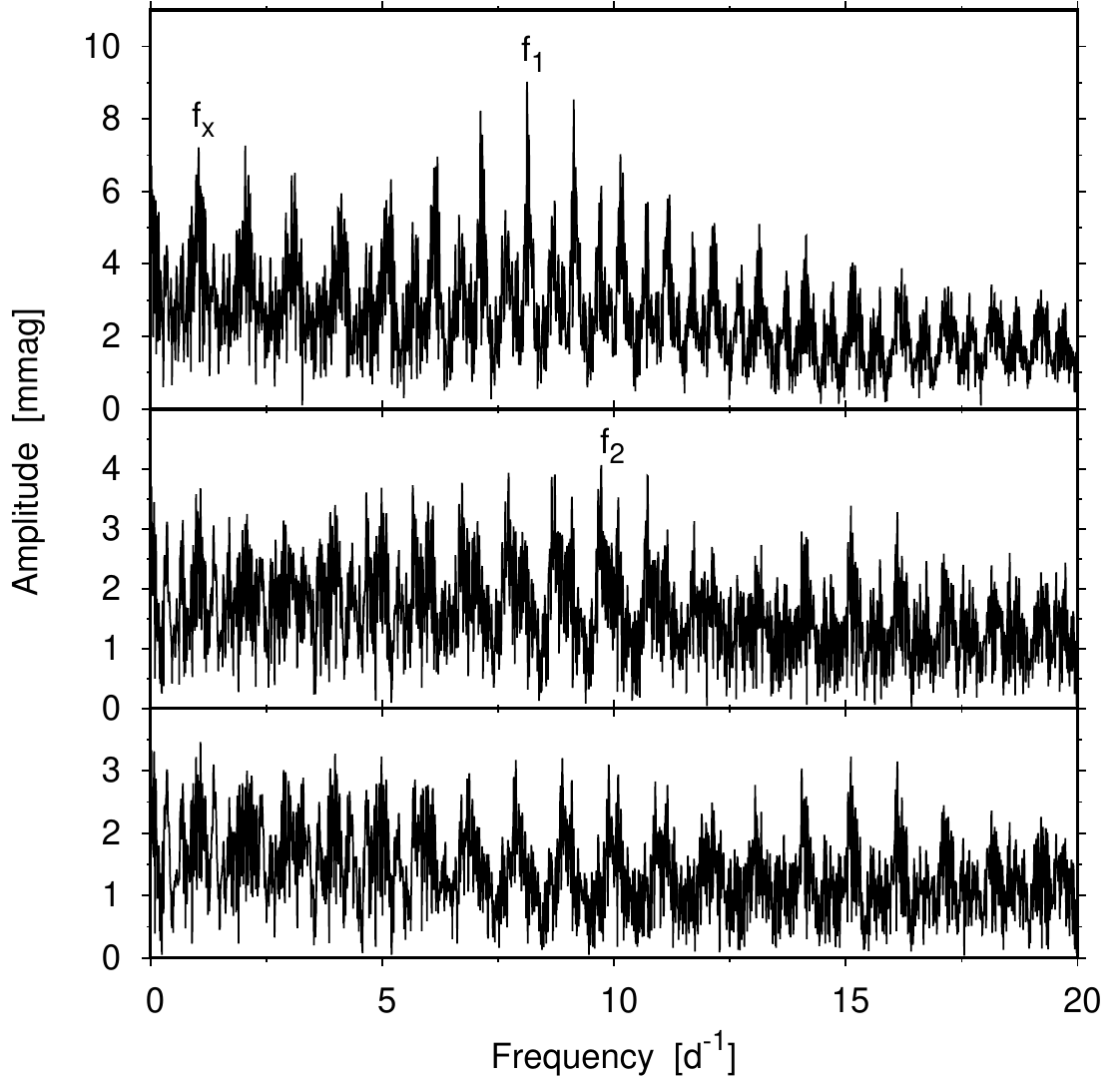}
\caption{Fourier frequency spectrum of the $V$-filter data of the $\delta$ Scuti candidate, star \#23: original data (top), after removing $f_1$ and $f_x$ (middle), after removing  $f_1$, $f_x$ and $f_2$ (bottom).} 
\label{f435}
\end{figure}

\begin{figure}
\centering
\includegraphics[width=76mm]{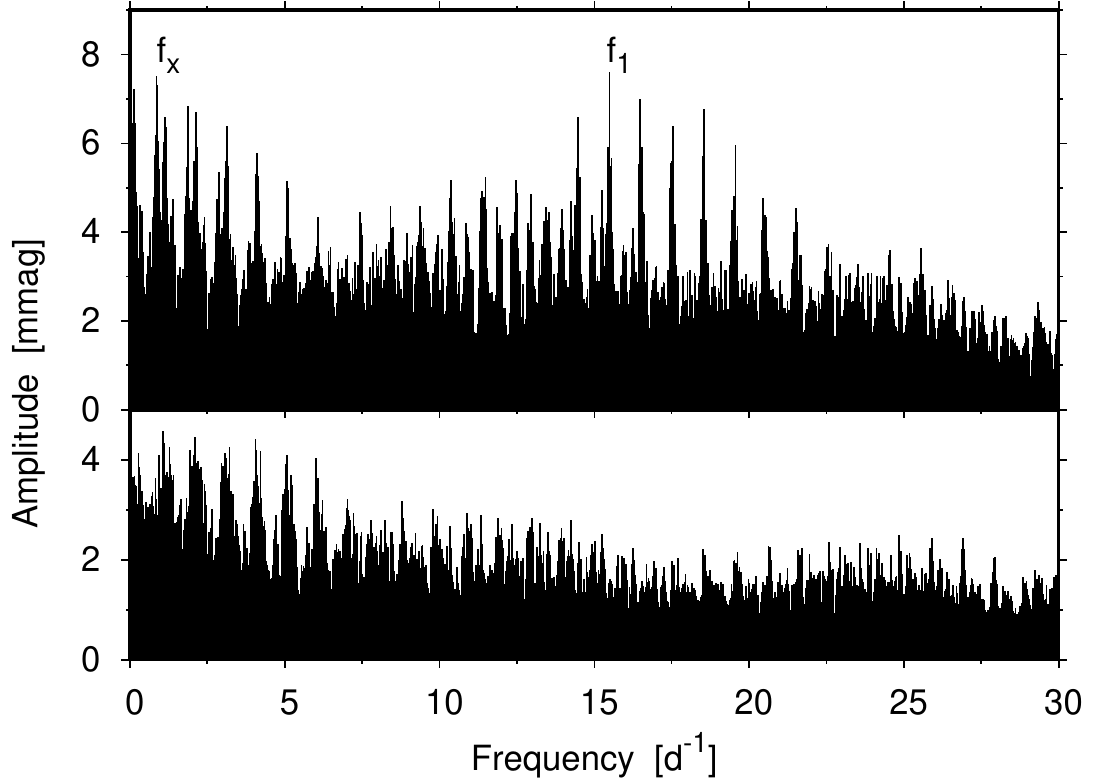}
\caption{Fourier frequency spectrum of the $V$-filter data of the $\delta$ Scuti candidate, star \#24: original data (top), after removing $f_1$ and $f_x$ (bottom).} 
\label{f759}
\end{figure}

\begin{figure}
\centering
\includegraphics[width=76mm]{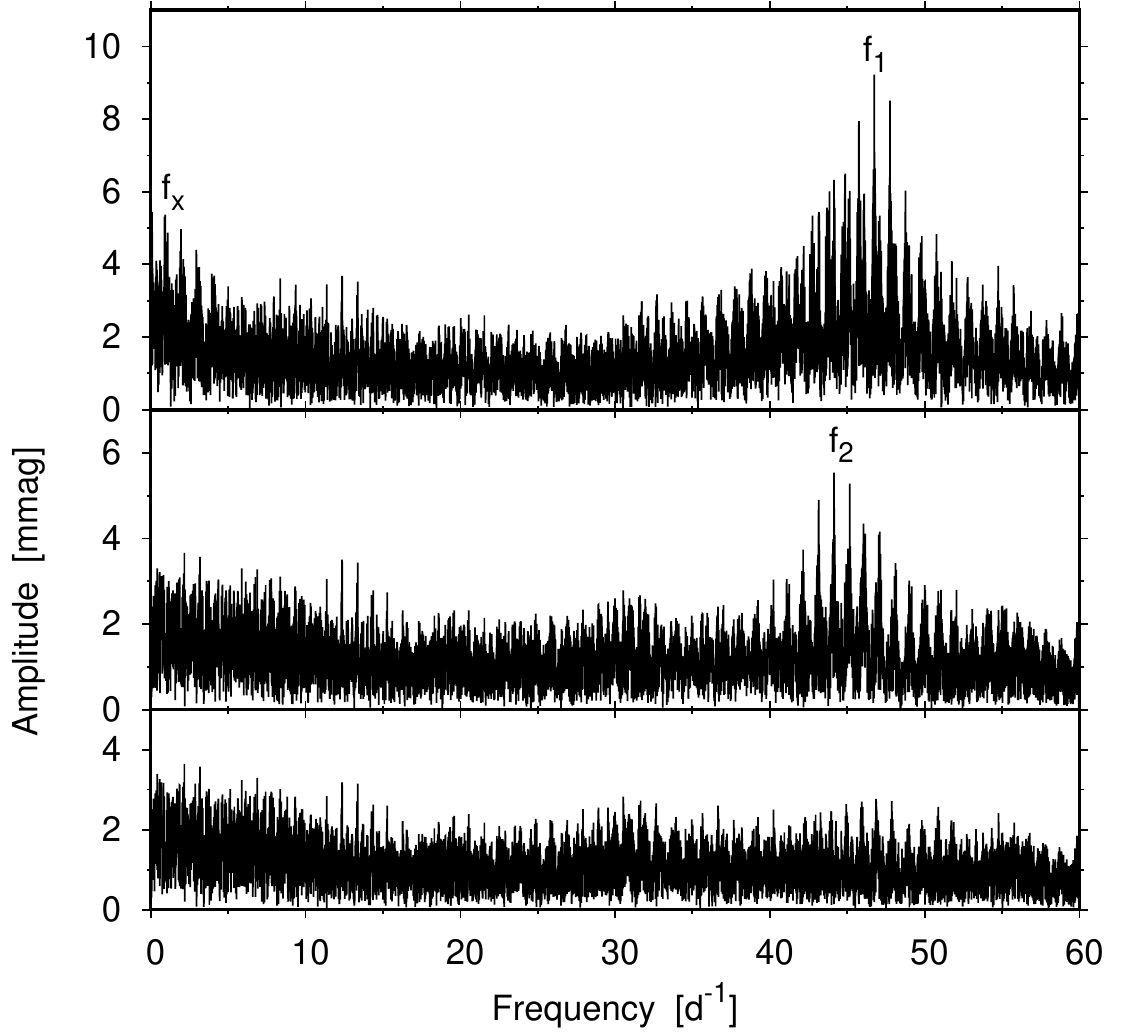}
\caption{Fourier frequency spectrum of the $V$-filter data of the $\delta$ Scuti candidate, star \#25: original data (top), after removing $f_1$ and $f_x$ (middle), after removing  $f_1$, $f_x$ and $f_2$ (bottom).} 
\label{f786}
\end{figure}

The other four candidate $\delta$ Scuti stars (\#22, \#23, \#24, and \#25) have no determined spectral types. Their Fourier spectra are shown in Figs.~\ref{f392}--\ref{f786}, and the parameters of the sinusoidal fits are given in Table \ref{tpar}. The frequencies that are not due to pulsations (likely instrumental) are labelled as $f_x$ and $f_y$.
We found two frequencies in Fourier spectra of variable \#23 and \#25 . The membership probability of stars \#22 and \#24 is 0, whilst for stars \#23 and \#25, it is more than 90\%. The variable \#24 has a visual companion separated by 0.872\arcsec \citep{2001A&A...379..147D}. This star was marked as `DSCT/GDOR/SXPHE' in the in the Gaia DR3 variability catalogue \citet{2022yCat.1358....0G}.


\begin{table*}
\centering
\caption{Parameters of sinusoidal fits to the $V$-filter differential magnitudes of eight $\delta$ Scuti-type candidates detected in NGC 6611. Numbers in parentheses denote the r.m.s.~errors of the preceding quantities with the leading zeroes omitted. The $S/N$ is the siglal-to-noise, $\sigma_{\rm res}$ is the standard deviation of residuals and $N_{\rm obs}$ is the number of observations. The frequencies not reliable due to aliasing problem are marked with an asterisk.}
\label{tpar}
\begin{tabular}{crcrrllrrcl}
\hline
Star & $N_{\rm obs}$ & \multicolumn{1}{c}{Frequency} & \multicolumn{1}{c}{$f_i$} & \multicolumn{1}{c}{$A_i$} & \multicolumn{1}{c}{$\phi _i$} & \multicolumn{1}{c}{$T_{\rm max} - T_0$} & $S/N$ & \multicolumn{1}{c}{$\sigma_{\rm res}$} & Spectal & Other\\ 
\#& & \multicolumn{1}{c}{[d$^{-1}$]}& & \multicolumn{1}{c}{[mmag]} & \multicolumn{1}{c}{[rad]} & \multicolumn{1}{c}{[d]} & & \multicolumn{1}{c}{[mmag]} & type & name  \\\hline\noalign{\vskip1pt} 
18   & 894& 5.899130(18)&$f_1$ & 28.5(3) & 1.18(1) & 898.0385(2) & 69.6 & 4.8& F0III/V & W135 \\
     &    & 6.683535(32)&$f_2$ & 15.0(2) & 1.32(2) & 898.1426(4) & 36.7 & & &\\
     &    & 6.089909(54)&$f_3$ & 10.4(2) & 5.01(2) & 898.0557(6) & 25.9 & & &\\
     &    &11.291034(54)&$f_4^{\star}$ &  6.7(2) & 1.37(3) & 898.0813(5) & 16.5 & & &\\ 
     &    & 0.67443(10) &$f_x$ &  5.2(2) & 1.35(5) & 898.636(12) & 12.9 & & &\\
     &    &10.998562(90)&$f_5^{\star}$ &  6.2(3) & 1.02(4) & 898.0952(6) & 15.2 & & &\\
     &    & 8.49109(13) &$f_6^{\star}$ &  4.0(2) & 5.05(6) & 898.0772(11) & 9.6& & &\\
     &    & 0.94842(14) &$f_y$ &  3.4(2) & 5.0(1)  & 897.981(16) & 8.1 & & &\\
     &    & 8.91439(14) &$f_7^{\star}$ &  3.6(3) & 0.87(7) & 898.1277(13) & 9.0& & &\\  
19   &1424& 7.747645(42)&$f_1$ &  9.4(3) & 3.78(3) & 777.0531(07) & 19.8 & 6.4&F2e&W374\\
     &    & 2.564905(89)&$f_x$ &  4.0(3) & 5.92(9) & 777.0190(44) & 8.4 & & &\\
     &    & 3.03159(13) &$f_y$ &  3.2(3) & 2.45(9) & 776.9201(49) & 6.7 & & &\\
     &    & 7.34435(11) &$f_2$ &  2.8(3) & 2.51(9) & 776.9944(21) & 6.1 & & &\\
     &    & 6.70819(11) &$f_3$ &  3.2(3) & 0.54(9) & 777.0949(21) & 7.0 & & &\\
20   &1708& 9.318127(46)&$f_1$ & 10.6(4) & 4.50(4) & 804.9010(7) & 13.2 &12.4 & B8Ve &W221\\
21   &1134&13.95823(11) &$f_1$ &  3.8(4) & 5.0(1)  & 828.4212(11) & 6.5 & 9.3& A7III &W347\\
     &    &23.07700(12) &$f_2$ &  3.6(4) & 2.36(11)& 828.4570(7) & 6.1 & & &\\ 
22   &1115&47.60165(8)  &$f_1$ &  5.7(4) & 5.35(7) & 829.2947(2) & 8.6 & 10.3& &W451\\
     &    & 3.01478(12) &$f_x$ &  6.0(4) & 1.01(8) & 829.3135(42) & 8.7 & & &\\
     &    & 0.89815(11) &$f_y$ &  5.8(5) & 3.63(9) & 829.428(16) & 8.6 & & &\\
23   &1132&8.126086(74) &$f_1$ &  8.9(5) & 4.17(6) & 828.7480(12) & 11.5 & & &W491\\
     &    &0.89012(11)  &$f_x$ &  5.9(5) & 5.63(9)& 828.499(16) & 8.1 & & &\\
     &    &9.724230(14) &$f_2^{\star}$ &  4.3(5) & 4.65(12)& 828.7291(20) & 6.0 & & &\\
24   &1844&15.486392(93)&$f_1$ &  7.6(6) & 5.55(8) & 807.6587(08) & 7.9 & 17.9 & &W270\\
     &    &0.866011(90))&$f_x$ &  7.8(6) & 5.51(8) & 807.485(14) & 8.1 & & & \\
25   &1858&46.749073(63)&$f_1$ &  9.5(5) & 3.26(5) & 808.7495(1) &12.2 &15.5 & &W435\\ 
     &    & 0.05769(13) &$f_x$ &  4.9(6) & 5.31(11)& 812.35(32) & 6.3 & & &\\ 
     &    &44.14812(10) &$f_2$ &  5.9(5) & 2.94(9) & 808.7390(3) & 7.6 & & &\\
\noalign{\vskip1pt}\hline
\end{tabular}
\end{table*}

\begin{figure}
\centering
\includegraphics[width=74mm]{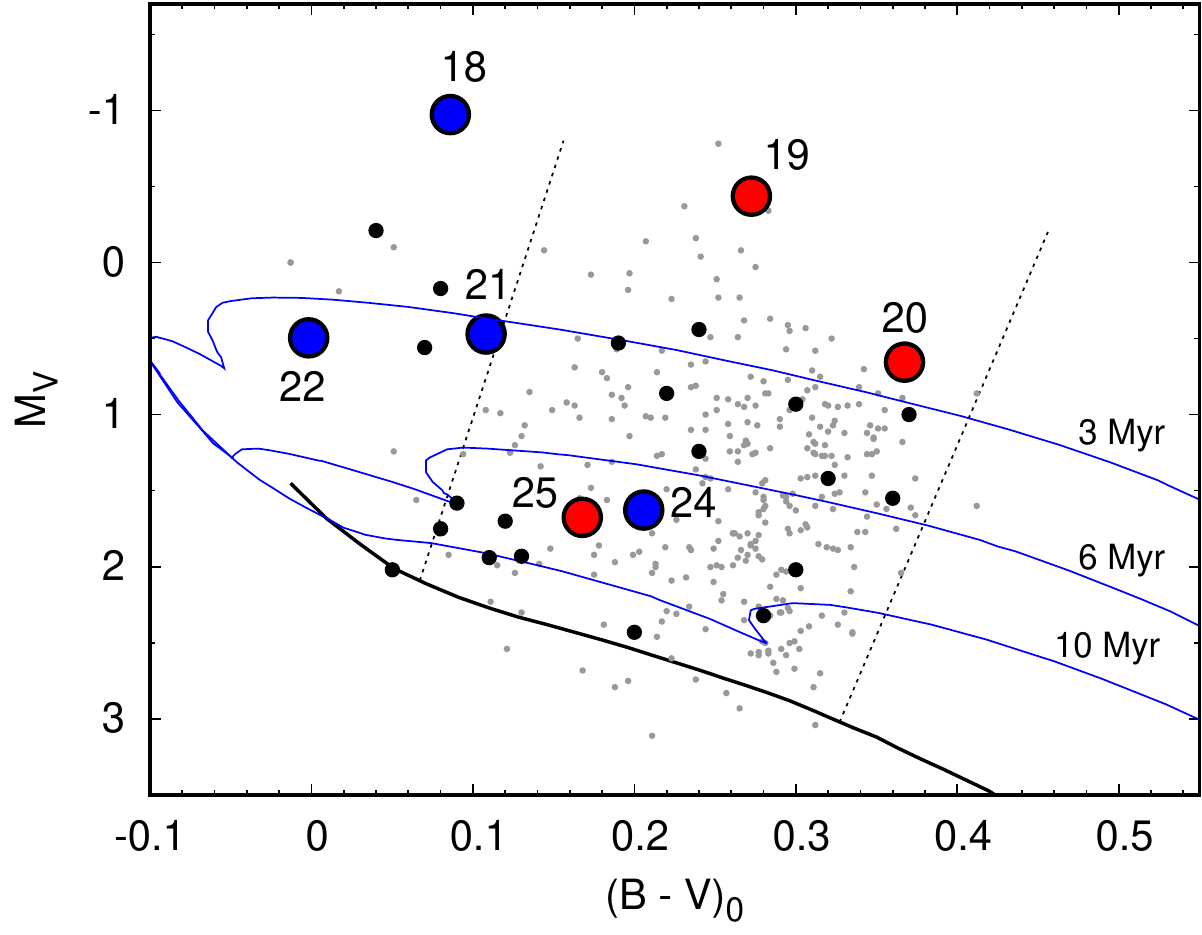}
\caption{Position of $\delta$ Scuti candidates in HR diagram: cluster members (red dots) and non-members (blue dots). Additionally, 20 pulsating PMS stars taken from \citet{zwintz2008} are plotted. The ZAMS (black solid line), the borders of the classical $\delta$ Scuti instability strip (dotted lines) and classical $\delta$ Scuti stars (grey dots) were transformed from the data of \citet{rod2001}. The blue line represents isochrones for 3, 6, and 10 Myr taken from \protect\cite{bress2012}.} 
\label{pms}
\end{figure}

If these four $\delta$ Scuti stars are cluster members, then they should be in the pre-main sequence (PMS) state of evolution. To verify this, we marked these stars in the Hertzsprung–Russell (HR) diagram together with the other $\delta$ Scuti stars (Fig.~\ref{pms}). 
The values of $M_V$ and $(B-V)_0$ were calculated assuming $A_V=2.6$ mag, $E(B-V)=0.795$ mag and a distance modulus equal to 11.215 mag. The black dots in Fig.~\ref{pms} represent PMS $\delta$ Scuti candidates taken from the work of  \citet{zwintz2008}. The author noted that the PMS pulsators occupy the same instability region in the HR diagram as classical $\delta$ Scuti stars, which are indicated by the grey dots in Fig.~\ref{pms}. We derived $(B-V)_0$ from the $(b-y)_0$ colours published by \citet{rod2001}\footnote{\url{http://www.iaa.es/~eloy/dsc00.html}} using transformation given by \citet{cald1993}. The instability strip (dotted line) and ZAMS (black solid line) were also transformed from the data of \citet{rod2001}. The membership probability of four $\delta$ Scuti candidates is greater than 90\%. Three of them, marked as red dots in Fig~\ref{pms}, lie inside the $\delta$ Scuti instability region. The fourth star, \#23, has $(B-V)_0=0.8578$ mag that places it far from the instability strip. 
If this star is indeed a $\delta$ Scuti-type variable, it is not a member of the cluster. 
At this time, we cannot exclude this star to be a $\delta$ Scuti-type star. The second frequency we found is questionable because it is close to the noise level. The variability of this star may be caused by spots.

\begin{figure}
\centering
\includegraphics[width=80mm]{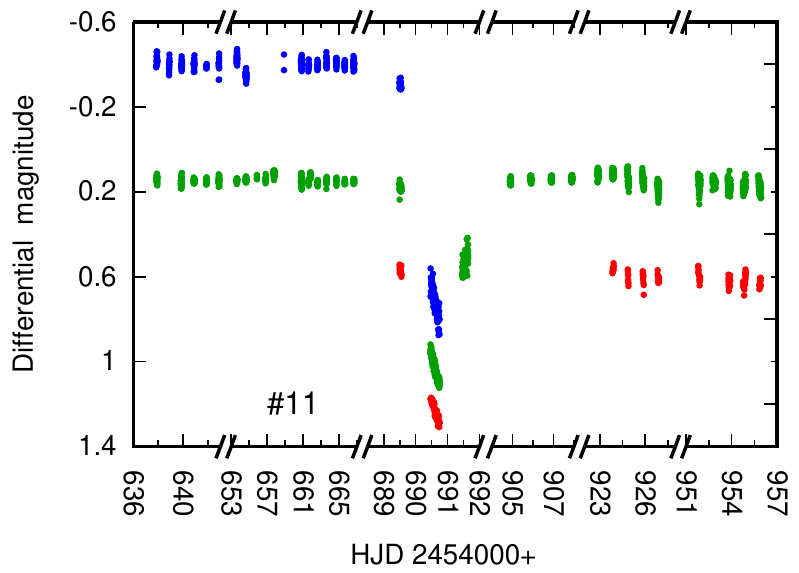}
\caption{Light curve of eclipsing system \#11. The blue, green and red dots represent $B$, $V$, and $I_{\rm C}$ data, respectively.} 
\label{f287}
\end{figure}

\begin{figure}
\centering
\includegraphics[width=76mm]{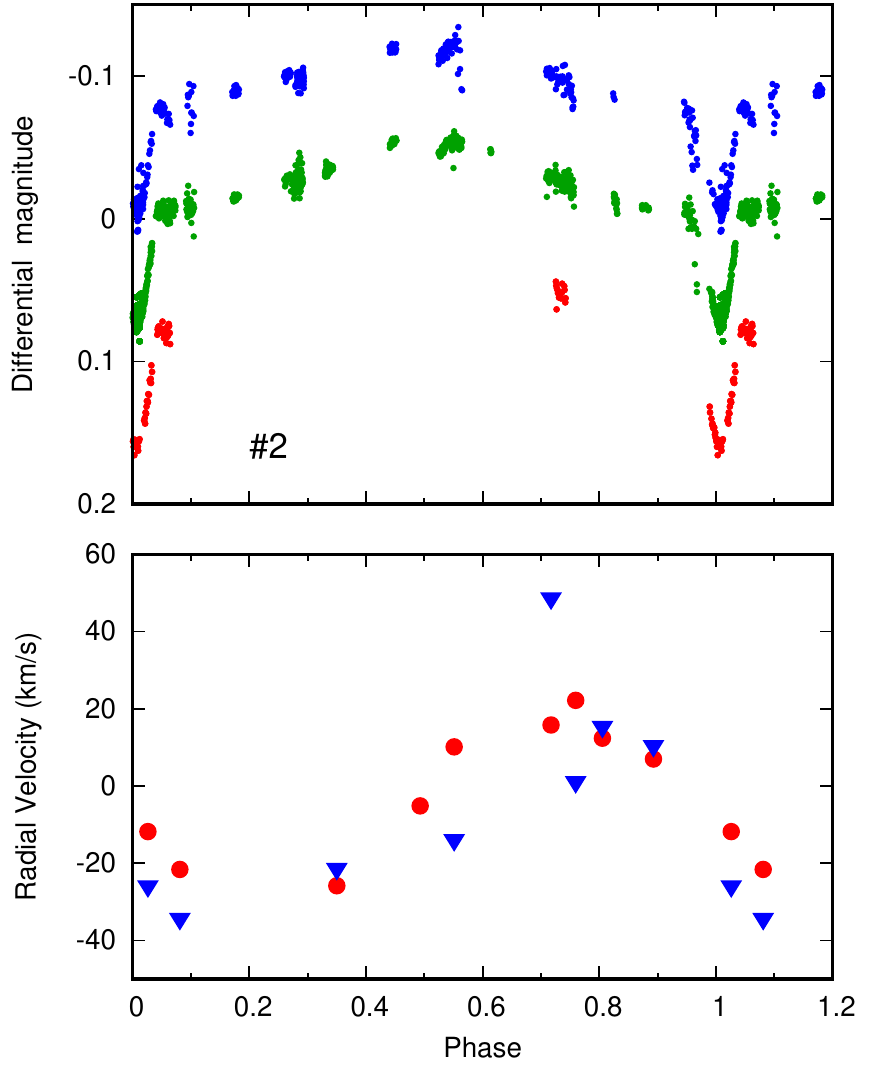}
\caption{{\it Top:} Phase diagrams of the $BVI_{\rm C}$ observations of the eclipsing system \#2, phased with $P=3.692046$ d. {\it Bottom:} Radial velocities for He I 4471 {\AA} (red dots) and He II 4686 {\AA} (blue triangles) derived by \protect\cite{2009MNRAS.400.1479S}.} 
\label{f19}
\end{figure}

\subsection{\label{secl}Binary stars}

In the observed field, we found 17 binary stars. Due to the saturation, we were unable to analyze a well-known spectroscopic binary, HD 186183 (W412, QR Ser), which is a probable eclipsing star \citep{2006SASS...25...47W}. The analysis of another binary star, BD $-13^{\circ}4923$ (W175), reveals no eclipse. This result confirms the low inclination predicted by  \cite{2009MNRAS.400.1479S}. The next interesting binary discovered recently by \cite{2021MNRAS.504.3203S}, BD $-13^{\circ}4937$ (W601), was outside the field of view in most of our images.

\begin{figure*}
\centering
\includegraphics[width=42pc]{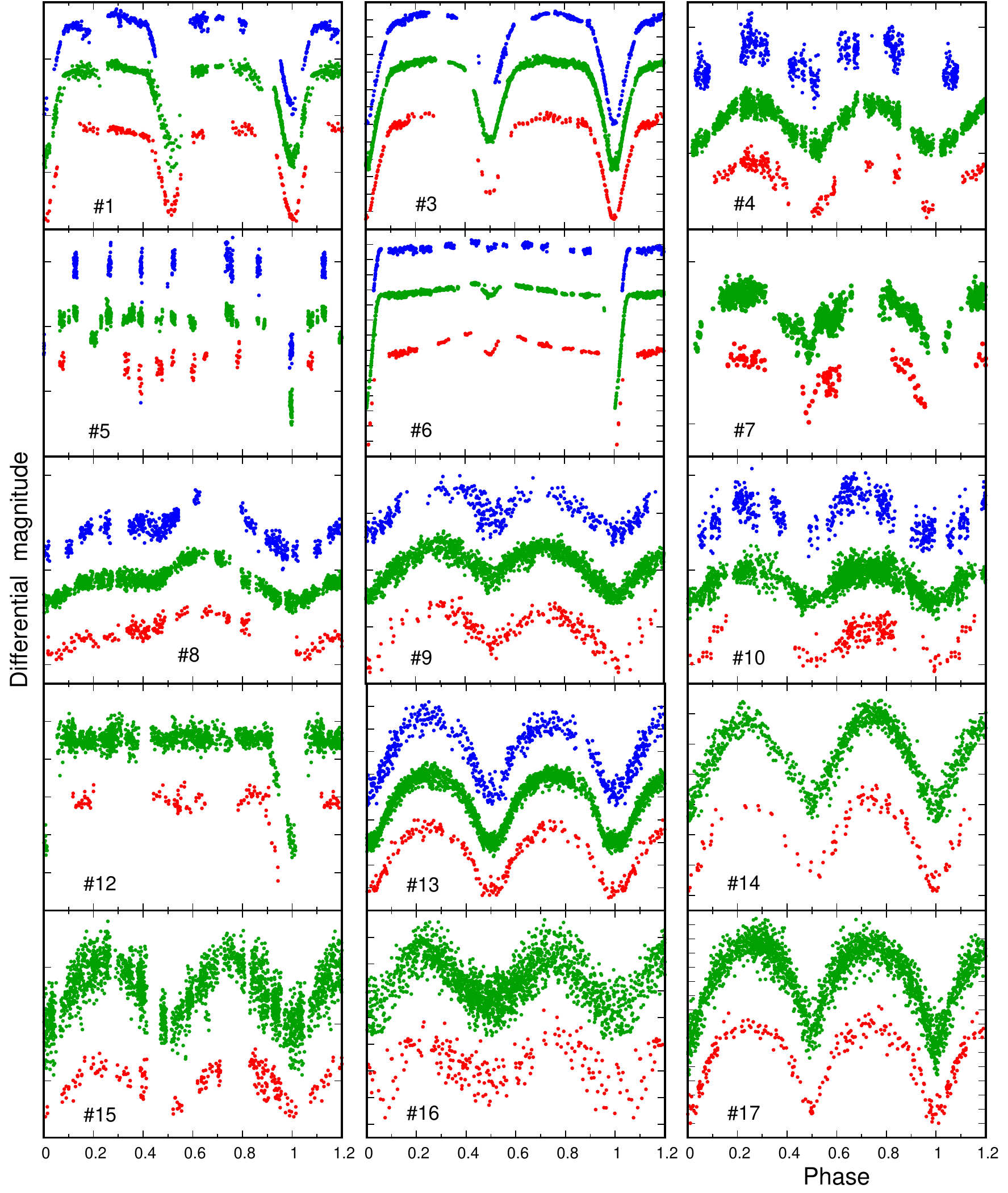}
\caption{Phased light curves of fifteen binary stars in NGC 6611. The blue, green and red dots represent the $B$, $V$, and $I_{\rm C}$ data, respectively. Ordinate ticks are separated by 0.1 mag.} 
\label{fecl}
\end{figure*}

The orbital periods, minimum light times, and variability ranges are listed in Table \ref{tecl}. We were unable to find the orbital period of one eclipsing star, \#11. This star's light curve is shown in Fig.~\ref{f287}. 
The phased light curves of the other fifteen binaries are shown in Fig.~\ref{fecl}.

The brightest eclipsing system in our sample is star \#1 (BD $-13^{\circ}4929$, W314). The binary nature was noted by Trumpler \citep{1961ApJ...133..438W} and was detected by \cite{1999RMxAA..35...85B} and \cite{2005A&A...437..467E}. This star was later discovered by \cite{2009MNRAS.400.1479S} to be a spectroscopic triple star (SB3) composed of a short-period binary (B0.5V+B0.5V) and an O7V star. The authors, however, found no evidence that the O-type star is gravitationally bound with the binary pair. We discovered it to be an eclipsing system with an orbital period equal to 1.475091 d.  

Star \#2 (BD $-13^{\circ}$4928, W280) was known as a Be star with a spectral type B0.5:V:ne \citep{1955ApJ...122..185H}. \cite{1993AJ....106.1906H} classified this star as O9.5 V with broad lines. Radial velocities derived by \cite{1999RMxAA..35...85B} showed no significant variations, and the authors considered the star to be constant.
We found this star to be an eclipsing system with an orbital period of $3.692046$ d (Fig.~\ref{f19}).
\cite{2009MNRAS.400.1479S} calculated the radial velocities for eleven lines, but they do not agree with each other. Using our period, we phased each set of radial velocities. We found the best agreement with the light curve for radial velocities derived from He I 4471 {\AA} and He II 4686 {\AA} lines. These radial velocities are shown in bottom panel of Fig.~\ref{f19}.

Another eclipsing system, star \#3 (W343), was classified as B1\,V by \cite{1993AJ....106.1906H} and \cite{2007AJ....133.1092W}.
This system was found to be an EA-type binary with an orbital period of $1.15023$ d \citep{2006SASS...25...47W}. 
\cite{2001A&A...379..147D} reported a visual companion of the star at a distance of 1.449$^{\prime\prime}$. The rotational velocity ($v\sin i$) of the star was found to be 300 km s$^{-1}$ \citep{2007AJ....133.1092W}.
The $BVI_C$ data phased with period of $1.1500665$ d are shown in Fig.~\ref{fecl}.

The next binary system we investigated was star \#4 (W444, ALS 15393). The spectral type B1.5\,V determined by \cite{1993AJ....106.1906H} was confirmed by \cite{2005A&A...437..467E} and \cite{2007AJ....133.1092W}.
The radial velocity measured by \cite{2005A&A...437..467E} is equal to 7 km\,s$^{-1}$, which is close to the average radial velocity of the cluster (10 km\,s$^{-1}$). The rotational velocity $v\sin i$ of the star \#4 is equal to 133 km\,s$^{-1}$ \citep{2007AJ....133.1092W}. \cite{2009A&A...496..453G} stated that the star is an X-ray source with no infrared excess, i.e.~it is a candidate for a star without a disk, while \cite{2012ApJ...753..117G}, based on optical and infrared colours, classified the star as a background object. The membership probability we found from the Gaia proper motions (Sect.~\ref{smem}) is 88\%. This probability is in good agreement with the value of 70\% determined by \cite{1999A&AS..134..525B}. In catalog of variable stars measured by ATLAS (Asteroid Terrestrial-impact Last Alert System), this star was classified as `dubious' with a period of 2.533454 d \citep{2018AJ....156..241H}. The light curve phased with a period 2.533757 d is shown in Fig.~\ref{fecl}. The shape and small amplitude indicate that the star may be an ellipsoidal variable. The system may be also EW-type binary with very small orbital inclination.

The next eclipsing binary is star \#5 (W188), classified as B0 V \citep{1993AJ....106.1906H}. \cite{2001A&A...379..147D} found a visual companion to this star at 0.543$^{\prime\prime}$ with an estimated mass of 2.4--2.6 M$_\odot$. Only a few points were detected during the two nights covering the eclipse, and thus we have determined a rather uncertain orbital period of 7.5956 d.

Star \#6 (W227) was classified as an emission-line star of spectral type B1.5 Ve \citep{1993AJ....106.1906H}. The star was found later to be a B2 V star \citep{2005A&A...437..467E}. \cite{2001A&A...379..147D} discovered a visual companion at the distance of 0.538$^{\prime\prime}$ with an estimated mass of 0.7 M$_\odot$. The variability of the star with period of 2.522625 d was announced by \cite{2018AJ....156..241H}. We found it to be an Algol type eclipsing system with an orbital period of 2.5224756 d. The phase diagram is shown in Fig.~\ref{fecl}.

Next binary system found in the observed field is star \#8 (W472). The star was classified as B3 Ve by \cite{1993AJ....106.1906H}. The emission in H$\alpha$ line has not been confirmed by other authors (\citealt{2001PASP..113..195H}, \citealt{2005A&A...437..467E}), but they suggest that the small emission in H$\alpha$ line may be due to the emission in nebula. \cite{2005A&A...437..467E} classified it as B3 V star with a companion. The binary nature of that star was confirmed by \cite{2008A&A...489..459M}. The authors derived the following fundamental parameters: $T_{\rm eff}=13000$ K, $\log g=3.2$ and $V\sin i=121$ km d$^{-1}$. They estimated the mass to be $M/M_{\odot}=5.8$, the luminosity $\log{L/L_{\odot}}=3.40$, and the radius $R/R_{\odot}=10.1$. The age of $73$ Myr that they obtained indicate that the star is not a member of the cluster. However, based on the proper motions from Gaia, we calculated that the membership probability is equal to 96.4\%. In the catalog of variable stars measured by ATLAS \citep{2018MNRAS.477.3145J}, the authors provide a variability period of 1.806230, although they classified the star as irregular. We found periodic variations of this star with a frequency of $f_1=0.553616$ d$^{-1}$, which corresponds to the period of 1.806307 d. The peculiar light curve, shown in Fig.~\ref{fecl}, reveals no eclipse. Since the secondary minimum is shifted from phase 0.5, the system most likely has an eccentric orbit.

Star \#9 (W310) is another binary system we have found. \cite{1993AJ....106.1906H} reported that the spectrum is a composite with features typical for B and G-type stars. The emission likely originates from circumstellar matter. The orbital period we derived is 0.392850 d, and we believe the star is not a cluster member.

For the star, \#10 (W262, ALS 15370), it was classified by \cite{1993AJ....106.1906H} as Ae, whilst \cite{1997A&AS..121..223D} classified it as a B7 star without any emission lines. The light curve (Fig.~\ref{fecl}) indicates, the star may be obscured by surrounding matter.

The orbital period of the EW-type binary \#13 (W258) has already been determined by \citet{2006SASS...25...47W}. The period is consistent with our value of $0.412470$ d. This star and four other eclipsing binaries, \#7, \#14, \#16, and \#17, are not cluster members. As can be see in Fig.~\ref{fecl}, the phase coverage of star \#5, \#7, and \#12 is incomplete, and therefore, their periods may be uncertain.

The frequencies corresponding to the periods of our five binary stars (star \#1, \#3, \#7, \#13, and \#17) were recently published in the Gaia DR3 variability catalogue \citet{2022yCat.1358....0G}. For two other binaries (star \#6 and \#16), the value of the frequency published by the authors was two times lower compared to our result.

\begin{table}
\centering
\caption{The parameters of binary stars in the field of NGC 6611. Numbers in parentheses denote the r.m.s.~errors of the preceding quantities with the leading zeroes omitted.}
\label{tecl}
\begin{tabular}{rlll}
\hline
\multicolumn{1}{c}{star}  &  \multicolumn{1}{c}{Period} & \multicolumn{1}{c}{$T_{\rm min}-T_0$} & \multicolumn{1}{c}{$\Delta V$} \\
 \multicolumn{1}{c}{\#}     & \multicolumn{1}{c}{[d]} &\multicolumn{1}{c}{[d]} &  [mag] \\
\noalign{\vskip1pt}\hline\noalign{\vskip1pt}
 1  & 1.475091(20) & \hspace{6pt}784.494(88) & \hspace{6pt}0.18 \\
 2  & 3.692046(56) & \hspace{6pt}775.428(14) & \hspace{6pt}0.12\\
 3  & 1.1500862(13)& \hspace{6pt}784.8521(46)& \hspace{6pt}0.61 \\
 4  & 2.533697(37) & \hspace{6pt}783.673(49) & \hspace{6pt}0.035\\
 5  & 7.59561(38)  & \hspace{6pt}784.21(56)  & \hspace{6pt}0.16\\
 6  & 2.522476(7)  & \hspace{6pt}786.0860(44)&$\sim$0.8  \\
 7  & 1.530347(39) & \hspace{6pt}785.715(91) & $\sim$0.05\\
 8  & 1.806307 (36)  & \hspace{6pt}785.5984(36) & \hspace{6pt}0.05\\
 9  & 0.392850(1)  & \hspace{6pt}784.2626(16)& \hspace{6pt}0.087 \\
10  & 1.169840(13) & \hspace{6pt}785.699(71) & \hspace{6pt}0.034\\
11  &  \multicolumn{1}{c}{--} & $\sim$690.76 & $\sim$1\\
12  & 1.581702(15) & \hspace{6pt}784.8095(56)& \hspace{6pt}0.28\\
13  & 0.4124697(4) & \hspace{6pt}784.550(20) & \hspace{6pt}0.31\\
14  & 0.348772(1)  & \hspace{6pt}784.029(12) & \hspace{6pt}0.30\\
15  & 2.206034(52)  & \hspace{6pt}783.50(23)  & \hspace{6pt}0.233\\
16  & 0.340387(2)  & \hspace{6pt}784.173(11) & \hspace{6pt}0.23\\ 
17  & 0.2721959(4) & \hspace{6pt}783.9597(26)& \hspace{6pt}0.70\\
\noalign{\vskip1pt}\hline\noalign{\vskip1pt}
\end{tabular}
\end{table}

\subsection{\label{sper}Other periodic variables}
We found 21 other periodic variables in the observed field. Unfortunately, these stars have no spectral classification. The periods, times of the maximum light and $V$-filter amplitudes are given in Table \ref{tsin}. The phased light curves of these stars are online in the electronic table in Appendix Fig.~A1.
Only four stars, \#26, \#27, \#28, and \#46 are not cluster members.
The periods, the amplitudes and the shape of light curves in combination with their position in colour-magnitude diagram indicate that these stars may be WTTS stars. The light variations of WTTS stars are mainly caused by cool spots rotating around the stars. In order to check for other type of light variations in these stars, such as flares, we subtracted the changes with main frequency from the original light curves. However, we did not find any additional variability.

Fifteen stars in this group have been identified as X-ray sources without infrared excess \citep{2009A&A...496..453G}. One of them, \#46, and three other stars (\#29, \#33, and \#44) were classified by \cite{2012ApJ...753..117G} as disk-bearing stars, while the rest as disk-less objects.
Similar to NGC 2244 \citep{ngc2244}, all periodic cluster members have periods larger than one day. However, in some young open clusters, such as NGC 2282 \citep{dutta2018} or Stock 8 \citep{2019AJ....158...68L}, a large fraction of the periodic variables have periods shorter than one day.


The distribution of rotation periods and their correlation with other stellar properties, such as mass, age, colour, have been studied by many authors. Some of them found that the distribution of rotation periods depends on mass and is unimodal for lower mass stars ($\leq 0.25 M_{\odot}$) rotating faster than higher mass stars, and bimodal for more massive stars ($\geq 0.25 M_{\odot}$)
(\citealt{2001ApJ...554L.197H} in Orion Nebula Cluster, \citealt{2005A&A...430.1005L} in NGC 2264). On the contrary, others (e.g.~\citealt{2012ApJ...747...51H} in NGC 6530) have found that the rotation periods have a unimodal distribution and that the lower mass stars rotate slower on average. To check the dependence of periods on mass for our stars, we have drawn our periodic variables in the colour-magnitude $V$ vs.~$(V-I)$ diagram. The objects are superimposed on the  evolutionary tracks taken from \cite{bress2012}. The colours of the symbols from light to dark green indicate the periods from the shortest to the longest. Three stars with $V-I<1.8$ mag are not cluster members. The remaining stars are distributed  uniformly in the wide range of masses.
We did not observe any correlation between variability period and age. 
However, the statistics of our sample is to poor to draw conclusions about correlation between mass and period. Moreover, the magnitudes and colours of the stars would need to be determined more precisely first, since (as we mentioned in Sect.\ref{sstransf}) we had some problems with the transformation to the standard system.
More properties of these variables are discussed in Sect.~\ref{sclass}.

\begin{table}
\centering
\caption{\label{tsin}Parameters of 21 periodic variable stars found in the observed field determined from the $V$-filter observations ($T_0={\rm HJD}2454000$). The numbers in parentheses denote the r.m.s.~errors of the preceding quantities with the leading zeroes omitted.}
  \begin{tabular}{@{}cllc}
  \hline
Star  & \multicolumn{1}{c}{Period} & \multicolumn{1}{c}{$T_{\rm max}-T_0$} & \multicolumn{1}{c}{$\Delta V$} \\
 & \multicolumn{1}{c}{[d]} & \multicolumn{1}{c}{[d]} & [mag] \\
\noalign{\vskip1pt}\hline\noalign{\vskip1pt}
26 & 0.462019(07) & 784.2858(27) & 0.024\\
27 & 1.19924(27)  & 784.5428(74) & 0.038\\
28 & 0.885414(30) & 784.3877(38) & 0.049\\
29 & 7.09957(94)  & 792.583(19)  & 0.163\\
30 & 4.96183(28)  & 787.6240(67) & 0.304\\
31 & 2.76215(15)  & 787.0732(80) & 0.143\\
32 & 2.86626(27)  & 786.896(14)  & 0.083\\
33 & 5.48703(56)  & 786.9072(14) & 0.221\\
34 & 2.313617(78) & 786.0491(47) & 0.196\\
35 & 1.262206(59) & 784.9040(44) & 0.102\\
36 & 1.219393(37) & 785.2882(36) & 0.251\\
37 & 2.69119(22)  & 785.072(10)  & 0.303\\
38 & 5.219092(91) & 782.543(25)  & 0.204\\
39 & 3.77175(39)  & 785.213(15)  & 0.306\\
40 & 5.29819(58)  & 787.854(15)  & 0.449\\
41 & 3.07131(56)  & 785.510(18)  & 0.313\\
42 & 1.154089(37) & 784.9332(43) & 0.357\\
43 & 1.134087(93) & 784.030(11)  & 0.174\\
44 & 2.17628(24)  & 786.529(12)  & 0.308\\
45 & 1.34165(11)  & 784.3147(77) & 0.418\\
46 & 1.56272(13)  & 784.8237(92) & 0.372\\
\hline
\end{tabular}
\end{table} 
\begin{figure}
\centering
\includegraphics[width=74mm]{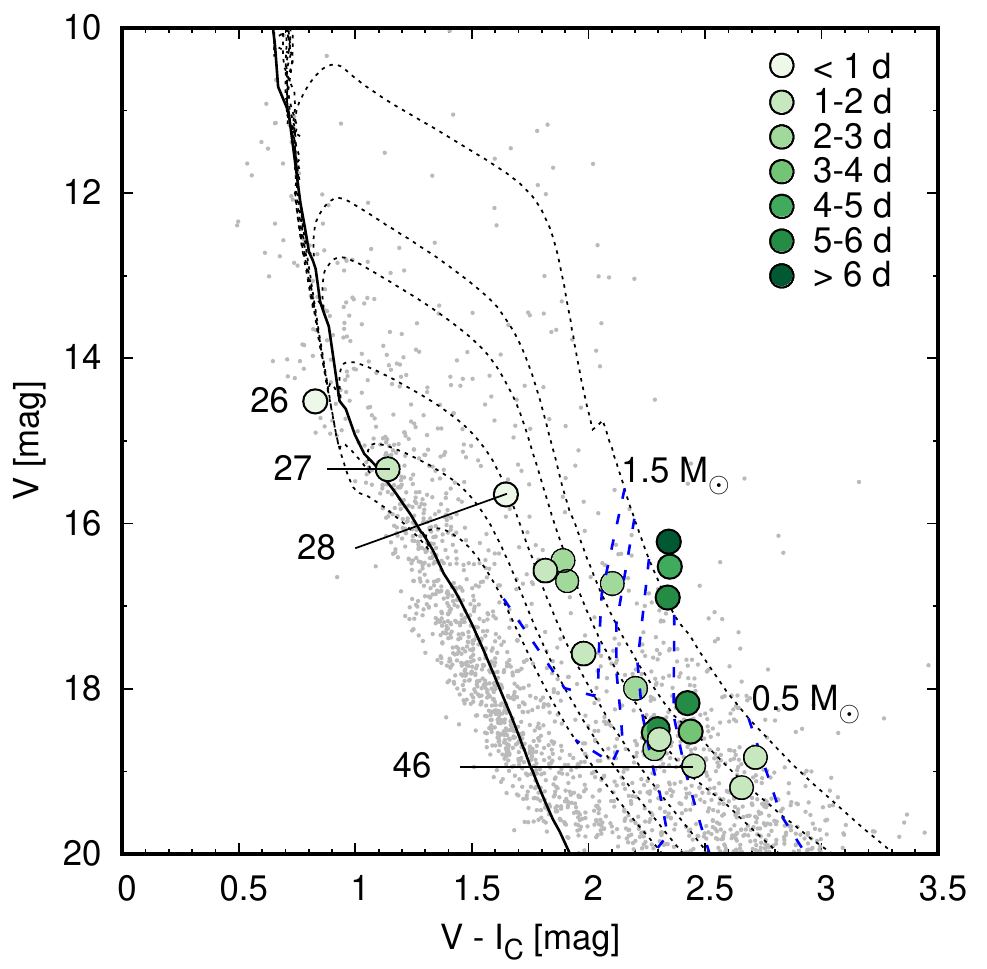}
\caption{Position of periodic variables in the colour-magnitude diagram. The ZAMS relation for Z=0.02 (thick line) was taken from \citet{pec2013}. The symbols denote the ranges of periods. Non-members are labeled with number. The isochrones (dotted lines) for 0.1, 0.5, 1, 3, 6, and 10\,Myr and mass tracks for 0.5, 0.75, 1, 1.25, and 1.5 M$_\odot$ (dashed blue lines) were taken from \protect\cite{bress2012}. We adopted the mean values of $E(V-I_{\rm C})=0.994$ mag, $A_V=2.6$ mag and $(m-M_{\rm V})_0=11.215$ mag.} 
\label{perio}
\end{figure}
\subsection{\label{sirr}Irregular variables}

In addition to the periodic variables described in Sects.~\ref{spuls}--\ref{sper}, we found 49 irregular variables. The membership probabilities determined from the Gaia proper motions (Sect.~\ref{smem}) indicate that ten of these stars are non-members. 
The light curves are available online in the electronic version in Appendix Fig.~A2. As can be seen in these figures, the range of variability is from 0.1 mag (\#47, \#48) to more than one magnitude (\#58, \#78, \#81, \#94, \#95). Their position in $V$ vs. $(V-I_{\rm C})$ diagram (Fig.~\ref{cmd}) indicates that many of them are young (0.1--3 Myr) PMS stars. As we mentioned in the introduction, the irregular light variations of PMS stars are associated with unstable accretion from the circumstellar disk, causing hot spots, obscuration by surrounding matter, and rotation of a star with asymmetrically distributed cool and hot spots \citep{herbst1994}. The photometric variability of PMS stars may be caused by flares. The typical feature of flares is fast rise of light and slow, exponential decay. In order to find this phenomenon, all light cures were carefully inspected by eyes prior and after rejection of outstanding data points. However, we found no evidence of flares in our data.

We cannot exclude that the variability of some of these stars are due to being members of binary or multiple systems. One star, \#49 (W494) turned out to be a Herbig Ae/Be variable \citep{1999AJ....118.1043H}. Three other stars (\#52, \#54, \#58) are known Orion-type variables \citep{2017ARep...61...80S}. 
About half of the irregular variables were classified as disk-bearing stars \citep{2012ApJ...753..117G}. They are indicated by `db' in Appendix Table A1. The classification of the identified irregular variables we found is discussed in Sect.~\ref{sclass}.

\section{IDENTIFICATION AND CLASSIFICATION OF PMS stars}\label{syso}

The PMS variable stars, described in Sect.~\ref{sper} and \ref{sirr}, can be easily classified on the basis of their infrared excess and the properties of the H$\alpha$ emission line.
We therefore had to use external near- and mid-infrared photometry and $riH\alpha$ photometry, which is described in the following subsections.

PMS stars and protostars are subgroups of Young Stellar Objects (YSOs). 
Based on their spectral index, YSOs have been divided into three main classes, I, II, and III \citep{lada1987}. This index is defined as the slope of the spectral energy distribution: 
\begin{equation}
\alpha=\frac{d\log(\lambda F_{\lambda})}{d\log\lambda},
\end{equation}
where $\lambda$ is in the range between 1 and 20\,$\mu$m.
Later, \citet{andre1993} discovered an extremely young object with a strong emission in the submillimeter domain. Such sources, called Class 0, are not detectable at $\lambda<10$ \,$\mu$m. 
Based on the $\alpha$ determinations in the range between 2.2 and 10\,$\mu$m, \cite{greene1994} proposed a fifth class, called the `flat spectrum' with $\alpha$ between $-$0.3 and 0.3. Objects with $\alpha>0.3$ were classified as Class I, with $-1.6<\alpha<-0.3$ as Class II, and with $\alpha<-1.6$ as Class III.

\citet{lada2006} noted that objects with thick disks have a spectral index of $\alpha > -$1.80, and objects with `anemic disks' (transition disk objects, objects with thin disks) have $-$2.56$<\alpha < -$1.8, and objects without disks have $\alpha <-$2.56. 
Following these authors, the spectral indices $\alpha$ of our variables were calculated for $\lambda$ between 3.6 and 8\,$\mu$m using the {\it Spitzer\/} photometry described in Sect.~\ref{sspitzer}. They are given in Table \ref{tubvi}. These indices were useful for classification of pre-main sequence objects, described in Sect.~\ref{sclass}.
\begin{figure}
\centering
\includegraphics[width=80mm]{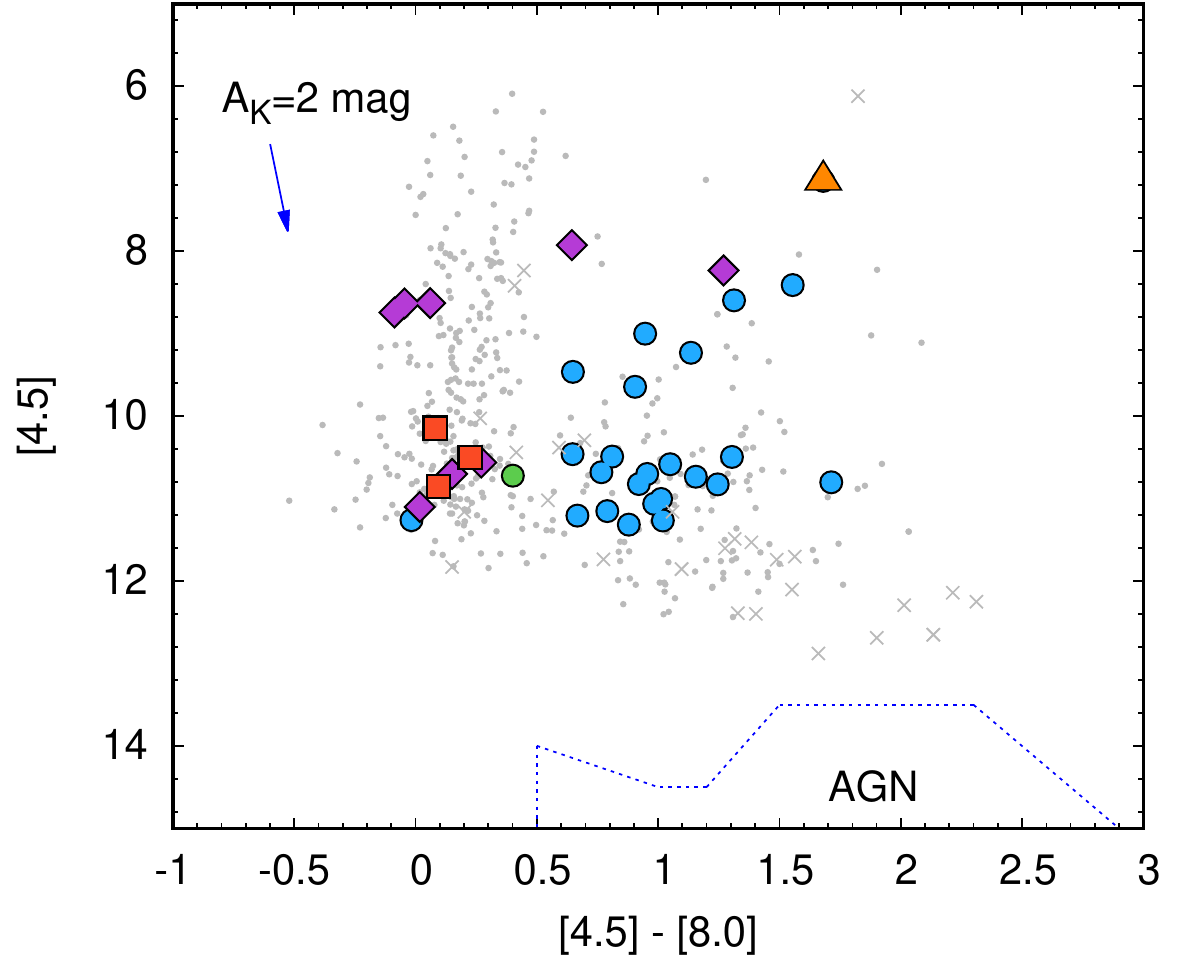}
\caption{IRAC colour-magnitude diagram. Symbols indicate variable stars found in this study: pulsating stars (red squares), eclipsing stars (pink diamonds), other periodic variables (green circles), HAeBe star (orange triangle), and the remaining variables (blue circles). The arrow shows the reddening vector \citep{flah2007}.} 
\label{cmdirac}
\end{figure}
\begin{figure}
\centering
\includegraphics[width=80mm]{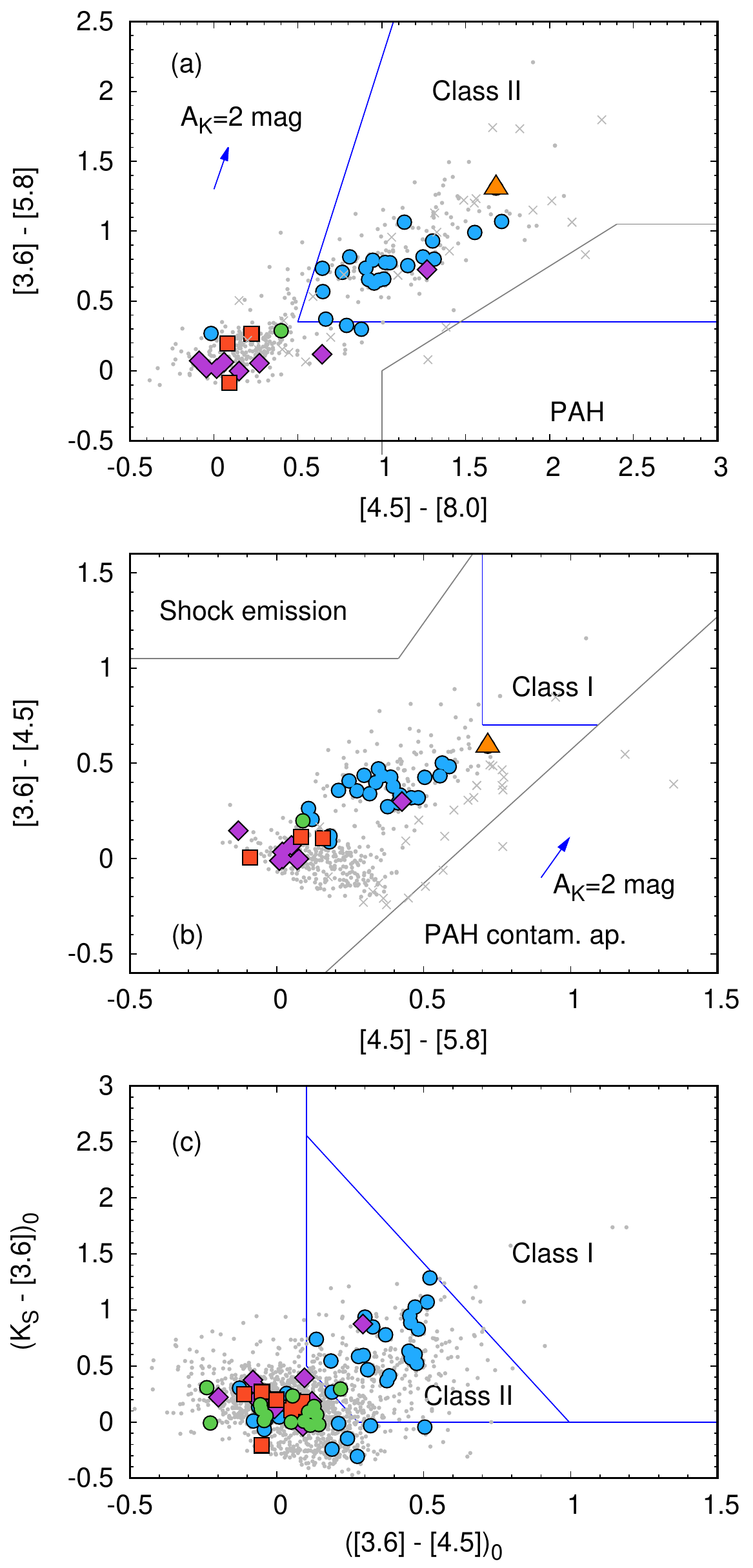}
\caption{IRAC colour-colour diagrams. The symbols are the same as in Fig.~\ref{cmdirac}.} 
\label{irac}
\end{figure}

\subsection{\label{sspitzer}\textbfit{Spitzer} and 2MASS photometry}
Mid-infrared photometry is very useful for understanding YSOs. Here, we used {\it Spitzer\/} photometry taken from the Galactic Legacy Infrared Mid-Plane Survey Extraordinaire (GLIMPSE) catalogue \citep{2009yCat.2293....0S} described by \cite{2003PASP..115..953B}. This four-band ([3.6], [4.5], [5.8] and [8.0] $\mu$m) photometry was used by \cite{2007ApJ...666..321I}, who found a large set of YSOs in the region of NGC 6611.
The data from {\it Spitzer\/} were also analyzed by \cite{2009A&A...496..453G}. Using $[3.6]-[4.5]$ {\it vs.}~$[5.8]-[8.0]$  colour-colour diagram and reddening-free colour indices, these authors published a catalogue of 474 candidates for pre-main sequence stars with disks and 790 without disks.

\begin{figure}
\centering
\includegraphics[width=80mm]{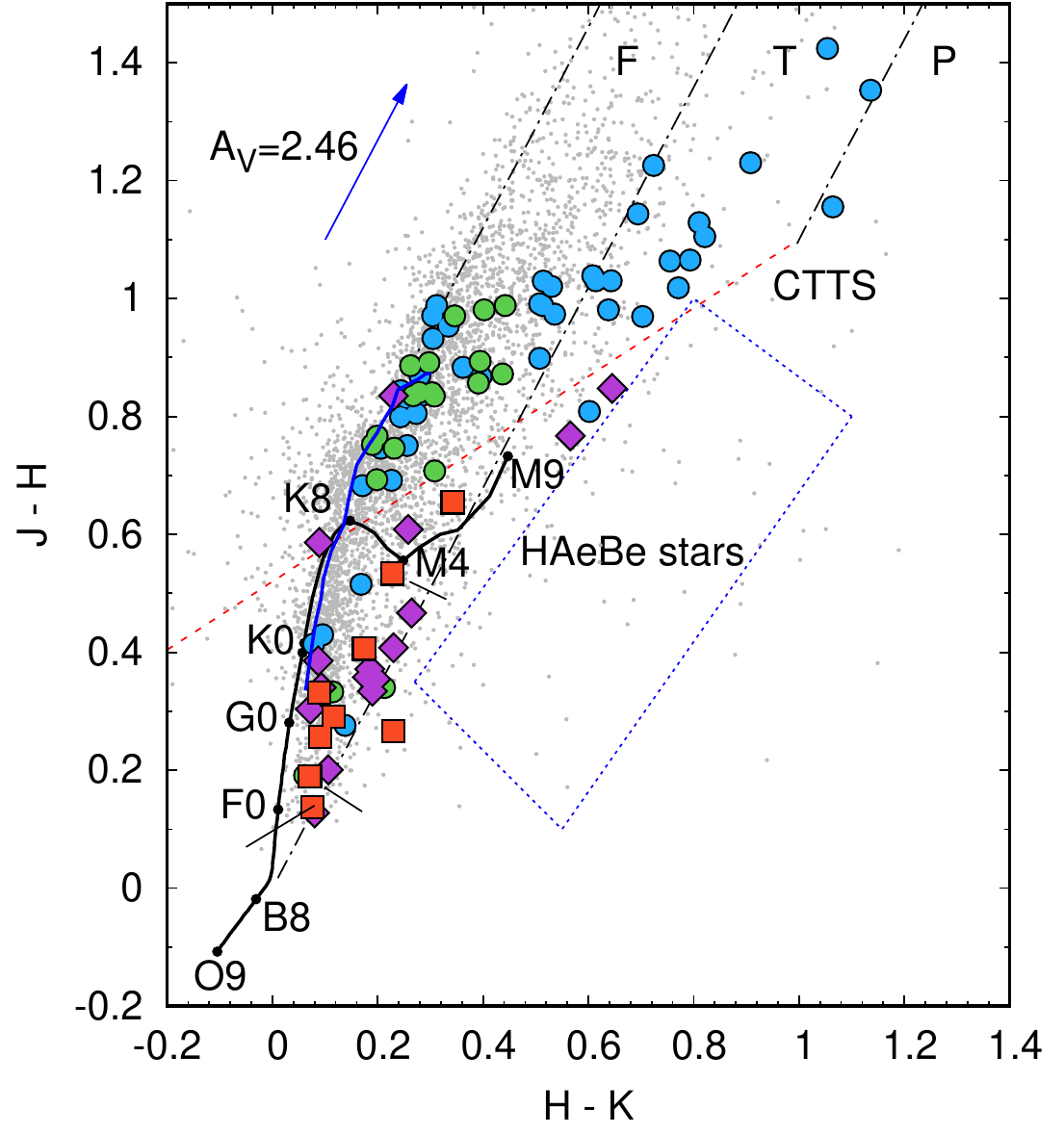}
\caption{$JHK$ UKIDSS colour-colour diagram for the observed field. The symbols are the same as in Fig.~\ref{cmd}. The dashed line is the position of CTTSs \citep{meyer1997}. The solid black line is the ZAMS \citep{pec2013}, and the solid blue line is late-type giants branch \citep{bebr1988}. The arrow shows the reddening vector for $A_V=2.6$ mag \citep{rieke1985}. The dot-and-dash lines are parallel to the reddening vector, and divided objects into three categories `F', `T', and `P' (see Sect.\ref{sukidss}).  The area delimited by the dotted blue line is the area of dereddened colours for HAeBe stars \citep{hern2005}.} 
\label{ukidss}
\end{figure}

To identify YSOs, we used the method described in Appendix A of \cite{guter2009}. Applying an offset of 1$\arcsec$ between our and {\it Spitzer} positions, we found about 3000 counterparts in the GLIMPSE catalogue. Only 573 of these have photometry in all four bands. First, we applied the colour and magnitude criteria of \cite{guter2009}. These allow us to identify extragalactic sources that may be misclassified as YSOs.  According to these criteria, two objects appear to be galaxies with bright polycyclic aromatic hydrocarbon (PAH) emission, and 29 sources have aperture photometry contaminated by PAH emission. These objects are marked with crosses in Figs.~\ref{cmdirac} and \ref{irac}. We did not find any AGN (active galactic nuclus) or shock emission knots.
We extracted Class II objects from the remaining sources, using the colour $[4.5]-[8.0]$ and $[3.6]-[5.8]$ constraints drawn in Fig.~\ref{irac}a.
If some of these objects had additional colours $[4.5]-[5.8]$ and $[3.6]-[4.5]$ greater than 0.7, we classified them as Class I objects (Fig.~\ref{irac}b).
Among the objects that lack photometry in [5.8] or [8.0] $\mu$m but have good-quality 2MASS photometry, we additionally found several YSOs using dereddened  $[3.6]-[4.5]$ and $K_S-[3.6]$ colours (Fig.~\ref{irac}c). 

None of the variables we found appears to be Class I object. Out of the 165 objects included in Class II, 20 stars are irregular variables and one is an eclipsing binary (star \#11). Most of these stars \cite{2009A&A...496..453G} are found to have an infrared excess and classified as candidates for objects with a disk. Only two objects, star \#71 and \#74, are classified as being diskless.

\subsection{\label{sukidss} UKIDSS photometry}
CTTSs can be easily recognized by their near-infrared excess. 
$JHK$ photometry was retrieved from the United Kingdom Infrared DeepSky Survey (UKIDSS), carried out with the United Kingdom Infrared Telescope (UKIRT) Wide-Field Camera \citep{2007A&A...467..777C}. For almost all the observed objects, we found counterparts in the UKIDSS-DR6 catalogue \citep{UKIDSS} within a search radius of 1$\arcsec$.
The $JHK$ data were converted to 2MASS photometry using the transformation relations of \cite{hewett2006}. For objects flagged as `close to saturated' in the UKIDSS-DR6 catalogue, we used magnitudes taken from the 2MASS All-Sky Catalog of Point Sources \citep{2003Curti}. The $(J-H)$ vs.~$(H-K)$ colour-colour diagram is shown in Fig.~\ref{ukidss}. The dashed red line indicates the intrinsic locus of CTTSs determined by \citet{meyer1997}. As this line was defined in the CIT (California Institute of Technology) system, the 2MASS magnitudes were transformed to this system using the relations described by \citet{2001AJ....121.2851C}.

Following \citet{sug2002}, we drew three lines parallel to the reddening vector: from the tip of the giant branch (left line), from the base (spectral type A0) of the MS stars (middle line), and from the tip of the intrinsic position of CTTS locus (right line).  
Objects located between the left and middle lines (`F' region) are either field stars (main-sequence or giants) or Class III/Class II sources with small near-infrared (NIR) excesses. Sources lying between the middle and right line (`T' region) are mainly CTTSs (Class II sources) with large NIR excesses or Herbig Ae/Be stars with small NIR excess. Sources lying redwards of the `T' region are most likely Class I objects (protostars) or Herbig Ae/Be stars. The blue rectangle denotes the typical dereddened $(J-H)$ and $(H-K)$ colours for HAeBe stars, as defined by \citet{hern2005}.
None of our variables fell into this area. The only HAeBe variable star in our data is flagged as saturated in the UKIDSS data.
The other known HAeBe star (W235) is not variable in our data and its UKIDSS photometry is  flagged as saturated. 
Out of 50 stars falling into the HAeBe area, only 9 have a membership probability greater than 50\%. They are all non-variable or constant in our data.

\begin{figure}
\centering
\includegraphics[width=76mm]{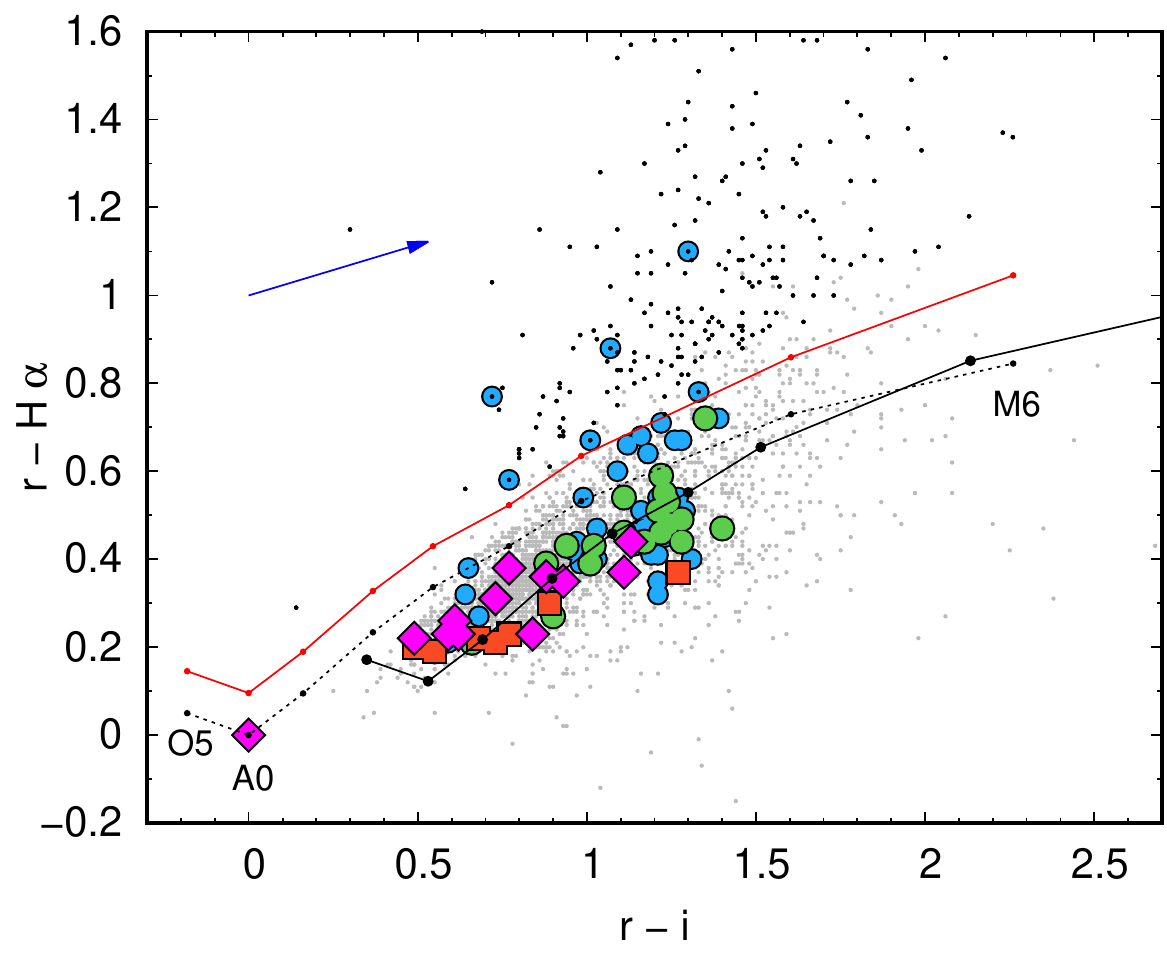}
\caption{VPHAS $(r-H{\alpha})$ vs.~$(r-i)$ colour-colour diagram for the observed field. Symbols denote variable stars found in this paper: pulsating (red squares), eclipsing (pink diamonds), HAeBe (orange triangles), other periodic (green circles) and irregular (blue circles).} 
\label{vphas}
\end{figure}
\begin{figure}
\centering
\includegraphics[width=80mm]{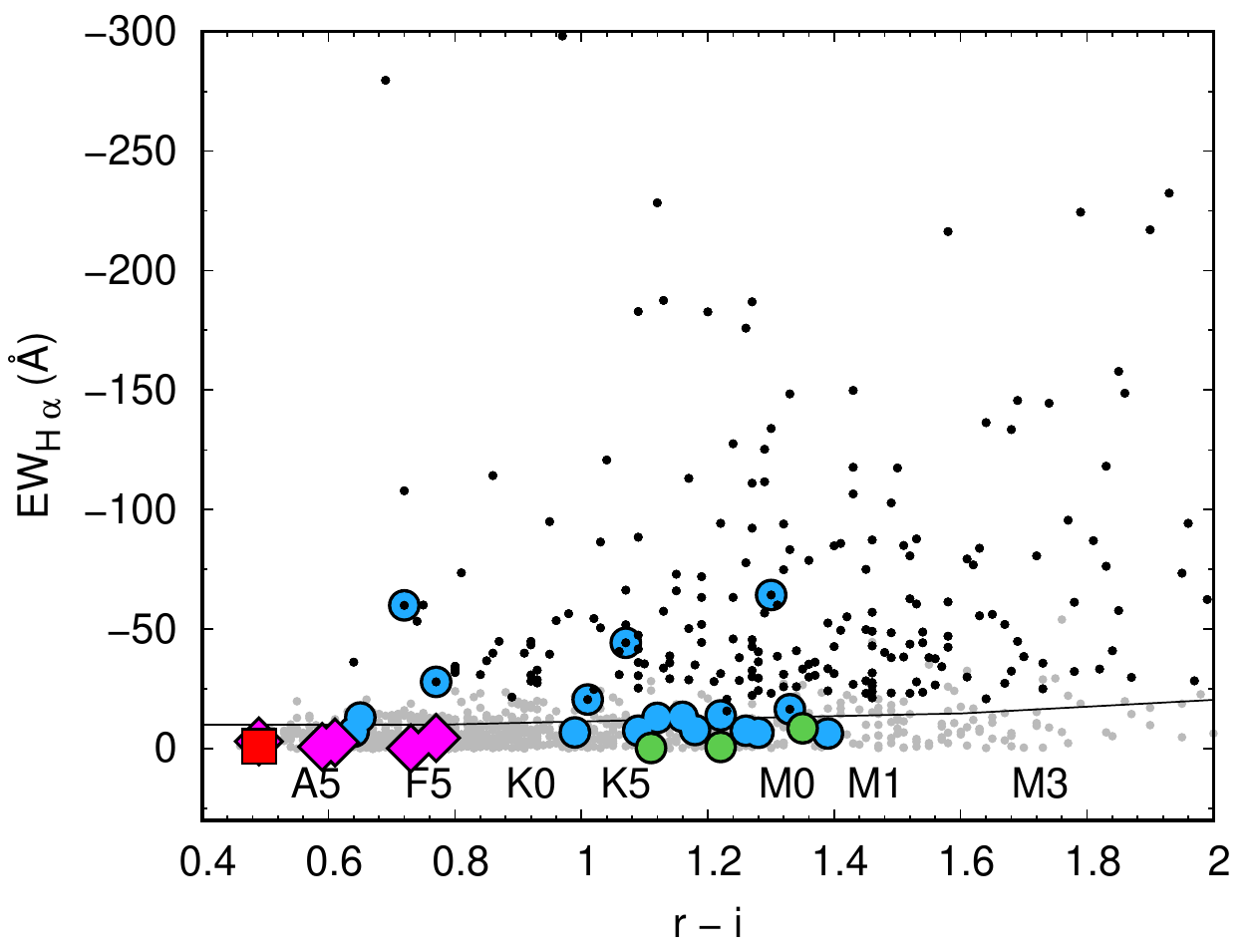}
\caption {Equivalent width $EW_{\rm H\alpha}$ vs.~$(r-i)$ colour for the observed field. The corresponding spectral types are labelled. The colour symbols are the same as in Fig.~\ref{vphas}. The black dots represent stars with emission in $H_\alpha$ (Sect.~\ref{svphas}).} 
\label{ew}
\end{figure}

\subsection{\label{svphas}VPHAS photometry}
A common feature of CTTSs is an equivalent width of the H$\alpha$ emission line greater than 10\,{\AA}. The intensity of the H$\alpha$ line can be measured by the colour index $(r-H\alpha)$ \citep{drew2005}. Therefore, we used the $riH\alpha$  photometry obtained during the VST Photometric H$\alpha$ Survey (VPHAS; \citealt{2014MNRAS.440.2036D}). We found counterparts in the search radius of 1$\arcsec$ for about 5300 observed stars. The $(r-H{\alpha})$ vs.~$(r-i)$ colour-colour  diagram for the observed field is shown in Fig.~\ref{vphas}. The black dashed line shows the position of the MS stars. The black continuous line is the same position shifted according to the reddening vector for $A_V=2.6$ mag.

The equivalent width ($EW_{\rm H\alpha}$) of each star was calculated by means of the code\footnote{\url{https://github.com/astroquackers/HaEW}} of \cite{2019MNRAS.484.5102K}. Following \citet{2010ApJ...715....1D}, the author defined $EW_{\rm H\alpha}$ as:

\begin{equation}
EW_{\rm H\alpha } = W\times [1-10^{0.4\times(r-H\alpha )_{\rm excess}}],
\end{equation}
where $W$ is the rectangular width of the $H\alpha$ filter and $(r-H\alpha)_{\rm excess}=(r-H\alpha)_{\rm observed}-(r-H\alpha)_{\rm model}$ is the colour excess caused by $H\alpha$ emission \citep{2015MNRAS.453.1026K}. 
To select a CTTSs candidate from the photometric $H\alpha$ emission, we used the same selection threshold as \citet{baren2011}. For early-type stars, this threshold (red line in Fig.~\ref{vphas}) is the same as a line for $EW_{\rm H\alpha}=10$~{\AA}, while for stars later than M0, the empirical threshold from \citet{barrado2003} was adopted. In total, we found 220 stars above the selection threshold, and about half of them are cluster members with a probability greater than 50\%. Six of stars are  variable. We cannot rule out that some stars may show emission in the $H\alpha$ line but were in quiescent phase of accretion (transitional objects) or the emission was not detected due to intrinsic variability or extinction.
The $EW_{\rm H\alpha}$ for the observed stars is shown in Fig.~\ref{ew}.

\subsection{\label{sclass}Classification of PMS stars}

The periodic and irregular variable stars, described in Sect.~\ref{sper} and \ref{sirr} are mainly PMS stars (HAeBe or TTSs). Based on the position in the colour-magnitude and colour-colour diagrams and the spectral index $\alpha$, we have classified 54 of them. The best candidates for CTTSs are five variables (\#52, \#56, \#85, \#86, and \#88) with an emission in $H{\alpha}$ (Sect.~\ref{svphas}). They all have a strong near-infrared excess and lie in the `T' region in the $(J-H)$ vs.~$(H-K)$ colour-colour diagram (Fig.~\ref{ukidss}). Due to the strong mid-infrared excesses, we classified them as Class II stars with a spectral index of $\alpha >-1.8$. The membership probabilities indicate that one of the CTTSs candidates, star \#85, is not a cluster member. Good candidates for CTTSs were 19 variables with near- and mid-infrared excess and $\alpha >-1.8$, but a photometric emission in $H{\alpha}$ line was not detected. These stars are marked as CTTS$^\star$ in Table \ref{tubvi}. Fifteen were classified as disk-bearing stars \citep{2012ApJ...753..117G}.

The stars with a near-infrared excess (lying above CTTS line in the $(J-H)$ vs.~$(H-K)$ colour-colour diagram in Fig.~\ref{ukidss}) and with a spectral index of $\alpha <-1.8$ (or without $\alpha$), we classified as WTTS. Seventeen are periodic variables, and 13 stars are irregular variables. 
Most of these stars were classified as disk-less stars \citep{2012ApJ...753..117G}.

The ratio of WTTS to CTTS stars counts is equal to 1.25, which is much higher than in case of the cluster of similar age, i.e., NGC 2244 \citep{ngc2244} that has a ratio of 0.7. This difference may be due to a less active star-forming process in NGC 6611 than in NGC 2244.


\section{Summary and conclusions}\label{ssum}
In this paper we present the results of a search for variable stars in the young open cluster, NGC 6611. Using combined data collected with three telescopes, we identified 95 variable stars. 
For 61 variables, the membership probability calculated from the Gaia proper motions is greater than 80\%.
Among 49 irregular variables, we found 24 CTTS and 13 WTTS candidates. It would be worthwhile to analyze the spectral energy distributions for these stars in the future to determine their stellar properties and of their disks, if they exist. An especially interesting issue worth exploring is if the variability of non-periodic WTTS stars are caused by short-lived irregular spots or by weak accretion from the surrounding thin disk. Out of the 21 periodic variables, we found 17 other WTTS candidates.
The large ratio of WTTS to CTTS stars suggests that the star-forming process is not very active in NGC 6611. As for NGC 2244 \citep{ngc2244}, we found no correlation between period of variability with mass or age.
However, our sample is too small for this conclusion to be definitive. More precise observations of NGC 6611 need to be carried out to find more periodic variables. The new data would be also used for the calculation of the standard magnitudes and colours of the stars. This future work would allow us to determine mass-period, age-period and amplitude-period relations, if they exist.


We detected seventeen eclipsing binaries in our field of view. Only two of them were previously known eclipsing binaries, and four other binary stars were newly identified as variables.
Eight other variables appear to be $\delta$ Scuti pulsating stars. Three are probably cluster members, which means they are pre-main sequence pulsators.

\section*{Data Availability}
The data underlying this article will be shared on reasonable request to the corresponding author.

\appendix
\section{}
In this section we present the coordinates, photometric data and some characteristics of variable stars. In Table \ref{tubvi} the individual columns are:\\
(1) identification number;\\
(2) and (3) right ascension and declination (J2000);\\
(4)--(6) average magnitudes and colours (see Sect.~\ref{sstransf});\\
(7)--(8) colours calculated from photometry carried out with 0.9-m CTIO telescope;\\
(9) classification taken from \citet{2009A&A...496..453G}:

-- A -- the source has an infrared excesses and is a candidate for having a circumstellar disk:

-- A01 -- member with a circumstellar disk classified based on its position in the IRAC colour-colour diagram;

-- A10 -- infrared excesses are detected only by Q indices ($Q_{VIJ[sp]} = \left( V-I \right) - \left( J-[sp] \right) \times E_{V-I}/E_{J-[sp]}$ where $[sp]$ is the magnitude in a given IRAC band);

-- A11 -- both diagnostics (IRAC colour-colour diagram and Q indices) detect emission from the disk;

-- B -- the star is an X-ray source without infrared excesses, candidate for a member without a disk;

-- C -- the star is an X-ray source without infrared excesses which optical colours are consistent with foreground main-sequence stars;\\
(10) Classification based on optical/infrared properties taken from \citet{2012ApJ...753..117G}:

-- db -- disk-bearing member;

-- dl -- disk-less member;\\
(11) $EW_{\rm H\alpha}$ -- equivalent width of H$\alpha$ emission computed by  the code of \citet{2019MNRAS.484.5102K} (Sect.~\ref{svphas});\\
(12) Classification based on $Spitzer$ photometry (Sect.~\ref{sspitzer}):

-- II -- Class II objects (II$^{\star}$ denote object with photometric uncertainty larger than required in method of \citet{guter2009} used to this classification);

-- III -- field stars and/or diskless YSOs (Class III);\\
(13) Area in $(J-H)-(H-K)$ colour-colour diagram in Fig.~\ref{ukidss};\\
(14) Spectral index $\alpha = d\log(\lambda F_{\lambda})/d\log(\lambda)$ calculated from the slope of the linear fit to the fluxes between the $K$ band and the IRAC [8] $\mu$m band \citep{lada2006};\\
(15) Membership probability determined in Sect.~\ref{smem};\\
(16) Variability type:

-- ECL -- eclipsing binary;

-- ELL -- ellipsoidal star;

-- CTTS -- the best candidates for CTTSs: Class II stars with photometric EW$_{\rm H\alpha}< -10$~{\AA} lying above CTTS line i in $(J-H)-(H-K)$ colour-colour diagram (Fig.~\ref{ukidss});

-- CTTS$^{\star}$ -- candidates for CTTSs lying above CTTS line (Class III stars with photometric EW$_{\rm H\alpha}<-10$~{\AA}  and Class II objects having spectral index $\alpha >-1.8$ with photometric EW$_{\rm H\alpha}>$ -10~{\AA}) or without EW$_{\rm H\alpha}$;

-- WTTS -- candidates for WTTSs: objects lying above the CTTS line for which no spectral index $\alpha$ was calculated or $\alpha <-1.8$;

-- PER -- periodic variable which was not classified as WTTS;

-- IRR -- irregular variable which was not classified as CTTS;\\
(17) object type taken from Simbad database;\\
(18) spectral type;\\
(19) WEBDA\footnote[2]{} ID number found within a search radius of 1$\arcsec$;\\
(20) Other name from GCVS, or HD catalogue.

Description of Table \ref{tir}:\\
(1) ID number;\\
(2) Gaia DR2 ID;\\
(3)--(5) Gaia DR2 magnitudes and colours;\\
(6)--(8) 2MASS photometry;\\
(9)--(11) UKIDSS photometry;\\
(12)--(15) IRAC photometry;\\
(16)--(18) VPHAS photometry.

\begin{landscape}
\begin{table}
\tiny
\centering
\caption{\label{tubvi}$UBVI_{\rm C}$ photometry and coordinates of variable stars found in NGC 6611 along with some important information about these stars.}
\begin{tabular}{@{}rccrccrrccccclrlccrc}
\hline
Star & R.Asc. & Dec. & \multicolumn{1}{c}{$V$}  & $V$--$I_{\rm C}$ & $B$--$V$ & \multicolumn{1}{c}{$B$--$V$} & \multicolumn{1}{c}{$U$--$B$} & G09 & G12 & $EW_{H\alpha}$ & Cl. & NIR & \multicolumn{1}{c}{$\alpha $}&  Prob. & Var. & Notes & Spectral & Webda & Other\\
  & (2000.0) & (2000.0) & \multicolumn{1}{c}{mag}  & mag & mag & \multicolumn{1}{c}{mag} & \multicolumn{1}{c}{mag} & & & & & reg. & & \multicolumn{1}{c}{[\%]} & type & & type & \multicolumn{1}{c}{ID} & name\\
 (1) & (2) & (3) & \multicolumn{1}{c}{(4)} &(5) & (6) & \multicolumn{1}{c}{(7)} & \multicolumn{1}{c}{(8)} & (9) & (10) & (11)\hspace{5pt} & (12) & (13) & \multicolumn{1}{c}{(14)} & \multicolumn{1}{c}{(15)} & (16) & (17) & (18) & \multicolumn{1}{c}{(19)} & (20) \\
\noalign{\vskip1pt}\hline\noalign{\vskip1pt}
\multicolumn{20}{l}{Eclipsing binaries}\\
\noalign{\vskip1pt}\noalign{\vskip1pt}
1 &  18:18:45.85 & -13:46:30.89 & 9.790 & 0.909 & 0.684 & 0.587 & -0.428 & B & dl & --\enspace & III & -- & -2.05 & 60.9 & ECL & YSO & B0.5V+B0.5V & 314 & BD -13$^\circ$4929\\
2 &  18:18:42.78 & -13:46:50.97 & 10.094 & 0.692 & 0.520 & 0.435 & -0.615 & B & dl & --\enspace & III & -- & -2.89 & 97.3 & ECL & Be$^\star$ & O9.5V & 280 & BD -13$^\circ$4928\\
3 &  18:18:49.38 & -13:46:50.04 & 11.792 & 1.331 & 0.854 & 0.841 & 0.233 & -- & -- & --\enspace & III & -- & -2.92 & 18.4 & ECL & Star & B1V & 343 & ALS 15387\\
4 &  18:19:00.43 & -13:42:41.02 & 12.603 & 1.341 & 0.925 & 0.811 & -0.019 & B & -- & --\enspace & -- & -- & -2.17 & 94.5 & ELL & YSO & B1.5V & 444 & ALS 15393\\
5 &  18:18:33.72 & -13:40:58.86 & 13.019 & 2.088 & 1.438 & --\hspace{6pt} & --\hspace{6pt} & B & dl & --\enspace & III & -- & -2.77 & 20.8 & ECL: & YSO & B0V & 188 & ALS 15355\\
6 &  18:18:38.40 & -13:47:09.06 & 12.794 & 0.958 & 0.693 & 0.604 & -0.229 & B & dl & -2.6 & III & -- & -2.50 & 95.5 & ECL & Em$^\star$ & B1.5Ve & 227 & ALS 15364\\
7 &  18:18:21.78 & -13:55:05.15 & 12.821 & 0.962 & 0.876 & 0.777 & 0.255 & -- & -- & -3.0 & III & -- & -2.69 & 0.0 & ECL & Star & - & 94 & \\
8 &  18:19:04.71 & -13:44:44.56 & 12.846 & 0.737 & 0.575 & 0.480 & -0.211 & -- & -- & -1.0 & -- & -- & -2.72$^\star$ & 96.4 & ELL & Em$^\star$ & B3Ve & 472 & \\
9 &  18:18:46.13 & -13:54:37.18 & 13.483 & 0.974 & 0.879 & 0.761 & 0.228 & B & dl & -0.7 & -- & -- & -2.95$^\star$ & 0.1 & ECL & YSO & B+G+emis. & 310 & ALS 15383\\
10 &  18:18:41.54 & -13:48:40.41 & 13.792 & 0.978 & 0.742 & 0.649 & 0.035 & -- & -- & --\enspace & III & -- & -2.71 & 81.6 & ECL & Em$^\star$ & B7 & 262 & ALS 15370\\
11 &  18:18:41.58 & -13:46:30.92 & 14.735 & 1.832 & 1.415 & 1.389 & -0.611 & A11 & -- & --\enspace & II & -- & -1.03 & 96.5 & ECL & Em$^\star$ & G: & 266 & \\
12 &  18:18:34.63 & -13:39:48.99 & 15.495 & 3.160 & 2.135 & --\hspace{6pt} & --\hspace{6pt} & -- & -- & --\enspace & -- & -- & \quad-- & 90.0 & ECL & YSOC & - & 2303 & \\
13 &  18:18:41.16 & -13:48:21.10 & 15.265 & 1.165 & 1.116 & 1.018 & 0.363 & -- & -- & -0.1 & -- & -- & -1.75$^\star$ & 0.0 & ECL & Star & - & 258 & \\
14 &  18:18:32.69 & -13:54:47.81 & 16.813 & 1.369 & 1.289 & --\hspace{6pt} & --\hspace{6pt} & -- & -- & -4.5 & -- & -- & \quad-- & 0.0 & ECL & -- & -- & 15536 & \\
15 &  18:18:41.43 & -13:47:08.33 & 17.599 & 2.243 & 1.654 & --\hspace{6pt} & --\hspace{6pt} & B & dl & --\enspace & -- & -- & -1.45$^\star$ & 96.9 & ECL & YSO & - & 16215 & \\
16 &  18:18:33.37 & -13:46:56.44 & 17.794 & 1.498 & 1.286 & --\hspace{6pt} & --\hspace{6pt} & C & dl & --\enspace & -- & -- & \quad-- & 0.0 & ELL & YSO & - & 17759 & \\
17 &  18:18:39.01 & -13:42:30.79 & 18.096 & 1.656 & 1.385 & --\hspace{6pt} & --\hspace{6pt} & -- & -- & --\enspace & -- & F & \quad-- & 0.0 & ECL & -- & -- & 6338 & \\
\noalign{\vskip1pt}\noalign{\vskip1pt}
\multicolumn{20}{l}{Pulsating stars}\\
\noalign{\vskip1pt}\noalign{\vskip1pt}
18 &  18:18:27.79 & -13:55:04.86 & 12.845 & 0.924 & 0.881 & 0.768 & 0.450 & -- & -- & --\enspace & III & -- & -2.76 & 0.0 & DSCT & Star & F0III/V & 135 & \\
19 &  18:18:53.27 & -13:46:07.18 & 13.363 & 1.360 & 1.067 & 0.951 & 0.625 & B & dl & --\enspace & III & -- & -2.62 & 94.8 & DSCT & Em$^\star$ & F2e & 374 & \\
20 &  18:18:37.64 & -13:45:13.22 & 14.461 & 1.584 & 1.162 & 1.040 & 0.683 & B & dl & --\enspace & III & -- & -2.45 & 97.0 & DSCT & Em$^\star$ & B8Ve & 221 & \\
21 &  18:18:49.36 & -13:39:08.36 & 14.266 & 1.053 & 0.903 & --\hspace{6pt} & --\hspace{6pt} & -- & -- & --\enspace & -- & -- & -2.15$^\star$ & 0.0 & DSCT & Star & A7III & 347 & \\
22 &  18:19:00.93 & -13:38:49.07 & 14.292 & 0.895 & 0.793 & --\hspace{6pt} & --\hspace{6pt} & -- & -- & --\enspace & -- & -- & \quad-- & 0.0 & DSCT & YSO & - & 451 & \\
23 &  18:19:07.43 & -13:40:37.93 & 15.275 & 2.439 & 1.653 & --\hspace{6pt} & --\hspace{6pt} & -- & dl & --\enspace & -- & -- & -2.50$^\star$ & 91.5 & DSCT & YSO & - & 491 & \\
24 &  18:18:42.04 & -13:47:14.02 & 15.431 & 1.198 & 1.001 & 0.890 & 0.456 & A10 & -- & --\enspace & -- & -- & \quad-- & 0.0 & DSCT & YSOC & - & 270 & \\
25 &  18:18:59.58 & -13:45:39.46 & 15.474 & 1.196 & 0.963 & 0.849 & 0.600 & -- & -- & --\enspace & -- & -- & \quad-- & 93.5 & DSCT & Star & - & 435 & \\
\noalign{\vskip1pt}\noalign{\vskip1pt}
\multicolumn{20}{l}{Other periodic variables}\\
\noalign{\vskip1pt}\noalign{\vskip1pt}
26 &  18:19:04.82 & -13:52:32.36 & 14.522 & 0.828 & 0.761 & 0.665 & 0.249 & -- & -- & --\enspace & -- & -- & \quad-- & 0.0 & PER & Star & - & 465 & \\
27 &  18:18:14.15 & -13:43:45.60 & 15.341 & 1.139 & 0.950 & --\hspace{6pt} & --\hspace{6pt} & -- & -- & --\enspace & -- & -- & \quad-- & 0.0 & PER & Star & - & 59 & \\
28 &  18:18:29.54 & -13:42:30.33 & 15.649 & 1.645 & 1.212 & 1.095 & 0.580 & -- & -- & --\enspace & -- & -- & \quad-- & 1.6 & PER & Star & - & 154 & \\
29 &  18:18:47.79 & -13:48:15.34 & 16.217 & 2.344 & 1.886 & --\hspace{6pt} & --\hspace{6pt} & A10 & db & --\enspace & -- & F & -2.75$^\star$ & --\enspace & WTTS & YSO & - & 16091 & \\
30 &  18:19:03.16 & -13:45:11.23 & 16.521 & 2.349 & 1.867 & --\hspace{6pt} & --\hspace{6pt} & B & dl & --\enspace & -- & F & -2.49$^\star$ & 96.4 & WTTS & YSO & - & 456 & \\
31 &  18:18:39.72 & -13:48:41.22 & 16.444 & 1.891 & 1.549 & --\hspace{6pt} & --\hspace{6pt} & B & dl & --\enspace & -- & F & -2.81$^\star$ & 93.9 & WTTS & YSO & - & 1182 & \\
32 &  18:18:36.23 & -13:46:19.16 & 16.724 & 2.103 & 1.639 & --\hspace{6pt} & --\hspace{6pt} & B & dl & --\enspace & -- & F & -2.01$^\star$ & 96.7 & WTTS & YSO & - & 16600 & \\
33 &  18:18:45.96 & -13:47:52.67 & 16.895 & 2.340 & 1.777 & --\hspace{6pt} & --\hspace{6pt} & A10 & db & --\enspace & III & F & -2.18 & 96.7 & WTTS & YSO & - & 16758 & \\
34 &  18:18:30.41 & -13:51:47.24 & 16.695 & 1.910 & 1.644 & --\hspace{6pt} & --\hspace{6pt} & B & dl & --\enspace & -- & F & -2.06$^\star$ & 96.7 & WTTS & YSO & - & 15742 & \\
35 &  18:18:45.92 & -13:55:22.03 & 16.569 & 1.815 & 1.510 & --\hspace{6pt} & --\hspace{6pt} & B & dl & --\enspace & -- & -- & \quad-- & 95.9 & PER & YSO & - & 15488 & \\
36 &  18:18:46.02 & -13:50:09.07 & 17.572 & 1.979 & 1.533 & --\hspace{6pt} & --\hspace{6pt} & B & dl & --\enspace & -- & F & \quad-- & 88.3 & WTTS & YSO & - & 15877 & \\
37 &  18:18:41.86 & -13:47:23.07 & 17.991 & 2.202 & 1.385 & --\hspace{6pt} & --\hspace{6pt} & B & dl & --\enspace & -- & F & \quad-- & 96.5 & WTTS & YSO & - & 17002 & \\
38 &  18:19:01.20 & -13:47:19.53 & 18.172 & 2.425 & 1.311 & --\hspace{6pt} & --\hspace{6pt} & A10 & dl & --\enspace & -- & F & \quad-- & 87.8 & WTTS & YSO & - & 13002 & \\
39 &  18:19:07.26 & -13:43:47.98 & 18.519 & 2.439 & -- & --\hspace{6pt} & --\hspace{6pt} & B & dl & --\enspace & -- & F & \quad-- & 55.9 & WTTS & YSO & - & 4610 & \\
40 &  18:18:41.93 & -13:48:59.90 & 18.486 & 2.297 & 1.136 & --\hspace{6pt} & --\hspace{6pt} & B & dl & --\enspace & -- & F & \quad-- & 96.8 & WTTS & Em$^\star$ & - & 17389 & \\
41 &  18:18:42.04 & -13:53:17.23 & 18.530 & 2.280 & 1.325 & --\hspace{6pt} & --\hspace{6pt} & B & dl & -0.3 & -- & F & \quad-- & 93.4 & WTTS & YSO & - & 15644 & \\
42 &  18:18:59.26 & -13:46:41.88 & 18.608 & 2.304 & -- & --\hspace{6pt} & --\hspace{6pt} & B & dl & --\enspace & -- & F & \quad-- & 87.4 & WTTS & YSO & - & 13088 & \\
43 &  18:18:26.05 & -13:47:35.88 & 18.826 & 2.717 & -- & --\hspace{6pt} & --\hspace{6pt} & B & dl & --\enspace & -- & F & \quad-- & 79.7 & WTTS & YSO & - & 16165 & \\
44 &  18:18:41.81 & -13:47:29.21 & 18.727 & 2.284 & -- & --\hspace{6pt} & --\hspace{6pt} & B & db & -0.7 & -- & F & \quad-- & 96.3 & WTTS & YSO & - & 16537 & \\
45 &  18:18:30.77 & -13:43:46.63 & 19.193 & 2.658 & -- & --\hspace{6pt} & --\hspace{6pt} & B & dl & --\enspace & -- & F & \quad-- & 83.2 & WTTS & YSO & - & 7906 & \\
46 &  18:18:58.14 & -13:54:32.03 & 18.933 & 2.452 & -- & --\hspace{6pt} & --\hspace{6pt} & B & db & -8.5 & -- & F & \quad-- & 11.4 & WTTS & YSO & - & 13244 & \\
\noalign{\vskip1pt}\noalign{\vskip1pt}
\noalign{\vskip1pt}\hline\noalign{\vskip1pt}
\end{tabular}
 \end{table} 
\end{landscape}
\begin{landscape}
\begin{table}
\tiny 
\centering
\contcaption{}
\begin{tabular}{@{}rccrccrrccccclrlccrc}
\hline
Star & R.Asc. & Dec. & \multicolumn{1}{c}{$V$}  & $V$--$I_{\rm C}$ & $B$--$V$ & \multicolumn{1}{c}{$B$--$V$} & \multicolumn{1}{c}{$U$--$B$} & G09 & G12 & $EW_{H\alpha}$ & Cl. & NIR & \multicolumn{1}{c}{$\alpha $}&  Prob. & Var. & Notes & Spectral & Webda & Other\\
  & (2000.0) & (2000.0) & \multicolumn{1}{c}{mag}  & mag & mag & \multicolumn{1}{c}{mag} & \multicolumn{1}{c}{mag} & & & & & reg. & & \multicolumn{1}{c}{[\%]} & type & & type & \multicolumn{1}{c}{ID} & name\\
 (1) & (2) & (3) & \multicolumn{1}{c}{(4)} &(5) & (6) & \multicolumn{1}{c}{(7)} & \multicolumn{1}{c}{(8)} & (9) & (10) & (11)\hspace{5pt} & (12) & (13) & \multicolumn{1}{c}{(14)} & \multicolumn{1}{c}{(15)} & (16) & (17) & (18) & \multicolumn{1}{c}{(19)} & (20) \\
 \noalign{\vskip1pt}\hline\noalign{\vskip1pt}
\multicolumn{20}{l}{Irregular variables}\\
\noalign{\vskip1pt}
47 &  18:18:46.61 & -13:46:59.89 & 13.572 & 0.954 & 0.725 & 0.626 & 0.010 & -- & dl & --\enspace & -- & -- & -3.73$^\star$ & 96.4 & IRR & Em$^\star$ & B8V:e & 322 & \\
48 &  18:18:59.16 & -13:39:55.32 & 14.204 & 1.219 & 1.079 & --\hspace{6pt} & --\hspace{6pt} & -- & -- & --\enspace & -- & -- & -2.41$^\star$ & 0.0 & IRR & YSO & - & 439 & \\
49 &  18:19:09.39 & -13:50:41.20 & 13.945 & 1.027 & 0.752 & 0.661 & 0.237 & A01 & -- & --\enspace & II & -- & -0.20 & 95.4 & Haebe & Ae$^\star$ & Be & 494 & EU  Ser\\
50 &  18:18:33.28 & -13:54:11.17 & 14.850 & 1.354 & 1.216 & 1.063 & 0.595 & B & dl & -7.0 & -- & -- & -3.01$^\star$ & 0.0 & IRR & YSO & - & 172 & \\
51 &  18:18:39.37 & -13:47:11.79 & 15.863 & 1.899 & 1.619 & --\hspace{6pt} & --\hspace{6pt} & -- & dl & --\enspace & III & -- & -2.73 & 97.0 & IRR & YSO & - & 642 & \\
52 &  18:19:13.82 & -13:58:44.86 & --\hspace{6pt} & -- & -- & --\hspace{6pt} & --\hspace{6pt} & A11 & -- & -60.0 & II & P & -0.94 & 95.5 & CTTS & Or$^\star$ & - & 1289 & EV  Ser\\
53 &  18:18:37.56 & -14:01:12.75 & --\hspace{6pt} & -- & -- & --\hspace{6pt} & --\hspace{6pt} & -- & -- & -13.0 & -- & -- & \quad-- & 0.0 & IRR & -- & -- & 16643 & \\
54 &  18:18:58.53 & -13:48:28.43 & 16.995 & 2.241 & 1.770 & --\hspace{6pt} & --\hspace{6pt} & A01 & db & --\enspace & II & -- & -1.71 & 96.0 & IRR & Or$^\star$ & - & 14021 & FN  Ser\\
55 &  18:18:48.23 & -13:49:08.27 & 16.953 & 2.272 & 1.633 & --\hspace{6pt} & --\hspace{6pt} & A11 & db & -13.3 & II & T & -1.27 & 95.7 & CTTS$^\star$ & Em$^\star$ & - & 16907 & \\
56 &  18:19:20.89 & -13:42:04.14 & 16.935 & -- & 1.263 & --\hspace{6pt} & --\hspace{6pt} & A11 & -- & -27.8 & II & T & -0.54 & 96.9 & CTTS & YSOC & - & 546 & \\
57 &  18:19:20.03 & -13:49:12.82 & 16.723 & -- & 1.480 & --\hspace{6pt} & --\hspace{6pt} & A11 & db & --\enspace & II & -- & -1.20 & 96.0 & IRR & YSO & - & 12804 & \\
58 &  18:18:39.83 & -13:47:44.40 & 16.661 & 2.114 & 1.868 & --\hspace{6pt} & --\hspace{6pt} & B & dl & -7.9 & -- & -- & \quad-- & 0.0 & IRR & Star & - & 1767 & FM  Ser\\
59 &  18:19:24.24 & -13:45:47.65 & 16.922 & -- & 1.598 & --\hspace{6pt} & --\hspace{6pt} & -- & dl & --\enspace & -- & F & \quad-- & 0.0 & WTTS & -- & -- & 10134 & \\
60 &  18:19:06.61 & -13:45:36.61 & 17.073 & 2.541 & 1.790 & --\hspace{6pt} & --\hspace{6pt} & A10 & db & --\enspace & -- & T & -1.38$^\star$ & 95.8 & CTTS$^\star$ & YSO & - & 4394 & \\
61 &  18:18:49.90 & -13:46:59.65 & 17.422 & 2.542 & 1.874 & --\hspace{6pt} & --\hspace{6pt} & A11 & db & --\enspace & II & T & -0.94 & 97.0 & CTTS$^\star$ & YSO & - & 13049 & \\
62 &  18:19:14.81 & -13:46:14.24 & 17.183 & 2.310 & 1.659 & --\hspace{6pt} & --\hspace{6pt} & B & dl & --\enspace & -- & F & -2.48$^\star$ & 5.3 & WTTS & YSO & - & 13134 & \\
63 &  18:18:49.84 & -13:48:17.42 & 17.376 & 2.289 & 1.804 & --\hspace{6pt} & --\hspace{6pt} & B & dl & --\enspace & -- & F & -2.85$^\star$ & 96.3 & WTTS & YSO & - & 13831 & \\
64 &  18:18:39.18 & -13:46:04.93 & 17.517 & 2.349 & 1.693 & --\hspace{6pt} & --\hspace{6pt} & A10 & dl & --\enspace & -- & F & -2.48$^\star$ & 96.8 & WTTS & YSO & - & 16587 & \\
65 &  18:18:59.83 & -13:46:10.57 & 17.462 & 2.232 & 1.742 & --\hspace{6pt} & --\hspace{6pt} & B & dl & --\enspace & -- & F & -2.11$^\star$ & 90.7 & WTTS & YSO & - & 13136 & \\
66 &  18:18:40.09 & -13:46:52.67 & 17.529 & 2.294 & 1.353 & --\hspace{6pt} & --\hspace{6pt} & A11 & db & --\enspace & II & F & -0.84 & 94.5 & CTTS$^\star$ & YSO & - & 18221 & \\
67 &  18:19:06.19 & -13:44:28.59 & 17.647 & 2.445 & 1.583 & --\hspace{6pt} & --\hspace{6pt} & A10 & db & --\enspace & III & F & -1.11 & 83.8 & CTTS$^\star$ & YSO & - & 4534 & \\
68 &  18:18:52.41 & -13:46:58.50 & 17.718 & 2.426 & 1.714 & --\hspace{6pt} & --\hspace{6pt} & A11 & db & --\enspace & II & T & -1.59 & 89.4 & CTTS$^\star$ & Em$^\star$ & - & 13051 & \\
69 &  18:18:38.55 & -13:51:10.31 & 17.599 & 2.054 & 1.615 & --\hspace{6pt} & --\hspace{6pt} & B & dl & --\enspace & -- & F & \quad-- & 96.0 & WTTS & YSO & - & 15777 & \\
70 &  18:18:52.13 & -13:48:10.18 & 17.459 & 2.132 & 1.614 & --\hspace{6pt} & --\hspace{6pt} & A11 & db & -7.6 & II & T & -1.38 & 96.3 & CTTS$^\star$ & Em$^\star$ & - & 14681 & \\
71 &  18:19:08.45 & -13:47:52.77 & 18.026 & 2.459 & 1.971 & --\hspace{6pt} & --\hspace{6pt} & B & dl & --\enspace & II & T & -1.12 & 13.9 & CTTS$^\star$ & YSO & - & 12927 & \\
72 &  18:18:55.24 & -13:47:59.55 & 17.781 & 2.039 & 1.625 & --\hspace{6pt} & --\hspace{6pt} & B & dl & --\enspace & -- & F & \quad-- & 96.7 & WTTS & YSO & - & 13730 & \\
73 &  18:18:37.94 & -13:51:56.24 & 18.090 & 2.584 & 1.705 & --\hspace{6pt} & --\hspace{6pt} & B & dl & -16.5 & -- & -- & -2.77$^\star$ & 73.1 & IRR & YSO & - & 15731 & \\
74 &  18:19:00.21 & -13:45:39.97 & 17.706 & 2.470 & 1.562 & --\hspace{6pt} & --\hspace{6pt} & B & db & --\enspace & II & F & -1.35 & 0.0 & CTTS$^\star$ & YSO & - & 4389 & \\
75 &  18:19:05.67 & -13:44:44.59 & 17.821 & 2.229 & 1.454 & --\hspace{6pt} & --\hspace{6pt} & A11 & dl & --\enspace & II & F & -1.36 & --\enspace & CTTS$^\star$ & YSOC & - & 6049 & \\
76 &  18:18:38.72 & -13:48:54.01 & 17.834 & 2.156 & 1.556 & --\hspace{6pt} & --\hspace{6pt} & B & dl & --\enspace & -- & F & -1.03$^\star$ & 93.2 & CTTS$^\star$ & YSO & - & 16012 & \\
77 &  18:19:05.90 & -13:47:51.79 & 18.116 & 2.369 & 1.664 & --\hspace{6pt} & --\hspace{6pt} & B & db & --\enspace & III & F & -1.73 & 94.2 & CTTS$^\star$ & YSO & - & 12931 & \\
78 &  18:18:42.71 & -13:47:30.24 & 17.948 & 2.289 & 1.322 & --\hspace{6pt} & --\hspace{6pt} & B & db & --\enspace & III & F & -1.81 & 93.8 & WTTS & YSO & - & 16173 & \\
79 &  18:18:55.31 & -13:46:51.72 & 17.963 & 2.454 & 1.454 & --\hspace{6pt} & --\hspace{6pt} & A11 & db & -7.5 & III & F & -1.61 & 86.0 & CTTS$^\star$ & Em$^\star$ & - & 13563 & \\
80 &  18:18:40.48 & -13:47:18.30 & 18.113 & 2.275 & 1.308 & --\hspace{6pt} & --\hspace{6pt} & A01 & db & --\enspace & II & F & -1.27 & 96.3 & CTTS$^\star$ & YSO & - & 16201 & \\
81 &  18:18:37.43 & -13:46:08.03 & 18.083 & 2.678 & -- & --\hspace{6pt} & --\hspace{6pt} & A11 & db & --\enspace & II & T & -1.35 & 90.0 & CTTS$^\star$ & YSO & - & 16593 & \\
82 &  18:18:39.47 & -14:00:34.05 & 17.777 & -- & 1.505 & --\hspace{6pt} & --\hspace{6pt} & A11 & -- & -6.8 & II & T & -0.93 & 96.8 & CTTS$^\star$ & YSOC & - & 16674 & \\
83 &  18:19:11.56 & -13:42:38.43 & 18.512 & 2.514 & -- & --\hspace{6pt} & --\hspace{6pt} & A10 & db & --\enspace & -- & F & -2.33$^\star$ & 96.4 & WTTS & YSO & - & 4725 & \\
84 &  18:19:06.85 & -13:52:40.45 & 18.344 & 2.334 & 1.610 & --\hspace{6pt} & --\hspace{6pt} & B & dl & --\enspace & -- & F & \quad-- & 96.1 & WTTS & YSO & - & 12569 & \\
85 &  18:18:58.61 & -13:49:56.84 & 18.261 & 2.194 & 1.183 & --\hspace{6pt} & --\hspace{6pt} & A10 & db & -20.5 & II & T & -1.38 & 14.4 & CTTS & YSO & - & 12743 & \\
86 &  18:18:47.98 & -13:48:36.25 & 18.797 & 2.675 & -- & --\hspace{6pt} & --\hspace{6pt} & A11 & db & -64.2 & II & T & -0.31 & 95.5 & CTTS & Em$^\star$ & - & 16508 & \\
87 &  18:18:35.56 & -13:45:58.16 & 18.787 & 2.544 & -- & --\hspace{6pt} & --\hspace{6pt} & B & dl & -6.6 & -- & F & \quad-- & 96.5 & WTTS & YSO & - & 10178 & \\
88 &  18:18:30.54 & -13:54:46.01 & 18.602 & 2.248 & -- & --\hspace{6pt} & --\hspace{6pt} & A01 & db & -44.2 & II$^\star$ & T & -1.23 & 85.7 & CTTS & YSO & - &  & \\
89 &  18:18:47.86 & -13:50:01.31 & 18.695 & 2.327 & -- & --\hspace{6pt} & --\hspace{6pt} & B & db & -13.9 & -- & F & -1.26$^\star$ & 0.0 & CTTS$^\star$ & YSO & - & 15894 & \\
90 &  18:18:37.14 & -13:45:05.83 & 18.942 & 2.557 & -- & --\hspace{6pt} & --\hspace{6pt} & B & dl & --\enspace & -- & F & \quad-- & 96.4 & WTTS & YSO & - & 7976 & \\
91 &  18:18:47.91 & -13:55:13.93 & 19.260 & 2.336 & -- & --\hspace{6pt} & --\hspace{6pt} & B & db & -12.9 & III & T & -1.70 & 95.3 & CTTS$^\star$ & YSO & - & 15501 & \\
92 &  18:18:38.69 & -13:44:57.53 & 18.858 & 2.645 & -- & --\hspace{6pt} & --\hspace{6pt} & A10 & db & -6.4 & II & T & -1.30 & 88.2 & CTTS$^\star$ & YSO & - & 8062 & \\
93 &  18:18:32.28 & -13:47:02.71 & 19.516 & 2.582 & -- & --\hspace{6pt} & --\hspace{6pt} & B & dl & --\enspace & -- & -- & \quad-- & 96.1 & WTTS & YSO & - & 16703 & \\
94 &  18:18:42.02 & -13:48:03.67 & --\hspace{6pt} & -- & -- & --\hspace{6pt} & --\hspace{6pt} & A11 & db & --\enspace & -- & -- & -0.70 & --\enspace & IRR & YSO & - & 16119 & \\
95 &  18:19:16.11 & -13:34:57.87 & --\hspace{6pt} & -- & -- & --\hspace{6pt} & --\hspace{6pt} & -- & -- & --\enspace & -- & -- & \quad-- & --\enspace & IRR & -- & -- & 5249 & \\
\noalign{\vskip1pt}\hline
\end{tabular}
 \end{table} 
\end{landscape}

\begin{landscape}
\begin{table}
\tiny 
\centering
\caption{\label{tir}Gaia DR2, 2MASS, UKIDSS and Spitzer infrared  photometry of variable stars.}
\begin{tabular}{@{}rccccccccccccccccc}
\hline
Star & Gaia DR2 ID & G & BP & RP & $J$ & $H$ & $K_S$  & $J$ & $H$ & $K$ & $[3.6]$ & $[4.5]$ & $[5.8]$ & $[8.0]$  & $r$ & $i$ & $H_{\alpha}$\\
 & & mag & mag & mag & mag & mag & mag & mag & mag & mag & mag & mag & mag & mag & mag & mag & mag\\
 (1) & (2) & (3) & (4) & (5) & (6) & (7) & (8) & (9) & (10) & (11) & (12) & (13) & (14) & (15) & (16) & (17) & (18)\\
\noalign{\vskip1pt}\hline\noalign{\vskip1pt}
\multicolumn{18}{l}{Eclipsing binaries}\\
\noalign{\vskip1pt}\noalign{\vskip1pt}
1 & 4146600781797073920 & 9.587 & 9.587 & 10.012 & 8.248 & 8.077 & 7.937 & -- & -- & 11.206 & 7.997 & 7.928 & 7.878 & 7.283 & -- & -- & --\\
2 & 4146600678717290624 & 9.922 & 9.922 & 10.219 & 8.961 & 8.869 & 8.755 & 11.346 & 12.087 & -- & 8.631 & 8.633 & 8.612 & 8.678 & -- & -- & --\\
3 & 4146600708779213312 & 11.304 & 11.304 & 11.920 & 9.476 & 9.119 & 8.900 & 10.436 & 10.764 & 9.691 & 8.742 & 8.744 & 8.669 & 8.831 & -- & -- & --\\
4 & 4152606172568933504 & 12.241 & 12.241 & 12.847 & 10.300 & 9.958 & 9.744 & 10.930 & 11.177 & 10.236 & -- & -- & -- & -- & -- & -- & --\\
5 & 4146614319533501440 & 12.271 & 12.271 & 13.328 & 9.534 & 9.074 & 8.776 & 10.622 & 11.615 & 9.692 & 8.624 & 8.631 & 8.561 & 8.571 & 11.93 & 12.30 & 11.19\\
6 & 4146600640059737088 & 12.526 & 12.526 & 12.945 & 11.048 & 10.725 & 10.598 & 11.175 & 11.181 & 10.643 & 10.597 & 10.562 & 10.543 & 10.291 & 12.14 & 12.40 & 11.79\\
7 & 4146592191861866112 & 12.647 & 12.647 & 13.105 & 11.231 & 10.948 & 10.842 & 11.377 & 11.678 & -- & 10.692 & 10.703 & 10.694 & 10.553 & 12.23 & 12.45 & 11.96\\
8 & 4146601022314634368 & 12.729 & 12.729 & 13.022 & 11.610 & 11.507 & 11.397 & 11.623 & 11.678 & 11.426 & -- & -- & -- & -- & 12.42 & 12.62 & 12.13\\
9 & 4146594219086377728 & 13.326 & 13.326 & 13.752 & 11.907 & 11.569 & 11.339 & 11.853 & 11.695 & 11.492 & -- & -- & -- & -- & 12.91 & 13.14 & 12.55\\
10 & 4146598887718489984 & 13.563 & 13.563 & 13.984 & 12.062 & 11.746 & 11.522 & 11.982 & 11.827 & 11.653 & 11.249 & 11.102 & 11.233 & 11.085 & 13.20 & 13.43 & 12.81\\
11 & 4146600678717298304 & 13.677 & 13.677 & 14.535 & 11.160 & 10.286 & 9.607 & 11.228 & 11.523 & 10.113 & 8.532 & 8.233 & 7.807 & 6.963 & 13.28 & 13.63 & 12.70\\
12 & 4146614353893249024 & 13.904 & 13.904 & 15.687 & 10.043 & 9.256 & 8.656 & 10.558 & 11.256 & 9.622 & -- & -- & -- & -- & -- & -- & --\\
13 & 4146599098169306624 & 14.925 & 14.925 & 15.477 & 13.168 & 12.738 & 12.613 & 13.076 & 12.729 & 12.588 & -- & -- & -- & -- & 14.67 & 14.98 & 14.25\\
14 & 4146597272809633152 & 16.560 & 16.560 & 16.831 & -- & -- & -- & 14.528 & 14.169 & 13.862 & -- & -- & -- & -- & 16.08 & 16.46 & 15.69\\
15 & 4146600640066272512 & 16.722 & 16.722 & 17.289 & 13.739 & 12.809 & 12.481 & 13.702 & 12.910 & 12.603 & -- & -- & -- & -- & 16.39 & 16.83 & 15.70\\
16 & 4146610982347466624 & 17.275 & 17.275 & 17.498 & -- & -- & -- & 15.151 & 14.592 & 14.251 & -- & -- & -- & -- & 16.86 & 17.09 & 16.25\\
17 & 4146613284443719552 & 17.628 & 17.628 & 18.370 & 15.186 & 14.419 & 14.226 & 15.193 & 14.643 & 14.499 & -- & -- & -- & -- & 16.93 & 17.29 & 16.41\\
\noalign{\vskip1pt}\noalign{\vskip1pt}
\multicolumn{18}{l}{Pulsating stars}\\
\noalign{\vskip1pt}\noalign{\vskip1pt}
18 & 4146591470307346944 & 12.690 & 12.690 & 13.138 & 11.316 & 11.084 & 10.960 & 11.452 & 11.738 & -- & 10.859 & 10.853 & 10.944 & 10.760 & 12.29 & 12.48 & 11.93\\
19 & 4146600541278311680 & 12.968 & 12.968 & 13.626 & 10.919 & 10.525 & 10.317 & 11.119 & 11.149 & 10.433 & 10.260 & 10.145 & 10.063 & 10.065 & 12.63 & 12.86 & 12.09\\
20 & 4146612498467320064 & 13.955 & 13.955 & 14.708 & 11.588 & 11.055 & 10.794 & 11.547 & 11.433 & 10.825 & 10.608 & 10.500 & 10.343 & 10.275 & 13.59 & 13.89 & 13.00\\
21 & 4152619061778260992 & 14.052 & 14.052 & 14.536 & 12.512 & 12.199 & 12.077 & 12.435 & 12.250 & 12.099 & -- & -- & -- & -- & -- & -- & --\\
22 & 4152618374583531008 & 14.124 & 14.124 & 14.545 & 12.762 & 12.472 & 12.380 & 12.701 & 12.551 & 12.429 & -- & -- & -- & -- & -- & -- & --\\
23 & 4152618099705619712 & 14.262 & 14.262 & 15.558 & 11.048 & 10.384 & 10.008 & 11.231 & 11.231 & 10.342 & -- & -- & -- & -- & 14.07 & 14.44 & 13.17\\
24 & 4146600644357549440 & 15.105 & 15.105 & 15.591 & 13.310 & 12.916 & 12.208 & 13.262 & 13.046 & 12.740 & -- & -- & -- & -- & 14.75 & 14.97 & 14.29\\
25 & 4146600953595160704 & 15.162 & 15.162 & 15.665 & -- & -- & -- & 13.281 & 13.032 & 12.857 & -- & -- & -- & -- & 14.86 & 15.07 & 14.34\\
\noalign{\vskip1pt}\noalign{\vskip1pt}
\multicolumn{18}{l}{Other periodic variables}\\
\noalign{\vskip1pt}\noalign{\vskip1pt}
26 & 4146596177591442688 & 14.385 & 14.385 & 14.750 & 13.097 & 12.932 & 12.824 & 13.045 & 12.893 & 12.782 & -- & -- & -- & -- & 14.04 & 14.23 & 13.68\\
27 & 4146612326668690816 & 15.073 & 15.073 & 15.604 & 13.338 & 12.973 & 12.785 & 13.287 & 12.995 & 12.822 & -- & -- & -- & -- & 14.71 & 14.92 & 14.26\\
28 & 4146613701058707200 & 15.077 & 15.077 & 15.878 & 12.799 & 12.476 & 12.229 & 12.729 & 12.435 & 12.209 & -- & -- & -- & -- & 14.79 & 15.06 & 14.16\\
29 & 4146600369479609344 & 15.151 & 15.151 & 16.179 & 12.091 & 11.167 & 10.739 & 12.025 & 11.371 & 10.853 & -- & -- & -- & -- & 14.71 & 15.22 & 14.01\\
30 & 4146600983663759104 & 15.557 & 15.557 & 16.575 & 12.462 & 11.540 & 11.209 & 12.282 & 11.628 & 11.158 & -- & -- & -- & -- & 15.26 & 15.72 & 14.50\\
31 & 4146599063809568000 & 15.718 & 15.718 & 16.434 & 13.013 & 12.291 & 11.949 & 13.103 & 12.450 & 12.177 & -- & -- & -- & -- & 15.22 & 15.65 & 14.63\\
32 & 4146612459816239232 & 15.845 & 15.845 & 16.687 & 13.016 & 12.245 & 12.021 & 13.151 & 12.349 & 11.960 & -- & -- & -- & -- & 15.40 & 15.86 & 14.75\\
33 & 4146600365188397568 & 15.872 & 15.872 & 16.708 & 12.735 & 11.851 & 11.426 & 12.625 & 11.816 & 11.407 & 10.922 & 10.724 & 10.635 & 10.323 & 15.48 & 15.96 & 14.72\\
34 & 4146598303597678336 & 15.905 & 15.905 & 16.793 & 13.498 & 12.712 & 12.479 & 13.348 & 12.734 & -- & -- & -- & -- & -- & 15.40 & 15.79 & 14.78\\
35 & 4146594180429043200 & 15.942 & 15.942 & 16.737 & -- & -- & -- & 13.325 & 12.748 & 12.449 & -- & -- & -- & -- & 15.48 & 15.87 & 14.99\\
36 & 4146598028725056128 & 16.799 & 16.799 & 17.070 & 14.231 & 13.382 & 13.132 & 14.086 & 13.384 & 13.075 & -- & -- & -- & -- & 16.40 & 16.83 & 15.89\\
37 & 4146600644357547648 & 17.068 & 17.068 & 17.384 & 13.608 & 12.636 & 12.178 & 13.875 & 13.225 & 12.954 & -- & -- & -- & -- & 16.64 & 17.08 & 15.91\\
38 & 4146600094601679360 & 17.182 & 17.182 & 17.537 & 13.978 & 12.945 & 12.440 & 13.862 & 13.021 & 12.676 & -- & -- & -- & -- & 16.87 & 17.40 & 16.16\\
39 & 4152605897695277568 & 17.389 & 17.389 & 17.998 & 14.180 & 13.255 & 12.942 & 14.074 & 13.280 & 12.915 & -- & -- & -- & -- & 17.11 & 17.55 & 16.27\\
40 & 4146598887718492416 & 17.473 & 17.473 & 17.765 & 14.269 & 13.398 & 13.087 & 14.236 & 13.450 & 13.053 & -- & -- & -- & -- & 17.26 & 17.75 & 16.47\\
41 & 4146597238451051648 & 17.490 & 17.490 & 17.464 & 14.327 & 13.440 & 13.141 & 14.250 & 13.459 & 13.066 & -- & -- & -- & -- & 17.10 & 17.64 & 16.53\\
42 & 4146600850510993664 & 17.593 & 17.593 & 17.870 & 14.593 & 13.681 & 13.431 & 14.403 & 13.614 & 13.263 & -- & -- & -- & -- & 17.25 & 17.80 & 16.57\\
43 & 4146610849199880960 & 17.640 & 17.640 & 18.441 & -- & -- & -- & 13.927 & 13.007 & 12.564 & -- & -- & -- & -- & 17.25 & 17.72 & 16.32\\
44 & 4146600575638068992 & 17.654 & 17.654 & 17.640 & -- & -- & -- & 14.514 & 13.588 & 13.080 & -- & -- & -- & -- & 17.26 & 17.85 & 16.63\\
45 & 4146613524968091392 & 17.959 & 17.959 & 18.139 & 14.175 & 13.245 & 12.723 & 14.202 & 13.271 & 12.717 & -- & -- & -- & -- & -- & -- & --\\
46 & 4146594352231975296 & 18.133 & 18.133 & 17.464 & -- & -- & -- & 14.561 & 13.747 & 13.198 & -- & -- & -- & -- & 17.26 & 17.98 & 16.63\\
\noalign{\vskip1pt}\hline\noalign{\vskip1pt}
\end{tabular}
\end{table} 
\end{landscape}
\begin{landscape}
\begin{table}
\tiny 
\centering
\contcaption{}
\begin{tabular}{@{}rccccccccccccccccc}
\hline
Star & Gaia DR2 & G & BP & RP & $J$ & $H$ & $K_S$  & $J$ & $H$ & $K$ & $[3.6]$ & $[4.5]$ & $[5.8]$ & $[8.0]$  & $r$ & $i$ & $H_{\alpha}$\\
 & & mag & mag & mag & mag & mag & mag & mag & mag & mag & mag & mag & mag & mag & mag & mag & mag\\
 (1) & (2) & (3) & (4) & (5) & (6) & (7) & (8) & (9) & (10) & (11) & (12) & (13) & (14) & (15) & (16) & (17) & (18)\\
\noalign{\vskip1pt}\hline\noalign{\vskip1pt}
\multicolumn{18}{l}{Irregular variables}\\
\noalign{\vskip1pt}
47 & 4146600609998380800 & 13.358 & 13.358 & 13.769 & 11.878 & 11.625 & 11.453 & 11.800 & 11.704 & 11.453 & -- & -- & -- & -- & 13.06 & 13.27 & 12.68\\
48 & 4152618340223791872 & 13.898 & 13.898 & 14.487 & 12.191 & 11.788 & 11.676 & 12.118 & 11.929 & 11.700 & -- & -- & -- & -- & 13.47 & 13.74 & 13.06\\
49 & 4146596315030403456 & 14.377 & 14.377 & 14.957 & -- & -- & -- & 11.616 & 10.891 & 9.394 & 7.736 & 7.146 & 6.428 & 5.465 & -- & -- & --\\
50 & 4146597277103128064 & 14.414 & 14.414 & 15.027 & 12.576 & 12.063 & 11.861 & 12.476 & 12.084 & 11.823 & -- & -- & -- & -- & 13.92 & 14.24 & 13.60\\
51 & 4146600644357558016 & 15.175 & 15.175 & 16.016 & -- & -- & -- & 12.418 & 11.848 & 11.546 & 11.350 & 11.262 & 11.082 & 11.279 & 14.77 & 15.17 & 14.14\\
52 & 4146405481046772224 & 15.248 & 15.248 & 15.726 & 12.462 & 11.253 & 10.155 & 12.563 & 11.488 & 10.272 & 8.916 & 8.597 & 8.115 & 7.284 & 14.73 & 15.50 & 14.78\\
53 & 4146586518207144320 & 15.382 & 15.382 & 15.931 & 13.657 & 13.237 & 13.108 & -- & -- & -- & -- & -- & -- & -- & 14.83 & 15.21 & 14.56\\
54 & 4146599613565335936 & 15.743 & 15.743 & 16.621 & -- & -- & -- & 12.613 & 11.678 & 10.892 & 9.822 & 9.464 & 9.254 & 8.815 & 15.62 & 16.09 & 14.92\\
55 & 4146600300760120704 & 15.922 & 15.922 & 16.639 & 12.661 & 11.550 & 10.723 & 12.695 & 11.802 & 10.889 & 9.437 & 9.002 & 8.644 & 8.055 & 15.30 & 15.98 & 14.82\\
56 & 4152606378727398656 & 15.955 & 15.955 & 16.517 & 12.982 & 11.558 & 10.388 & 12.840 & 11.729 & 10.444 & 8.846 & 8.411 & 7.855 & 6.857 & 15.31 & 15.89 & 15.12\\
57 & 4146596864786208768 & 16.034 & 16.034 & 16.807 & 13.385 & 12.553 & 11.917 & 13.291 & 12.512 & 11.771 & 10.963 & 10.584 & 10.188 & 9.535 & 15.48 & 15.92 & 14.95\\
58 & 4146599166888799488 & 16.113 & 16.113 & 16.785 & -- & -- & -- & -- & 12.555 & -- & -- & -- & -- & -- & 15.34 & 15.98 & 14.80\\
59 & 4152604763821562624 & 16.135 & 16.135 & 16.978 & 13.649 & 12.986 & 12.745 & 13.654 & 13.013 & 12.773 & -- & -- & -- & -- & 15.69 & 16.11 & 15.15\\
60 & 4146600227749527296 & 16.136 & 16.136 & 17.111 & 12.323 & 11.143 & 10.299 & 12.339 & 11.463 & 10.432 & -- & -- & -- & -- & -- & -- & --\\
61 & 4146600506918571392 & 16.227 & 16.227 & 17.231 & 12.826 & 11.819 & 11.082 & 12.831 & 11.885 & 11.063 & 9.735 & 9.233 & 8.670 & 8.098 & 15.93 & 16.39 & 15.13\\
62 & 4146599991522442368 & 16.334 & 16.334 & 17.201 & -- & -- & -- & 13.202 & 12.415 & 12.080 & -- & -- & -- & -- & 16.06 & 16.51 & 15.28\\
63 & 4146600335119864832 & 16.365 & 16.365 & 16.991 & -- & -- & -- & 13.353 & 12.553 & 12.229 & -- & -- & -- & -- & 16.19 & 16.54 & 15.33\\
64 & 4146612429747830144 & 16.401 & 16.401 & 17.213 & 13.341 & 12.399 & 12.011 & 13.225 & 12.437 & 12.064 & -- & -- & -- & -- & 16.14 & 16.55 & 15.36\\
65 & 4146600846224792448 & 16.459 & 16.459 & 17.252 & 13.529 & 12.668 & 12.387 & 13.442 & 12.687 & 12.364 & -- & -- & -- & -- & 16.09 & 16.52 & 15.38\\
66 & 4146600640066263424 & 16.524 & 16.524 & 17.182 & 13.539 & 12.457 & 11.815 & 13.198 & 12.308 & 11.688 & 10.922 & 10.495 & 9.991 & 9.191 & 16.15 & 16.56 & 15.35\\
67 & 4146601022314633856 & 16.524 & 16.524 & 17.371 & 13.509 & 12.611 & 12.183 & 13.404 & 12.574 & 12.112 & -- & -- & -- & -- & 16.09 & 16.57 & 15.38\\
68 & 4146600502621005184 & 16.557 & 16.557 & 17.418 & 13.417 & 12.398 & 11.726 & 13.377 & 12.464 & 11.857 & 10.974 & 10.682 & 10.269 & 9.916 & 16.24 & 16.78 & 15.57\\
69 & 4146598651492680832 & 16.719 & 16.719 & 17.407 & 14.004 & 13.292 & 13.022 & 13.953 & 13.306 & 13.003 & -- & -- & -- & -- & 16.35 & 16.74 & 15.76\\
70 & 4146600335120472576 & 16.750 & 16.750 & 17.115 & 13.466 & 12.484 & 11.745 & 13.821 & 12.752 & 11.903 & 10.961 & 10.490 & 10.144 & 9.680 & 15.79 & 16.39 & 15.30\\
71 & 4146599854083486976 & 16.780 & 16.780 & 17.457 & 13.648 & 12.588 & 11.783 & 13.406 & 12.493 & 11.899 & 11.069 & 10.735 & 10.315 & 9.580 & -- & -- & --\\
72 & 4146599677993670016 & 16.887 & 16.887 & 17.322 & 14.255 & 13.477 & 13.205 & 14.232 & 13.528 & 13.249 & -- & -- & -- & -- & 16.46 & 16.93 & 15.90\\
73 & 4146597517621308160 & 16.900 & 16.900 & 17.684 & 13.584 & 12.576 & 12.238 & 13.469 & 12.690 & -- & -- & -- & -- & -- & 16.35 & 17.13 & 15.80\\
74 & 4146600949304008704 & 16.934 & 16.934 & 17.555 & 13.435 & 12.348 & 11.631 & 13.529 & 12.562 & 11.923 & 11.233 & 10.825 & 10.579 & 9.904 & 16.53 & 17.04 & 15.75\\
75 & 4146601022314630912 & 16.935 & 16.935 & 17.210 & 13.465 & 12.422 & 11.726 & 13.440 & 12.531 & 11.866 & 11.057 & 10.701 & 10.428 & 9.746 & -- & -- & --\\
76 & 4146599063809568512 & 16.938 & 16.938 & 17.547 & 14.236 & 13.320 & 13.159 & 13.972 & 13.267 & 12.929 & -- & -- & -- & -- & 16.61 & 17.05 & 15.91\\
77 & 4146600025882184448 & 17.039 & 17.039 & 17.473 & -- & -- & -- & 13.827 & 12.898 & 12.268 & 11.356 & 11.151 & 11.031 & 10.360 & -- & -- & --\\
78 & 4146600571346805376 & 17.108 & 17.108 & 17.486 & 13.758 & 12.828 & 12.287 & -- & 13.849 & -- & 11.468 & 11.205 & 11.097 & 10.538 & 16.73 & 17.05 & 15.84\\
79 & 4146600468267652736 & 17.124 & 17.124 & 17.603 & 13.681 & 12.546 & 11.851 & 13.492 & 12.535 & 11.878 & 10.896 & 10.459 & 10.162 & 9.812 & 16.55 & 17.22 & 15.96\\
80 & 4146600640066266624 & 17.200 & 17.200 & 17.495 & -- & -- & -- & 14.030 & 13.105 & 12.467 & 11.586 & 11.267 & 10.809 & 10.247 & 16.75 & 17.26 & 16.10\\
81 & 4146612391096769920 & 17.206 & 17.206 & 17.660 & 14.081 & 12.580 & 11.492 & 13.242 & 12.269 & 11.559 & 10.044 & 9.644 & 9.307 & 8.739 & 16.26 & 16.80 & 15.53\\
82 & 4146587652078618752 & 17.211 & 17.211 & 18.091 & 14.260 & 13.106 & 12.251 & -- & -- & -- & 11.256 & 10.828 & 10.440 & 9.582 & 16.22 & 16.76 & 15.77\\
83 & 4152606039437447936 & 17.231 & 17.231 & 18.177 & 14.073 & 13.116 & 12.700 & 13.870 & 13.047 & 12.681 & -- & -- & -- & -- & -- & -- & --\\
84 & 4146595971427931904 & 17.323 & 17.323 & 17.859 & 14.324 & 13.470 & 13.201 & 14.320 & 13.562 & 13.204 & -- & -- & -- & -- & 17.00 & 17.46 & 16.32\\
85 & 4146599300035455488 & 17.423 & 17.423 & 17.505 & 14.182 & 13.062 & 12.272 & 14.147 & 13.189 & 12.399 & 11.335 & 11.063 & 10.686 & 10.080 & 16.71 & 17.38 & 16.37\\
86 & 4146600296467802880 & 17.449 & 17.449 & 17.409 & 13.476 & 12.367 & 11.578 & 13.582 & 12.442 & 11.510 & 11.287 & 10.804 & 10.217 & 9.090 & 16.41 & 17.51 & 16.21\\
87 & 4146612464107579136 & 17.542 & 17.542 & 17.704 & 14.244 & 13.234 & 12.843 & 14.147 & 13.263 & 12.868 & -- & -- & -- & -- & 17.21 & 17.88 & 16.60\\
88 & 4146591397294365184 & 17.708 & 17.708 & 17.953 & 14.417 & 13.127 & 12.185 & 14.424 & -- & -- & -- & -- & -- & -- & 16.98 & 17.86 & 16.79\\
89 & 4146598067372166272 & 17.761 & 17.761 & 17.387 & -- & -- & -- & 14.446 & 13.634 & 13.132 & -- & -- & -- & -- & 17.09 & 17.80 & 16.58\\
90 & 4146612494175982976 & 17.858 & 17.858 & 18.024 & 14.492 & 13.482 & 13.094 & 14.322 & 13.419 & 12.990 & -- & -- & -- & -- & 17.58 & 17.98 & 16.67\\
91 & 4146594184726628992 & 18.001 & 18.001 & 18.092 & 14.636 & 13.468 & 12.539 & 14.457 & 13.496 & 12.741 & 11.432 & 11.314 & 11.133 & 10.435 & 16.81 & 17.47 & 16.35\\
92 & 4146612597255207168 & 18.039 & 18.039 & 17.989 & 13.874 & 12.732 & 12.114 & 14.422 & 13.272 & 12.389 & 11.345 & 11.004 & 10.688 & 9.992 & 17.41 & 18.13 & 16.74\\
93 & 4146610947987721600 & 18.289 & 18.289 & 18.198 & -- & -- & -- & 14.878 & 13.938 & 13.535 & -- & -- & -- & -- & -- & -- & --\\
94 & 4146599093876923264 & 19.608 & 19.608 & 100.000 & 14.241 & 13.210 & 12.484 & -- & -- & -- & -- & -- & -- & -- & -- & -- & --\\
95 & 4152620504873901056 & 20.111 & 20.111 & 20.616 & -- & -- & -- & -- & -- & -- & -- & -- & -- & -- & -- & -- & --\\
\noalign{\vskip1pt}\hline\noalign{\vskip1pt}
\end{tabular}
\end{table} 
\end{landscape}

\setcounter{section}{2}
\begin{figure*}
\centering
\includegraphics[width=42pc]{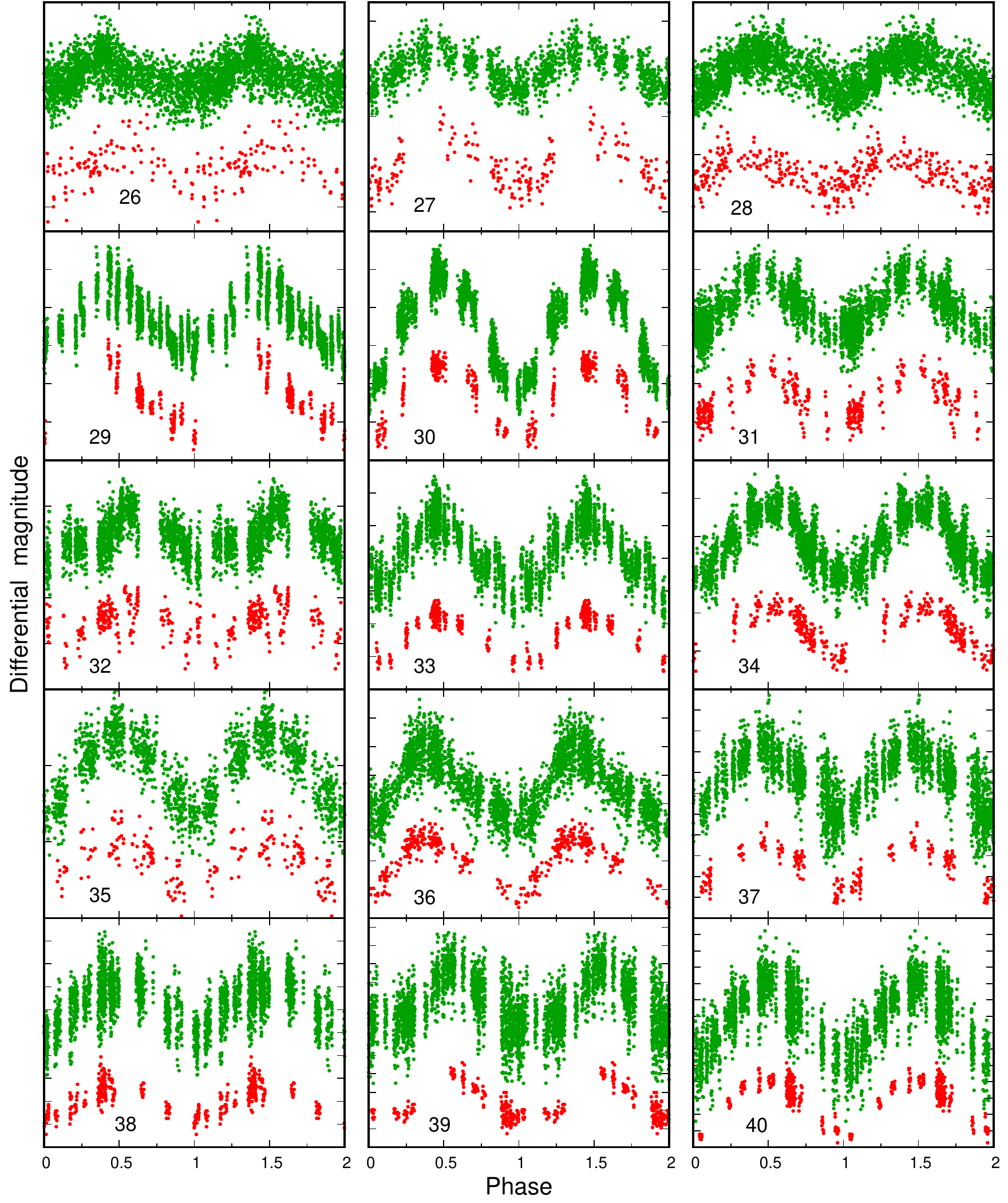}
\caption{Phase diagrams of $B$ (blue), $V$ (green) and $I$ (red) observations of other periodic variables found in the observed field of NGC 6611 (Sect.~4.3). Star numbers are given in the panels. The ordinate ticks are separated by 0.1 mag.} 
\label{figper}
\end{figure*}
\begin{figure*}
\centering
\includegraphics[width=42pc]{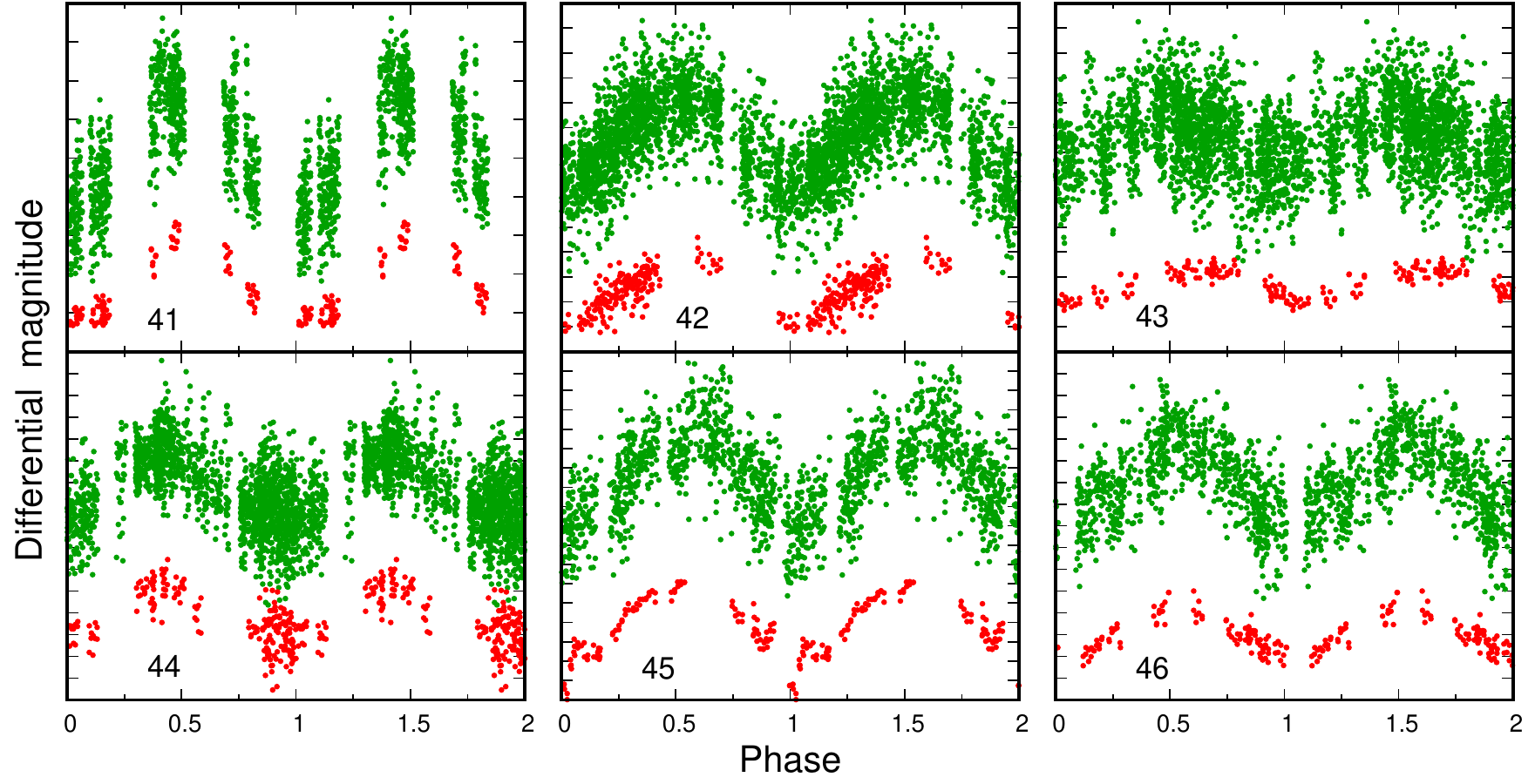}
  \contcaption{}
\end{figure*}

\begin{figure*}
\centering
\includegraphics[width=42pc]{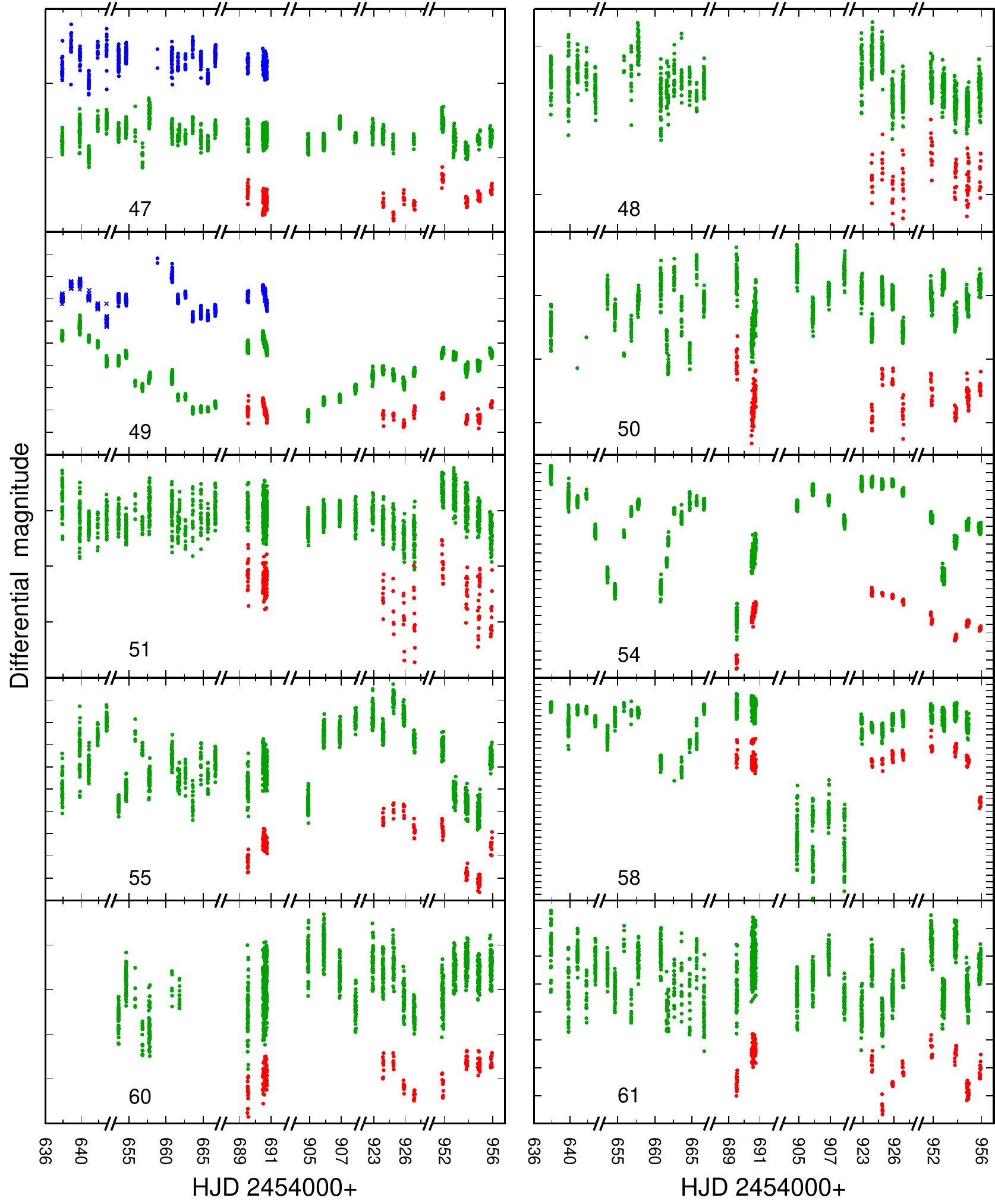}
  \caption{Light curves of irregular variables found in the observed field of NGC6611 (Sect.~4.4) in $B$ (blue), $V$ (green) and $I$ (red) passbands. Star numbers are given in the numbers. The ordinate ticks are separated by 0.1 mag.}
\label{figother1}
\end{figure*}
\begin{figure*}
\centering
\includegraphics[width=42pc]{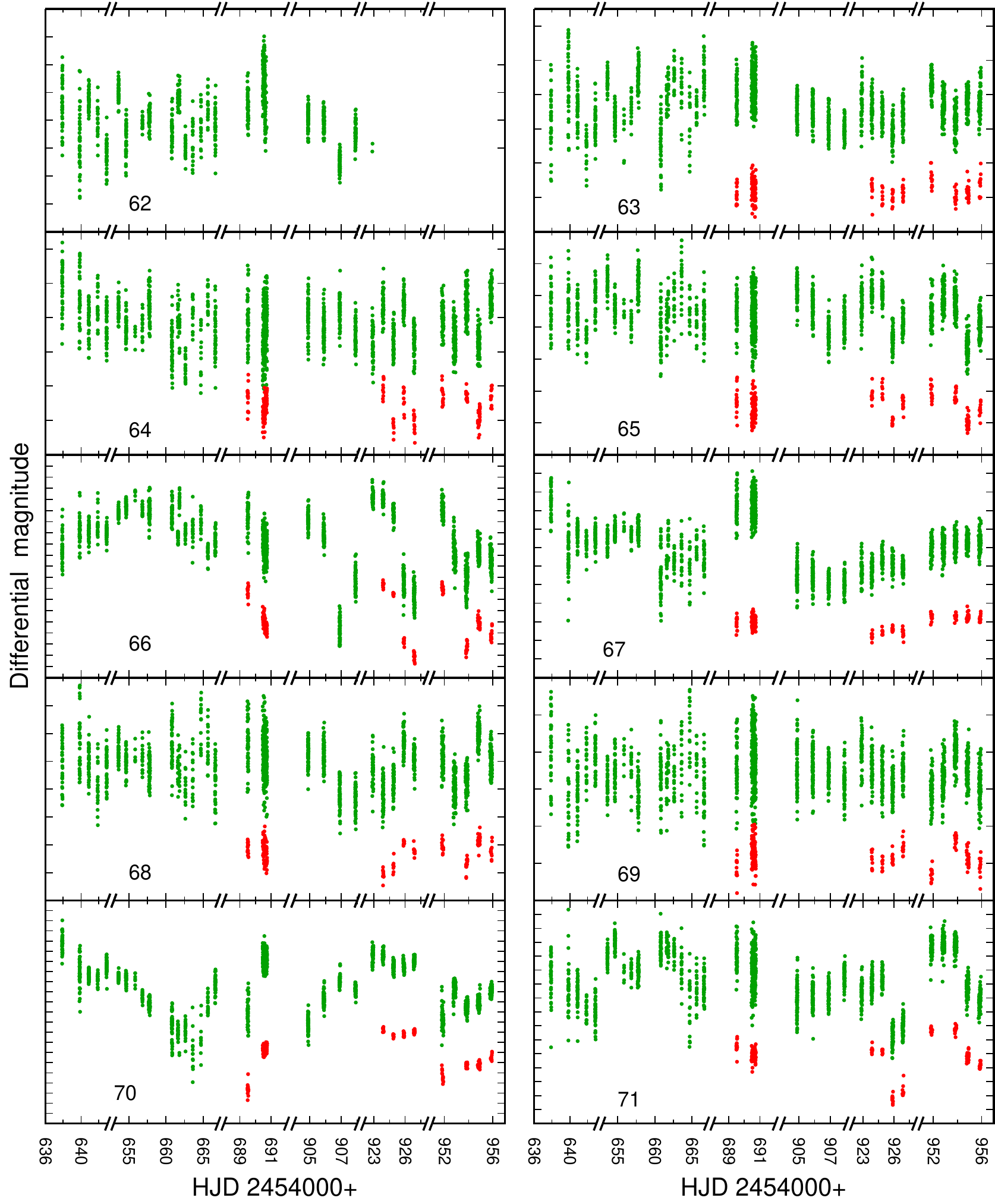}
  \contcaption{}
\end{figure*}
\begin{figure*}
\centering
\includegraphics[width=42pc]{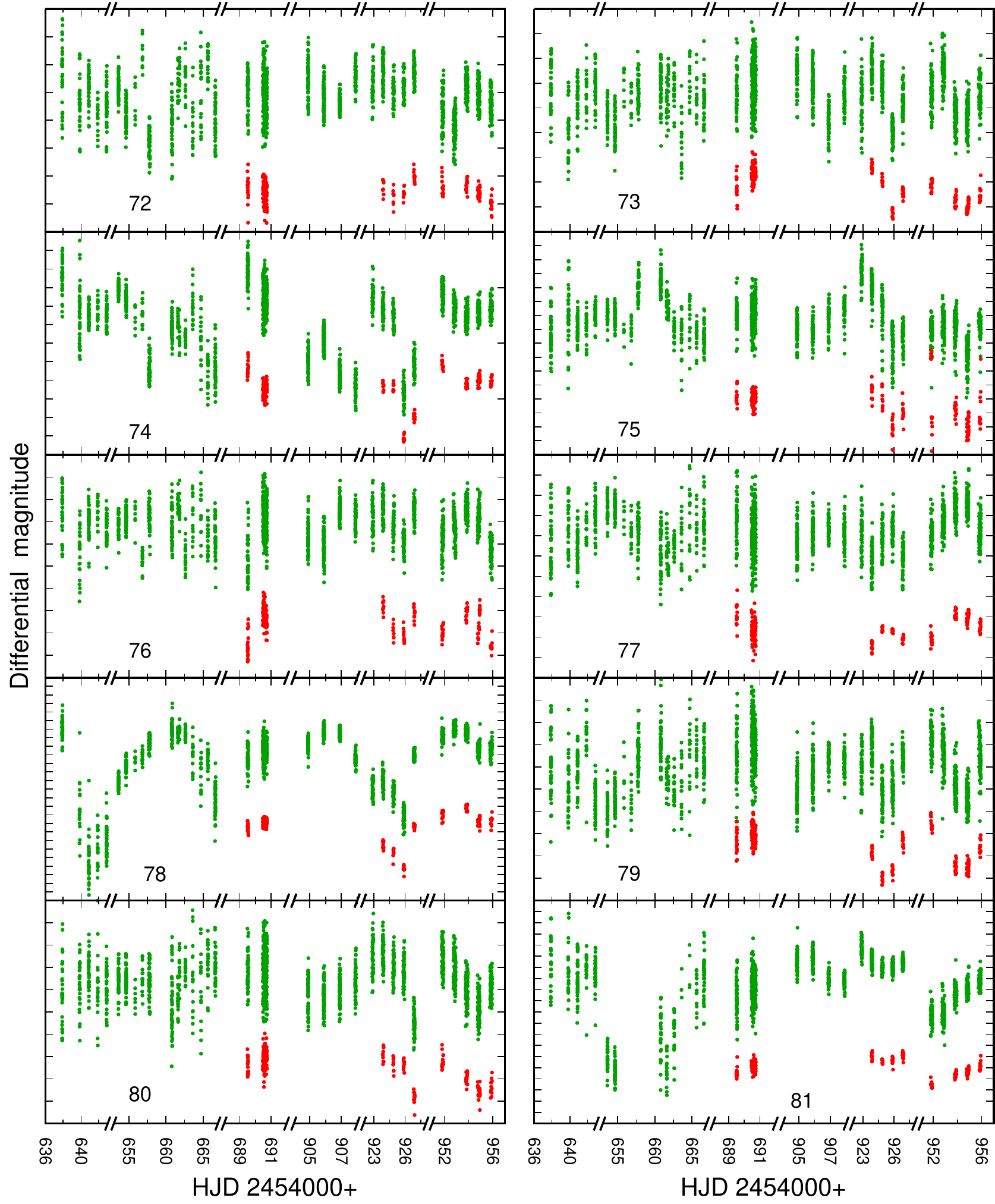}
  \contcaption{}
\end{figure*}
\begin{figure*}
\centering
\includegraphics[width=42pc]{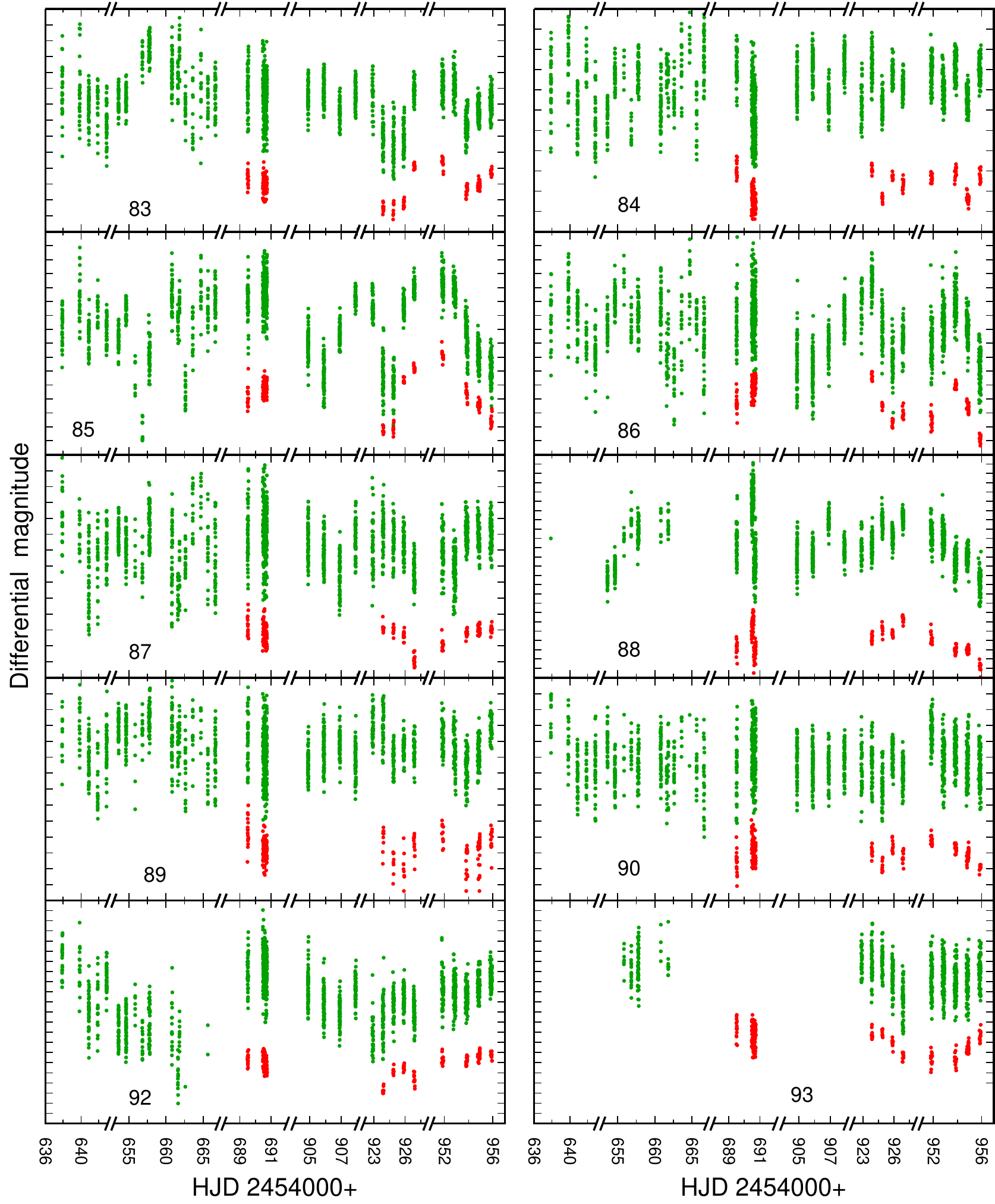}
  \contcaption{}
\end{figure*}
\begin{figure*}
\centering
\includegraphics[width=42pc]{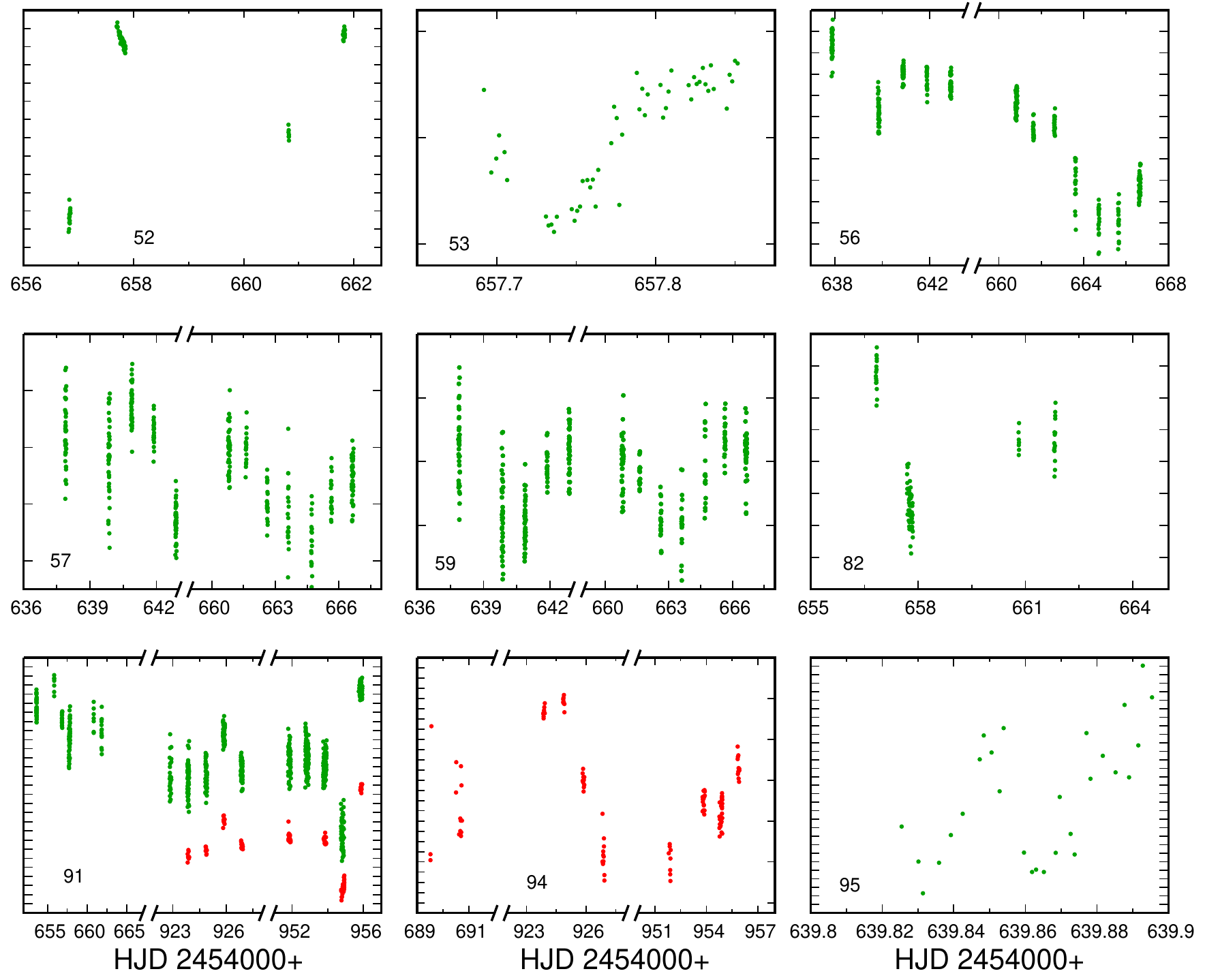}
  \caption{Light curves of $V$ observations of irregular variables observed during only few days field of NGC6611. The ordinate ticks are separated by 0.1 mag.}
\label{figother2}
\end{figure*}
\clearpage

\section*{Acknowledgements}
This research has made use of the VizieR catalogue access tool, CDS,  Strasbourg, France (DOI : 10.26093/cds/vizier). The original description of the VizieR service was published in 2000, A\&AS 143, 23.

This research has made use of the WEBDA database, operated at the Department of Theoretical Physics and Astrophysics of the Masaryk University.

This work has made use of data from the European Space Agency (ESA) mission
{\it Gaia} (\url{https://www.cosmos.esa.int/gaia}), processed by the {\it Gaia}
Data Processing and Analysis Consortium (DPAC,
\url{https://www.cosmos.esa.int/web/gaia/dpac/consortium}). Funding for the DPAC
has been provided by national institutions, in particular the institutions
participating in the {\it Gaia} Multilateral Agreement.

This publication makes use of data products from the Two Micron All Sky Survey, which is a joint project of the University of Massachusetts and the Infrared Processing and Analysis Center/California Institute of Technology, funded by the National Aeronautics and Space Administration and the National Science Foundation.

This research has made use of the NASA/IPAC Infrared Science Archive, which is funded by the National Aeronautics and Space Administration and operated by the California Institute of Technology.

This work is based in part on observations made with the Spitzer Space Telescope, which is operated by the Jet Propulsion Laboratory, California Institute of Technology under a contract with NASA.

Based on data products from observations made with ESO Telescopes
at the La Silla Paranal Observatory under public survey programme ID,
177.D-3023.

This work is based in part on data obtained as part of the UKIRT Infrared Deep Sky Survey.

The observations were supported by Chilean Fondecyt Project 3085010 realised at the Universidad de Concepci\'on.

We are indebted to Prof.~Andrzej Pigulski for his comments made upon reading the manuscript.

\bibliographystyle{mnras}
\bibliography{ngc6611} 

\bsp
\label{lastpage}
\end{document}